\documentclass[english]{article}
\usepackage{mathpazo}
\usepackage[latin9]{inputenc}
\usepackage[letterpaper]{geometry}
\usepackage{babel}
\usepackage{verbatim}
\usepackage{amsmath}
\usepackage{amssymb}
\usepackage{graphicx}
\usepackage[unicode=true,
 bookmarks=true,bookmarksnumbered=false,bookmarksopen=false,
 breaklinks=false,pdfborder={0 0 1},backref=false,colorlinks=false]
 {hyperref}

\makeatletter

\let\SF@@footnote\footnote
\def\footnote{\ifx\protect\@typeset@protect
    \expandafter\SF@@footnote
  \else
    \expandafter\SF@gobble@opt
  \fi
}
\expandafter\def\csname SF@gobble@opt \endcsname{\@ifnextchar[
  \SF@gobble@twobracket
  \@gobble
}
\edef\SF@gobble@opt{\noexpand\protect
  \expandafter\noexpand\csname SF@gobble@opt \endcsname}
\def\SF@gobble@twobracket[#1]#2{}

\numberwithin{equation}{section}

\renewcommand\[{\begin{equation}}
\renewcommand\]{\end{equation}}

\usepackage[margin=1cm]{caption}

\usepackage[svgnames]{xcolor}
\usepackage{tikz}
\usetikzlibrary{decorations.markings}
\usetikzlibrary{shapes.geometric}

\pgfdeclarelayer{edgelayer}
\pgfdeclarelayer{nodelayer}
\pgfsetlayers{edgelayer,nodelayer,main}

\DeclareMathOperator{\diag}{diag}

\DeclareMathOperator{\e}{e}

\DeclareMathOperator{\ii}{i}

\DeclareMathOperator{\tr}{tr}

\makeatother

\begin{document}

\global\long\def\A{\mathbf{A}}%
\global\long\def\B{\mathbf{B}}%
\global\long\def\C{\mathbf{C}}%
\global\long\def\D{\mathbf{D}}%
\global\long\def\E{\mathbf{E}}%
\global\long\def\F{\mathbf{F}}%
\global\long\def\G{\mathbf{G}}%
\global\long\def\H{\mathbf{H}}%
\global\long\def\I{\mathbf{I}}%
\global\long\def\J{\mathbf{J}}%
\global\long\def\K{\mathbf{K}}%
\global\long\def\LL{\mathbf{L}}%
\global\long\def\M{\mathbf{M}}%
\global\long\def\N{\mathbf{N}}%
\global\long\def\OO{\mathbf{O}}%
\global\long\def\P{\mathbf{P}}%
\global\long\def\Q{\mathbf{Q}}%
\global\long\def\RR{\mathbf{R}}%
\global\long\def\SS{\mathbf{S}}%
\global\long\def\T{\mathbf{T}}%
\global\long\def\U{\mathbf{U}}%
\global\long\def\V{\mathbf{V}}%
\global\long\def\W{\mathbf{W}}%
\global\long\def\X{\mathbf{X}}%
\global\long\def\Y{\mathbf{Y}}%
\global\long\def\Z{\mathbf{Z}}%

\global\long\def\a{\mathbf{a}}%
\global\long\def\b{\mathbf{b}}%
\global\long\def\c{\mathbf{c}}%
\global\long\def\dd{\mathbf{d}}%
\global\long\def\ee{\mathbf{e}}%
\global\long\def\f{\mathbf{f}}%
\global\long\def\g{\mathbf{g}}%
\global\long\def\h{\mathbf{h}}%
\global\long\def\iii{\mathbf{i}}%
\global\long\def\j{\mathbf{j}}%
\global\long\def\k{\mathbf{k}}%
\global\long\def\l{\boldsymbol{l}}%
\global\long\def\el{\boldsymbol{\ell}}%
\global\long\def\m{\mathbf{m}}%
\global\long\def\n{\mathbf{n}}%
\global\long\def\o{\mathbf{o}}%
\global\long\def\p{\mathbf{p}}%
\global\long\def\q{\mathbf{q}}%
\global\long\def\r{\mathbf{r}}%
\global\long\def\s{\mathbf{s}}%
\global\long\def\t{\mathbf{t}}%
\global\long\def\u{\mathbf{u}}%
\global\long\def\v{\mathbf{v}}%
\global\long\def\w{\mathbf{w}}%
\global\long\def\x{\mathbf{x}}%
\global\long\def\y{\mathbf{y}}%
\global\long\def\z{\mathbf{z}}%

\global\long\def\Ga{\boldsymbol{\Gamma}}%
\global\long\def\De{\boldsymbol{\Delta}}%
\global\long\def\Th{\boldsymbol{\Theta}}%
\global\long\def\La{\boldsymbol{\Lambda}}%
\global\long\def\Xii{\boldsymbol{\Xi}}%
\global\long\def\Pii{\boldsymbol{\Pi}}%
\global\long\def\Si{\boldsymbol{\Sigma}}%
\global\long\def\Ph{\boldsymbol{\Phi}}%
\global\long\def\Ps{\boldsymbol{\Psi}}%
\global\long\def\Om{\boldsymbol{\Omega}}%

\global\long\def\al{\boldsymbol{\alpha}}%
\global\long\def\be{\boldsymbol{\beta}}%
\global\long\def\ga{\boldsymbol{\gamma}}%
\global\long\def\de{\boldsymbol{\delta}}%
\global\long\def\ep{\boldsymbol{\epsilon}}%
\global\long\def\vep{\boldsymbol{\varepsilon}}%
\global\long\def\ze{\boldsymbol{\zeta}}%
\global\long\def\et{\boldsymbol{\eta}}%
\global\long\def\th{\boldsymbol{\theta}}%
\global\long\def\io{\boldsymbol{\iota}}%
\global\long\def\ka{\boldsymbol{\kappa}}%
\global\long\def\la{\boldsymbol{\lambda}}%
\global\long\def\muu{\boldsymbol{\mu}}%
\global\long\def\nuu{\boldsymbol{\nu}}%
\global\long\def\xii{\boldsymbol{\xi}}%
\global\long\def\pii{\boldsymbol{\pi}}%
\global\long\def\rhh{\boldsymbol{\rho}}%
\global\long\def\si{\boldsymbol{\sigma}}%
\global\long\def\ta{\boldsymbol{\tau}}%
\global\long\def\ups{\boldsymbol{\upsilon}}%
\global\long\def\ph{\boldsymbol{\phi}}%
\global\long\def\vph{\boldsymbol{\varphi}}%
\global\long\def\ch{\boldsymbol{\chi}}%
\global\long\def\ps{\boldsymbol{\psi}}%
\global\long\def\om{\boldsymbol{\omega}}%

\global\long\def\AAb{\boldsymbol{\mathcal{A}}}%
\global\long\def\BBb{\boldsymbol{\mathcal{B}}}%
\global\long\def\CCb{\boldsymbol{\mathcal{C}}}%
\global\long\def\DDb{\boldsymbol{\mathcal{D}}}%
\global\long\def\EEb{\boldsymbol{\mathcal{E}}}%
\global\long\def\FFb{\boldsymbol{\mathcal{F}}}%
\global\long\def\GGb{\boldsymbol{\mathcal{G}}}%
\global\long\def\HHb{\boldsymbol{\mathcal{H}}}%
\global\long\def\IIb{\boldsymbol{\mathcal{I}}}%
\global\long\def\JJb{\boldsymbol{\mathcal{J}}}%
\global\long\def\KKb{\boldsymbol{\mathcal{K}}}%
\global\long\def\LLb{\boldsymbol{\mathcal{L}}}%
\global\long\def\MMb{\boldsymbol{\mathcal{M}}}%
\global\long\def\NNb{\boldsymbol{\mathcal{N}}}%
\global\long\def\OOb{\boldsymbol{\mathcal{O}}}%
\global\long\def\PPb{\boldsymbol{\mathcal{P}}}%
\global\long\def\QQb{\boldsymbol{\mathcal{Q}}}%
\global\long\def\RRb{\boldsymbol{\mathcal{R}}}%
\global\long\def\SSb{\boldsymbol{\mathcal{S}}}%
\global\long\def\TTb{\boldsymbol{\mathcal{T}}}%
\global\long\def\UUb{\boldsymbol{\mathcal{U}}}%
\global\long\def\VVb{\boldsymbol{\mathcal{V}}}%
\global\long\def\WWb{\boldsymbol{\mathcal{W}}}%
\global\long\def\XXb{\boldsymbol{\mathcal{X}}}%
\global\long\def\YYb{\boldsymbol{\mathcal{Y}}}%
\global\long\def\ZZb{\boldsymbol{\mathcal{Z}}}%

\global\long\def\Ab{\bar{A}}%
\global\long\def\Bb{\bar{B}}%
\global\long\def\Cb{\bar{C}}%
\global\long\def\Db{\bar{D}}%
\global\long\def\Eb{\bar{E}}%
\global\long\def\Fb{\bar{F}}%
\global\long\def\Gb{\bar{G}}%
\global\long\def\Hb{\bar{H}}%
\global\long\def\Ib{\bar{I}}%
\global\long\def\Jb{\bar{J}}%
\global\long\def\Kb{\bar{K}}%
\global\long\def\Lb{\bar{L}}%
\global\long\def\Mb{\bar{M}}%
\global\long\def\Nb{\bar{N}}%
\global\long\def\Ob{\bar{O}}%
\global\long\def\Pb{\bar{P}}%
\global\long\def\Qb{\bar{Q}}%
\global\long\def\Rb{\bar{R}}%
\global\long\def\Sb{\bar{S}}%
\global\long\def\Tb{\bar{T}}%
\global\long\def\Ub{\bar{U}}%
\global\long\def\Vb{\bar{V}}%
\global\long\def\Wb{\bar{W}}%
\global\long\def\Xb{\bar{X}}%
\global\long\def\Yb{\bar{Y}}%
\global\long\def\Zb{\bar{Z}}%

\global\long\def\ab{\bar{a}}%
\global\long\def\bb{\bar{b}}%
\global\long\def\cb{\bar{c}}%
\global\long\def\db{\bar{d}}%
\global\long\def\eb{\bar{e}}%
\global\long\def\fb{\bar{f}}%
\global\long\def\gb{\bar{g}}%
\global\long\def\hb{\bar{h}}%
\global\long\def\ib{\bar{i}}%
\global\long\def\jb{\bar{j}}%
\global\long\def\kb{\bar{k}}%
\global\long\def\lb{\bar{l}}%
\global\long\def\elb{\bar{\ell}}%
\global\long\def\mb{\bar{m}}%
\global\long\def\nb{\bar{n}}%
\global\long\def\ob{\bar{o}}%
\global\long\def\pb{\bar{p}}%
\global\long\def\qb{\bar{q}}%
\global\long\def\rb{\bar{r}}%
\global\long\def\ssb{\bar{s}}%
\global\long\def\tb{\bar{t}}%
\global\long\def\ub{\bar{u}}%
\global\long\def\vb{\bar{v}}%
\global\long\def\wb{\bar{w}}%
\global\long\def\xb{\bar{x}}%
\global\long\def\yb{\bar{y}}%
\global\long\def\zb{\bar{z}}%

\global\long\def\Gab{\bar{\Gamma}}%
\global\long\def\Deb{\bar{\Delta}}%
\global\long\def\Thb{\bar{\Theta}}%
\global\long\def\Lab{\bar{\Lambda}}%
\global\long\def\Xib{\bar{\Xi}}%
\global\long\def\Pib{\bar{\Pi}}%
\global\long\def\Sib{\bar{\Sigma}}%
\global\long\def\Phb{\bar{\Phi}}%
\global\long\def\Psb{\bar{\Psi}}%
\global\long\def\Thb{\bar{\Theta}}%

\global\long\def\alb{\bar{\alpha}}%
\global\long\def\beb{\bar{\beta}}%
\global\long\def\gab{\bar{\gamma}}%
\global\long\def\deb{\bar{\delta}}%
\global\long\def\epb{\bar{\epsilon}}%
\global\long\def\vepb{\bar{\varepsilon}}%
\global\long\def\zeb{\bar{\zeta}}%
\global\long\def\etb{\bar{\eta}}%
\global\long\def\thb{\bar{\theta}}%
\global\long\def\iob{\bar{\iota}}%
\global\long\def\kab{\bar{\kappa}}%
\global\long\def\lab{\bar{\lambda}}%
\global\long\def\mub{\bar{\mu}}%
\global\long\def\nub{\bar{\nu}}%
\global\long\def\xib{\bar{\xi}}%
\global\long\def\pib{\bar{\pi}}%
\global\long\def\rhb{\bar{\rho}}%
\global\long\def\sib{\bar{\sigma}}%
\global\long\def\tab{\bar{\tau}}%
\global\long\def\upb{\bar{\upsilon}}%
\global\long\def\phb{\bar{\phi}}%
\global\long\def\vphb{\bar{\varphi}}%
\global\long\def\chb{\bar{\chi}}%
\global\long\def\psb{\bar{\psi}}%
\global\long\def\omb{\bar{\omega}}%

\global\long\def\adt{\dot{a}}%
\global\long\def\add{\ddot{a}}%
\global\long\def\bd{\dot{b}}%
\global\long\def\bdd{\ddot{b}}%
\global\long\def\cd{\dot{c}}%
\global\long\def\cdd{\ddot{c}}%
\global\long\def\ddd{\dot{d}}%
\global\long\def\dddd{\ddot{d}}%
\global\long\def\ed{\dot{e}}%
\global\long\def\edd{\ddot{e}}%
\global\long\def\fd{\dot{f}}%
\global\long\def\fdd{\ddot{f}}%
\global\long\def\gd{\dot{g}}%
\global\long\def\gdd{\ddot{g}}%
\global\long\def\hd{\dot{h}}%
\global\long\def\hdd{\ddot{h}}%
\global\long\def\kd{\dot{k}}%
\global\long\def\kdd{\ddot{k}}%
\global\long\def\ld{\dot{l}}%
\global\long\def\ldd{\ddot{l}}%
\global\long\def\eld{\dot{\ell}}%
\global\long\def\eldd{\ddot{\ell}}%
\global\long\def\md{\dot{m}}%
\global\long\def\mdd{\ddot{m}}%
\global\long\def\nd{\dot{n}}%
\global\long\def\ndd{\ddot{n}}%
\global\long\def\od{\dot{o}}%
\global\long\def\odd{\ddot{o}}%
\global\long\def\pd{\dot{p}}%
\global\long\def\pdd{\ddot{p}}%
\global\long\def\qd{\dot{q}}%
\global\long\def\qdd{\ddot{q}}%
\global\long\def\rd{\dot{r}}%
\global\long\def\rdd{\ddot{r}}%
\global\long\def\sd{\dot{s}}%
\global\long\def\sdd{\ddot{s}}%
\global\long\def\td{\dot{t}}%
\global\long\def\tdd{\ddot{t}}%
\global\long\def\ud{\dot{u}}%
\global\long\def\udd{\ddot{u}}%
\global\long\def\vd{\dot{v}}%
\global\long\def\vdd{\ddot{v}}%
\global\long\def\wdt{\dot{w}}%
\global\long\def\wdd{\ddot{w}}%
\global\long\def\xd{\dot{x}}%
\global\long\def\xdd{\ddot{x}}%
\global\long\def\yd{\dot{y}}%
\global\long\def\ydd{\ddot{y}}%
\global\long\def\zd{\dot{z}}%
\global\long\def\zdd{\ddot{z}}%

\global\long\def\Adt{\dot{A}}%
\global\long\def\Add{\ddot{A}}%
\global\long\def\Bd{\dot{B}}%
\global\long\def\Bdd{\ddot{B}}%
\global\long\def\Cd{\dot{C}}%
\global\long\def\Cdd{\ddot{C}}%
\global\long\def\Dd{\dot{D}}%
\global\long\def\Ddd{\ddot{D}}%
\global\long\def\Ed{\dot{E}}%
\global\long\def\Edd{\ddot{E}}%
\global\long\def\Fd{\dot{F}}%
\global\long\def\Fdd{\ddot{F}}%
\global\long\def\Gd{\dot{G}}%
\global\long\def\Gdd{\ddot{G}}%
\global\long\def\Hd{\dot{H}}%
\global\long\def\Hdd{\ddot{H}}%
\global\long\def\Id{\dot{I}}%
\global\long\def\Idd{\ddot{I}}%
\global\long\def\Jd{\dot{J}}%
\global\long\def\Jdd{\ddot{J}}%
\global\long\def\Kd{\dot{K}}%
\global\long\def\Kdd{\ddot{K}}%
\global\long\def\Ld{\dot{L}}%
\global\long\def\Ldd{\ddot{L}}%
\global\long\def\Md{\dot{M}}%
\global\long\def\Mdd{\ddot{M}}%
\global\long\def\Nd{\dot{N}}%
\global\long\def\Ndd{\ddot{N}}%
\global\long\def\Od{\dot{O}}%
\global\long\def\Odd{\ddot{O}}%
\global\long\def\Pd{\dot{P}}%
\global\long\def\Pdd{\ddot{P}}%
\global\long\def\Qd{\dot{Q}}%
\global\long\def\Qdd{\ddot{Q}}%
\global\long\def\Rd{\dot{R}}%
\global\long\def\Rdd{\ddot{R}}%
\global\long\def\Sd{\dot{S}}%
\global\long\def\Sdd{\ddot{S}}%
\global\long\def\Td{\dot{T}}%
\global\long\def\Tdd{\ddot{T}}%
\global\long\def\Ud{\dot{U}}%
\global\long\def\Udd{\ddot{U}}%
\global\long\def\Vd{\dot{R}}%
\global\long\def\Vdd{\ddot{R}}%
\global\long\def\Wd{\dot{W}}%
\global\long\def\Wdd{\ddot{W}}%
\global\long\def\Xd{\dot{X}}%
\global\long\def\Xdd{\ddot{X}}%
\global\long\def\Yd{\dot{Y}}%
\global\long\def\Ydd{\ddot{Y}}%
\global\long\def\Zd{\dot{Z}}%
\global\long\def\Zdd{\ddot{Z}}%

\global\long\def\Gad{\dot{\Gamma}}%
\global\long\def\Gadd{\ddot{\Gamma}}%
\global\long\def\Ded{\dot{\Delta}}%
\global\long\def\Dedd{\ddot{\Delta}}%
\global\long\def\Thd{\dot{\Theta}}%
\global\long\def\Thdd{\ddot{\Theta}}%
\global\long\def\Lad{\dot{\Lambda}}%
\global\long\def\Ladd{\ddot{\Lambda}}%
\global\long\def\Xid{\dot{\Xi}}%
\global\long\def\Xidd{\ddot{\Xi}}%
\global\long\def\Pid{\dot{\Pi}}%
\global\long\def\Pidd{\ddot{\Pi}}%
\global\long\def\Sid{\dot{\Sigma}}%
\global\long\def\Sidd{\ddot{\Sigma}}%
\global\long\def\Phd{\dot{\Phi}}%
\global\long\def\Phdd{\ddot{\Phi}}%
\global\long\def\Psd{\dot{\Psi}}%
\global\long\def\Psdd{\ddot{\Psi}}%
\global\long\def\Thd{\dot{\Theta}}%
\global\long\def\Thdd{\ddot{\Theta}}%

\global\long\def\ald{\dot{\alpha}}%
\global\long\def\aldd{\ddot{\alpha}}%
\global\long\def\bed{\dot{\beta}}%
\global\long\def\bedd{\ddot{\beta}}%
\global\long\def\gad{\dot{\gamma}}%
\global\long\def\gadd{\ddot{\gamma}}%
\global\long\def\ded{\dot{\delta}}%
\global\long\def\dedd{\ddot{\delta}}%
\global\long\def\epd{\dot{\epsilon}}%
\global\long\def\epdd{\ddot{\epsilon}}%
\global\long\def\vepd{\dot{\varepsilon}}%
\global\long\def\vepdd{\ddot{\varepsilon}}%
\global\long\def\zed{\dot{\zeta}}%
\global\long\def\zedd{\ddot{\zeta}}%
\global\long\def\etd{\dot{\eta}}%
\global\long\def\etdd{\ddot{\eta}}%
\global\long\def\thd{\dot{\theta}}%
\global\long\def\thdd{\ddot{\theta}}%
\global\long\def\iod{\dot{\iota}}%
\global\long\def\iodd{\ddot{\iota}}%
\global\long\def\kad{\dot{\kappa}}%
\global\long\def\kadd{\ddot{\kappa}}%
\global\long\def\lad{\dot{\lambda}}%
\global\long\def\ladd{\ddot{\lambda}}%
\global\long\def\mud{\dot{\mu}}%
\global\long\def\mudd{\ddot{\mu}}%
\global\long\def\nud{\dot{\nu}}%
\global\long\def\nudd{\ddot{\nu}}%
\global\long\def\xid{\dot{\xi}}%
\global\long\def\xidd{\ddot{\xi}}%
\global\long\def\pid{\dot{\pi}}%
\global\long\def\pidd{\ddot{\pi}}%
\global\long\def\rhod{\dot{\rho}}%
\global\long\def\rhodd{\ddot{\rho}}%
\global\long\def\sid{\dot{\sigma}}%
\global\long\def\sidd{\ddot{\sigma}}%
\global\long\def\tad{\dot{\tau}}%
\global\long\def\tadd{\ddot{\tau}}%
\global\long\def\upd{\dot{\upsilon}}%
\global\long\def\updd{\ddot{\upsilon}}%
\global\long\def\phd{\dot{\phi}}%
\global\long\def\phdd{\ddot{\phi}}%
\global\long\def\vpd{\dot{\varphi}}%
\global\long\def\vpdd{\ddot{\varphi}}%
\global\long\def\chd{\dot{\chi}}%
\global\long\def\chdd{\ddot{\chi}}%
\global\long\def\psd{\dot{\psi}}%
\global\long\def\psdd{\ddot{\psi}}%
\global\long\def\omd{\dot{\omega}}%
\global\long\def\omdd{\ddot{\omega}}%

\global\long\def\BBA{\mathbb{A}}%
\global\long\def\BBB{\mathbb{B}}%
\global\long\def\BBC{\mathbb{C}}%
\global\long\def\BBD{\mathbb{D}}%
\global\long\def\BBE{\mathbb{E}}%
\global\long\def\BBF{\mathbb{F}}%
\global\long\def\BBG{\mathbb{G}}%
\global\long\def\BBH{\mathbb{H}}%
\global\long\def\BBI{\mathbb{I}}%
\global\long\def\BBJ{\mathbb{J}}%
\global\long\def\BBK{\mathbb{K}}%
\global\long\def\BBL{\mathbb{L}}%
\global\long\def\BBM{\mathbb{M}}%
\global\long\def\BBN{\mathbb{N}}%
\global\long\def\BBO{\mathbb{O}}%
\global\long\def\BBP{\mathbb{P}}%
\global\long\def\BBQ{\mathbb{Q}}%
\global\long\def\BBR{\mathbb{R}}%
\global\long\def\BBS{\mathbb{S}}%
\global\long\def\BBT{\mathbb{T}}%
\global\long\def\BBU{\mathbb{U}}%
\global\long\def\BBV{\mathbb{V}}%
\global\long\def\BBW{\mathbb{W}}%
\global\long\def\BBX{\mathbb{X}}%
\global\long\def\BBY{\mathbb{Y}}%
\global\long\def\BBZ{\mathbb{Z}}%

\global\long\def\AA{\mathcal{A}}%
\global\long\def\BB{\mathcal{B}}%
\global\long\def\CC{\mathcal{C}}%
\global\long\def\DD{\mathcal{D}}%
\global\long\def\EE{\mathcal{E}}%
\global\long\def\FF{\mathcal{F}}%
\global\long\def\GG{\mathcal{G}}%
\global\long\def\HH{\mathcal{H}}%
\global\long\def\II{\mathcal{I}}%
\global\long\def\JJ{\mathcal{J}}%
\global\long\def\KK{\mathcal{K}}%
\global\long\def\LLL{\mathcal{L}}%
\global\long\def\MM{\mathcal{M}}%
\global\long\def\NN{\mathcal{N}}%
\global\long\def\OOO{\mathcal{O}}%
\global\long\def\PP{\mathcal{P}}%
\global\long\def\QQ{\mathcal{Q}}%
\global\long\def\RRR{\mathcal{R}}%
\global\long\def\SSS{\mathcal{S}}%
\global\long\def\TT{\mathcal{T}}%
\global\long\def\UU{\mathcal{U}}%
\global\long\def\VV{\mathcal{V}}%
\global\long\def\WW{\mathcal{W}}%
\global\long\def\XX{\mathcal{X}}%
\global\long\def\YY{\mathcal{Y}}%
\global\long\def\ZZ{\mathcal{Z}}%

\global\long\def\At{\tilde{A}}%
\global\long\def\Bt{\tilde{B}}%
\global\long\def\Ct{\tilde{C}}%
\global\long\def\Dt{\tilde{D}}%
\global\long\def\Et{\tilde{E}}%
\global\long\def\Ft{\tilde{F}}%
\global\long\def\Gt{\tilde{G}}%
\global\long\def\Ht{\tilde{H}}%
\global\long\def\It{\tilde{I}}%
\global\long\def\Jt{\tilde{J}}%
\global\long\def\Kt{\tilde{K}}%
\global\long\def\Lt{\tilde{L}}%
\global\long\def\Mt{\tilde{M}}%
\global\long\def\Nt{\tilde{N}}%
\global\long\def\Ot{\tilde{O}}%
\global\long\def\Pt{\tilde{P}}%
\global\long\def\Qt{\tilde{Q}}%
\global\long\def\Rt{\tilde{R}}%
\global\long\def\St{\tilde{S}}%
\global\long\def\Tt{\tilde{T}}%
\global\long\def\Ut{\tilde{U}}%
\global\long\def\Vt{\tilde{V}}%
\global\long\def\Wt{\tilde{W}}%
\global\long\def\Xt{\tilde{X}}%
\global\long\def\Yt{\tilde{Y}}%
\global\long\def\Zt{\tilde{Z}}%

\global\long\def\at{\tilde{a}}%
\global\long\def\bt{\tilde{b}}%
\global\long\def\ct{\tilde{c}}%
\global\long\def\dt{\tilde{d}}%
\global\long\def\eet{\tilde{e}}%
\global\long\def\ft{\tilde{f}}%
\global\long\def\gt{\tilde{g}}%
\global\long\def\hht{\tilde{h}}%
\global\long\def\it{\tilde{i}}%
\global\long\def\jt{\tilde{j}}%
\global\long\def\kt{\tilde{k}}%
\global\long\def\lt{\tilde{l}}%
\global\long\def\elt{\tilde{\ell}}%
\global\long\def\mt{\tilde{m}}%
\global\long\def\nt{\tilde{n}}%
\global\long\def\ot{\tilde{o}}%
\global\long\def\pt{\tilde{p}}%
\global\long\def\qt{\tilde{q}}%
\global\long\def\rt{\tilde{r}}%
\global\long\def\st{\tilde{s}}%
\global\long\def\tt{\tilde{t}}%
\global\long\def\ut{\tilde{u}}%
\global\long\def\vt{\tilde{v}}%
\global\long\def\wt{\tilde{w}}%
\global\long\def\xt{\tilde{x}}%
\global\long\def\yt{\tilde{y}}%
\global\long\def\zt{\tilde{z}}%

\global\long\def\mfA{\mathfrak{A}}%
\global\long\def\mfB{\mathfrak{B}}%
\global\long\def\mfC{\mathfrak{C}}%
\global\long\def\mfD{\mathfrak{D}}%
\global\long\def\mfE{\mathfrak{E}}%
\global\long\def\mfF{\mathfrak{F}}%
\global\long\def\mfG{\mathfrak{G}}%
\global\long\def\mfH{\mathfrak{H}}%
\global\long\def\mfI{\mathfrak{I}}%
\global\long\def\mfJ{\mathfrak{J}}%
\global\long\def\mfK{\mathfrak{K}}%
\global\long\def\mfL{\mathfrak{L}}%
\global\long\def\mfM{\mathfrak{M}}%
\global\long\def\mfN{\mathfrak{N}}%
\global\long\def\mfO{\mathfrak{O}}%
\global\long\def\mfP{\mathfrak{P}}%
\global\long\def\mfQ{\mathfrak{Q}}%
\global\long\def\mfR{\mathfrak{R}}%
\global\long\def\mfS{\mathfrak{S}}%
\global\long\def\mfT{\mathfrak{T}}%
\global\long\def\mfU{\mathfrak{U}}%
\global\long\def\mfV{\mathfrak{V}}%
\global\long\def\mfW{\mathfrak{W}}%
\global\long\def\mfX{\mathfrak{X}}%
\global\long\def\mfY{\mathfrak{Y}}%
\global\long\def\mfZ{\mathfrak{Z}}%
\global\long\def\mfa{\mathfrak{a}}%
\global\long\def\mfb{\mathfrak{b}}%
\global\long\def\mfc{\mathfrak{c}}%
\global\long\def\mfd{\mathfrak{d}}%
\global\long\def\mfe{\mathfrak{e}}%
\global\long\def\mff{\mathfrak{f}}%
\global\long\def\mfg{\mathfrak{g}}%
\global\long\def\mfh{\mathfrak{h}}%
\global\long\def\mfi{\mathfrak{i}}%
\global\long\def\mfj{\mathfrak{j}}%
\global\long\def\mfk{\mathfrak{k}}%
\global\long\def\mfl{\mathfrak{l}}%
\global\long\def\mfm{\mathfrak{m}}%
\global\long\def\mfn{\mathfrak{n}}%
\global\long\def\mfo{\mathfrak{o}}%
\global\long\def\mfp{\mathfrak{p}}%
\global\long\def\mfq{\mathfrak{q}}%
\global\long\def\mfr{\mathfrak{r}}%
\global\long\def\mfs{\mathfrak{s}}%
\global\long\def\mft{\mathfrak{t}}%
\global\long\def\mfu{\mathfrak{u}}%
\global\long\def\mfv{\mathfrak{v}}%
\global\long\def\mfw{\mathfrak{w}}%
\global\long\def\mfx{\mathfrak{x}}%
\global\long\def\mfy{\mathfrak{y}}%
\global\long\def\mfz{\mathfrak{z}}%

\global\long\def\d{\mathrm{d}}%
\global\long\def\DDD{\mathrm{D}}%
\global\long\def\EEE{\mathrm{E}}%
\global\long\def\i{\ii}%
\global\long\def\MMM{\mathrm{M}}%
\global\long\def\OOOO{\mathrm{O}}%
\global\long\def\RRRR{\mathrm{R}}%
\global\long\def\TTT{\mathrm{T}}%
\global\long\def\UUU{\mathrm{U}}%

\global\long\def\GL{\mathrm{GL}}%
\global\long\def\ISU{\mathrm{ISU}}%
\global\long\def\ISUT{\mathrm{ISU}\left(2\right)}%
\global\long\def\SL{\mathrm{SL}}%
\global\long\def\SO{\mathrm{SO}}%
\global\long\def\SOH{\mathrm{SO}\left(3\right)}%
\global\long\def\SOT{\mathrm{SO}\left(2\right)}%
\global\long\def\Sp{\mathrm{Sp}}%
\global\long\def\SU{\mathrm{SU}}%
\global\long\def\SUT{\mathrm{SU}\left(2\right)}%
\global\long\def\UO{\mathrm{U}\left(1\right)}%
\global\long\def\gl{\mathfrak{gl}}%
\global\long\def\sl{\mathfrak{sl}}%
\global\long\def\sso{\mathfrak{so}}%
\global\long\def\soh{\mathfrak{so}\left(3\right)}%
\global\long\def\su{\mathfrak{su}}%
\global\long\def\sut{\mathfrak{su}\left(2\right)}%
\global\long\def\isut{\mathfrak{isu}\left(2\right)}%

\global\long\def\so{\Rightarrow}%
\global\long\def\os{\Leftarrow}%
\global\long\def\to{\rightarrow}%
\global\long\def\ot{\leftarrow}%
\global\long\def\soo{\Longrightarrow}%
\global\long\def\oos{\Longleftarrow}%
\global\long\def\too{\longrightarrow}%
\global\long\def\oot{\longleftarrow}%
\global\long\def\sos{\Leftrightarrow}%
\global\long\def\tot{\leftrightarrow}%
\global\long\def\soos{\Longleftrightarrow}%
\global\long\def\toot{\longleftrightarrow}%
\global\long\def\mt{\mapsto}%
\global\long\def\mtt{\longmapsto}%
\global\long\def\dn{\downarrow}%
\global\long\def\up{\uparrow}%
\global\long\def\updn{\updownarrow}%
\global\long\def\sea{\searrow}%
\global\long\def\nea{\nearrow}%
\global\long\def\nwa{\nwarrow}%
\global\long\def\swa{\swarrow}%
\global\long\def\hk{\hookrightarrow}%
\global\long\def\kh{\hookleftarrow}%
\global\long\def\soosp{\quad\Longrightarrow\quad}%
\global\long\def\oossp{\quad\Longleftarrow\quad}%
\global\long\def\soossp{\quad\Longleftrightarrow\quad}%

\global\long\def\multibrl#1{\left(#1\right.}%
\global\long\def\multibrr#1{\left.#1\right)}%
\global\long\def\multisql#1{\left[#1\right.}%
\global\long\def\multisqr#1{\left.#1\right]}%
\global\long\def\multicul#1{\left\{  #1\right.}%
\global\long\def\multicur#1{\left.#1\right\}  }%

\global\long\def\bl{\bigl|}%
\global\long\def\bll{\Bigl|}%
\global\long\def\blll{\biggl|}%
\global\long\def\bllll{\Biggl|}%

\global\long\def\ma#1#2{\left\langle #1\thinspace\middle|\thinspace#2\right\rangle }%
\global\long\def\mma#1#2#3{\left\langle #1\thinspace\middle|\thinspace#2\thinspace\middle|\thinspace#3\right\rangle }%
\global\long\def\mc#1#2{\left\{  #1\thinspace\middle|\thinspace#2\right\}  }%
\global\long\def\mmc#1#2#3{\left\{  #1\thinspace\middle|\thinspace#2\thinspace\middle|\thinspace#3\right\}  }%
\global\long\def\mr#1#2{\left(#1\thinspace\middle|\thinspace#2\right) }%
\global\long\def\mmr#1#2#3{\left(#1\thinspace\middle|\thinspace#2\thinspace\middle|\thinspace#3\right)}%

\global\long\def\pr{\parallel}%
\global\long\def\xx{\times}%
\global\long\def\dg{\lyxmathsym{\textdegree}}%
\global\long\def\sp{,\qquad}%
\global\long\def\sq{\square}%
\global\long\def\pt{\propto}%
\global\long\def\lrc{\lrcorner\thinspace}%
\global\long\def\pexp{\overrightarrow{\exp}}%
\global\long\def\dui#1#2#3{#1_{#2}{}^{#3}}%
\global\long\def\udi#1#2#3{#1^{#2}{}_{#3}}%
\global\long\def\pab{\bar{\partial}}%
\global\long\def\zr{\mathbf{0}}%
\global\long\def\on{\mathbf{1}}%
\global\long\def\na{\boldsymbol{\nabla}}%
\global\long\def\hf{\frac{1}{2}}%
\global\long\def\trd{\frac{1}{3}}%
\global\long\def\fr{\frac{1}{4}}%
\global\long\def\ap{\approx}%
\global\long\def\eqm{\overset{?}{=}}%
\global\long\def\fa{\forall}%
\global\long\def\ex{\exists}%
\global\long\def\xxd{\dot{\boldsymbol{x}}}%
\global\long\def\xxdd{\ddot{\boldsymbol{x}}}%
\global\long\def\ept{\tilde{\epsilon}}%

\title{Lectures on Faster-than-Light Travel and Time Travel}
\author{Barak Shoshany\thanks{bshoshany@perimeterinstitute.ca}\\
\\
\emph{Perimeter Institute for Theoretical Physics}\\
\emph{31 Caroline St. N., Waterloo, ON, Canada, N2L 2Y5}}
\maketitle
\begin{abstract}
These lecture notes were prepared for a 25-hour course for advanced
undergraduate students participating in Perimeter Institute's Undergraduate
Summer Program. The lectures cover some of what is currently known
about the possibility of superluminal travel and time travel within
the context of established science, that is, general relativity and
quantum field theory. Previous knowledge of general relativity at
the level of a standard undergraduate-level introductory course is
recommended, but all the relevant material is included for completion
and reference. No previous knowledge of quantum field theory, or anything
else beyond the standard undergraduate curriculum, is required. Advanced
topics in relativity, such as causal structures, the Raychaudhuri
equation, and the energy conditions are presented in detail. Once
the required background is covered, concepts related to faster-than-light
travel and time travel are discussed. After introducing tachyons in
special relativity as a warm-up, exotic spacetime geometries in general
relativity such as warp drives and wormholes are discussed and analyzed,
including their limitations. Time travel paradoxes are also discussed
in detail, including some of their proposed resolutions.
\end{abstract}
\tableofcontents{}

\section{Introduction}
\begin{quote}
``Space is big. Really big. You just won't believe how vastly, hugely,
mind-bogglingly big it is. I mean, you may think it's a long way down
the road to the chemist, but that's just peanuts to space.''

-- Douglas Adams, \emph{The Hitchhiker's Guide to the Galaxy}
\end{quote}
In science fiction, whenever the plot encompasses more than one solar
system, faster-than-light travel is an almost unavoidable necessity.
In reality, however, it seems that space travel is limited by the
speed of light. In fact, even just accelerating to a significant fraction
of the speed of light is already a hard problem by itself, e.g. due
to the huge energy requirements and the danger of high-speed collisions
with interstellar dust. However, this is a problem of engineering,
not physics -- and we may assume that, given sufficiently advanced
technology, it could eventually be solved.

Unfortunately, even once we are able to build and power a spaceship
that can travel at close to the speed of light, the problem still
remains that interstellar distances are measured in light years --
and therefore will take years to complete, no matter how close we
can get to the speed of light. The closest star system, Alpha Centauri,
is roughly 4.4 light years away, and thus the trip would take at least
4.4 years to complete. Other star systems with potentially habitable
exoplanets are located tens, hundreds, or even thousands of light
years away; the diameter of the Milky Way galaxy is estimated at $175\pm25$
thousand light years\footnote{And, for intergalactic travel, the Andromeda galaxy, for example,
is $2.54\pm0.11$ \textbf{millions }of light years away. But let's
solve one problem at a time...}.

For a one-way trip, the long time it takes to reach the destination
may not be an insurmountable obstacle. First of all, humanity is already
planning to send people to Mars, a journey which is estimated to take
around 9 months. Thus it is not inconceivable to send people on a
journey which will take several years, especially if technological
advances make the trip more tolerable.

Second, relativistic time dilation means that, while for an observer
on Earth it would seem that the spaceship takes at least 4.4 years
to complete the trip to Alpha Centauri, an observer on the spaceship
will measure an arbitrarily short proper time on their own clock,
which will get shorter the closer they get to the speed of light\footnote{Recall that the time dilation factor is given (in units where $c\equiv1$)
by $\gamma=1/\sqrt{1-v^{2}}$, which is equal to 1 for $v=0$ but
approaches infinity in the limit $v\to1$, that is, as the velocity
approaches the speed of light.}.

Furthermore, both scientists and science fiction authors have even
imagined journeys lasting hundreds or even thousands of years, while
the passengers are in suspended animation in order for them to survive
the long journey. Others have envisioned generation ships, where only
the distant descendants of the original passengers actually make it
to the destination\footnote{Another possibility is that, by the time humanity develops interstellar
travel, humans will also have much longer lifespans and/or will be
able to inhabit artificial bodies, which would make a journey lasting
decades not seem so long any more.}.

However, while a trip lasting many years may be possible -- and,
indeed, might well be the \textbf{only }way humankind could ever realistically
reach distant star systems, no matter how advanced our technology
becomes -- this kind of trip is only feasible for initial colonization
of distant planets. It is hard to imagine going on vacation to an
exotic resort on the planet Proxima Centauri b in the Alpha Centauri
system, when the journey takes 4.4 years or more. Moreover, due to
the relativistic time dilation mentioned above, when the tourists
finally arrive back on Earth they will discover that all of their
friends and relatives have long ago died of old age -- not a fun
vacation at all!

So far, the scenarios mentioned are safely within the realm of established
science. However, science fiction writers often find them to be too
restrictive. In a typical science-fiction scenario, the captain of
a spaceship near Earth might \textbf{instantaneously }receives news
of an alien attack on a human colony on Proxima Centauri b and, after
a quick 4.4-light-year journey using the ship's warp drive, will arrive
at the exoplanet just in time to stop the aliens\footnote{And, speaking of aliens, most scenarios where aliens from a distant
planet visit Earth assume the aliens are capable of superluminal travel.}. Such scenarios require faster-than-light communication and travel,
both of which are considered by most to be disallowed by the known
laws of physics.

Another prominent staple of science fiction is the concept of time
travel\footnote{In these notes, by ``time travel'' we will always mean travel to
the \textbf{past}. Travel to the future is trivial -- you can just
sit and wait to time to pass, or use special or general relativistic
time dilation to make it pass faster -- and does not violate causality
or create any paradoxes.}. However, any use of a time machine seems to bluntly violate the
principle of causality, and inherently bring upon irreconcilable paradoxes
such as the so-called grandfather paradox. Unfortunately, works of
science fiction which treat time travel paradoxes in a logical and
consistent manner are extremely rare, and this is no surprise given
that even us physicists don't really understand how to make it consistent.
As we will see below, time travel paradoxes may, in fact, be resolved,
but a concrete mathematical model of paradox-free time travel has
not yet been constructed.

In writing these notes, I relied heavily on the three excellent books
on the subject by Visser \cite{Visser}, Lobo \cite{Lobo} and Krasnikov
\cite{Krasnikov}, as well as the popular general relativity textbooks
by Carroll \cite{Carroll}, Wald \cite{Wald}, Hawking \& Ellis \cite{HawkingEllis},
and Poisson \cite{Poisson}. Many of the definitions, proofs and discussions
are based on the material in these books. The reader is also encouraged
to read Baez \& Muniain \cite{BaezMuniain} for a great introduction
to relevant concepts in differential geometry.

Importantly, throughout these notes we will be using \emph{Planck
units}\footnote{Another popular convention in the literature is the \emph{reduced
Planck units} where instead of $G=1$ one takes $8\pi G=1$. This
simplifies some equations (e.g. the Einstein equation becomes $G_{\mu\nu}=T_{\mu\nu}$
instead of $G_{\mu\nu}=8\pi T_{\mu\nu}$) but complicated others (e.g.
the coefficient of $-\d t^{2}$ in the Schwarzschild metric becomes
$1-M/4\pi r$ instead of $1-2M/r$). Here we will take $G=1$ which
seems like the more natural choice, since it continues the trend of
setting fundamental dimensionful constant to 1.} , where $c=G=\hbar=1$, and our metric signature of choice will be
$\left(-,+,+,+\right)$.

\section{An Outline of General Relativity}

\subsection{Basic Concepts}

\subsubsection{Manifolds and Metrics}

Let spacetime be represented by a 4-dimensional \emph{pseudo-Riemannian
manifold }$M$ equipped with a \emph{metric }$\g$ of signature\footnote{This means that $\g$ has one negative and three positive eigenvalues.}
$\left(-,+,+,+\right)$. The metric is a symmetric tensor with components
$g_{\mu\nu}$ in some coordinate system\footnote{We will use a popular abuse of notation where $g_{\mu\nu}$ will be
called ``the metric'', even though the metric is actually a tensor
$\g$ which happens to have these components in some coordinate system.
Similarly, $x^{\mu}$ will be called a ``vector'' even though the
vector is actually $\x$, and so on.} $\left\{ x^{\mu}\right\} $, where $\mu,\nu\in\left\{ 0,1,2,3\right\} $.
It has the \emph{line element}\footnote{\label{fn:Summation}Here we are, of course, using the \emph{Einstein
summation convention}, where we automatically sum over any index which
is repeated twice: once as an \textbf{upper }index and once as a \textbf{lower
}index. So for example, $x_{\mu}y^{\mu}\equiv\sum_{\mu=0}^{3}x_{\mu}y^{\mu}$
and $g_{\mu\nu}x^{\nu}\equiv\sum_{\nu=0}^{3}g_{\mu\nu}x^{\nu}$. The
index which appears twice is said to be \emph{contracted}.}
\[
\d s^{2}=g_{\mu\nu}\thinspace\d x^{\mu}\otimes\d x^{\nu},
\]
and determinant $g\equiv\det\g$.

The expression $\d s^{2}$ may be understood as the infinitesimal
and curved version of the expression $\Delta s^{2}=-\Delta t^{2}+\Delta x^{2}+\Delta y^{2}+\Delta z^{2}$
for the \emph{spacetime interval }between two points in special relativity,
where $\Delta$ represents the coordinate difference between the points.
In a curved spacetime, the difference between points loses its meaning
since one cannot compare vectors at two separate tangent spaces without
performing a parallel transport (see Sec. \ref{subsec:Parallel-Transport}).

The simplest metric is the \emph{Minkowski metric}, which is diagonal:
\[
\eta_{\mu\nu}\equiv\diag\left(-1,1,1,1\right)=\left(\begin{array}{cccc}
-1 & 0 & 0 & 0\\
0 & 1 & 0 & 0\\
0 & 0 & 1 & 0\\
0 & 0 & 0 & 1
\end{array}\right).
\]
It describes a flat spacetime, and its line element, in the Cartesian
coordinates $\left\{ t,x,y,z\right\} $, is simply
\[
\d s^{2}=-\d t^{2}+\d x^{2}+\d y^{2}+\d z^{2}.
\]

\subsubsection{Tangent Spaces and Vectors}

At every point $p\in M$ there is a \emph{tangent space }$T_{p}M$,
consisting of \emph{tangent vectors }to the manifold at that particular
point. For example, if $M$ is a sphere, the tangent space is the
plane which intersects that sphere at exactly one point $p$, and
the tangent vectors are vectors in that plane.

Given two tangent vectors $u^{\mu}$ and $v^{\mu}$, the metric imposes
the \emph{inner product }and \emph{norm squared}
\begin{equation}
\left\langle \u,\v\right\rangle \equiv g_{\mu\nu}u^{\mu}v^{\nu}\sp\left|\v\right|^{2}\equiv\left\langle \v,\v\right\rangle =g_{\mu\nu}v^{\mu}v^{\nu}.\label{eq:inner-product}
\end{equation}
Since the manifold is not Riemannian, the norm squared is not positive-definite;
it can be negative, positive, or even zero for a non-zero vector.
Thus it doesn't really make sense to talk about ``the norm'' $\left|\v\right|$,
since it can be imaginary, and when we say ``the norm'' we will
always mean the norm squared. Given a tangent vector $v^{\mu}$, if
$\left|\v\right|^{2}<0$ it is called \emph{timelike}, if $\left|\v\right|^{2}>0$
it is called \emph{spacelike}, and if $\left|\v\right|^{2}<0$ it
is called \emph{lightlike} or \emph{null}\footnote{Some people use the convention where the metric signature has the
opposite sign, $\left(+,-,-,-\right)$. In this case, the definitions
for timelike and spacelike have the opposite signs as well. Basically,
``timelike'' means ``has a norm squared with the same sign as the
time dimension in the metric signature'', and similarly for ``spacelike''.}.

Note that we have been using upper indices for vectors. We now define
\emph{covectors} (or \emph{1-forms}), which have components with lower
indices, and act on vectors to produce a scalar. The metric relates
a vector to a covector by \emph{lowering an index}, for example $u_{\nu}=g_{\mu\nu}u^{\mu}.$
We then find that the inner product (\ref{eq:inner-product}) may
be written as the product\footnote{This explains why, in Footnote \ref{fn:Summation}, we said that the
summation convention only works if the same index appears once as
an upper index and once as a lower index. The summation simply gives
us the inner product between the contracted vector and covector, or
more generally, between the corresponding component of two tensors.
One cannot contract two upper indices, since the inner product in
that case would require first adding the metric to the expression.} of a vector with a covector: $\left\langle \u,\v\right\rangle \equiv g_{\mu\nu}u^{\mu}v^{\nu}=u_{\nu}v^{\nu}$.
Similarly, for the metric, we have used two lower indices, and wrote
it as a matrix. The corresponding inverse matrix gives us the components
of the \emph{inverse metric} $g^{\mu\nu}$, which satisfies $g^{\mu\lambda}g_{\lambda\nu}=\delta_{\nu}^{\mu}$.

Finally, if we assign a tangent vector to every point in the manifold,
then we have a \emph{vector field} $v^{\mu}\left(\x\right)$. Similarly,
we may talk about a \emph{scalar field} $\phi\left(\x\right)$, which
assigns a number to every point, a covector field $v_{\mu}\left(\x\right)$,
a tensor field $g_{\mu\nu}\left(\x\right)$, and so on.

\subsubsection{Curves and Proper Time}

Let $x^{\mu}\left(\lambda\right)$ be a \emph{curve} (or \emph{path},
or \emph{worldline}) parametrized by some real parameter $\lambda$.
For each value of $\lambda$, $x^{\mu}\left(\lambda\right)$ is a
point on the manifold. This point has a tangent space, and the tangent
vector to the curve, defined as
\[
\xd^{\mu}\equiv\frac{\d x^{\mu}}{\d\lambda},
\]
is a vector in this tangent space.

If $x^{\mu}\left(\lambda\right)$ describes the worldline of a massive
particle, then its tangent vector is timelike everywhere; $\left|\xxd\left(\lambda\right)\right|^{2}<0$
for all $\lambda$. In this case, we may calculate the \emph{proper
time }along the path, which is the \emph{coordinate-independent} time
as measured by a clock moving with the particle, and thus is an observable;
in contrast, the \emph{coordinate time }$x^{0}\equiv t$ is not an
observable since it, of course, depends on the arbitrary choice of
coordinates.

The differential of proper time is defined as minus the line element,
that is, $\d\tau^{2}\equiv-\d s^{2}$ or\footnote{The minus sign comes from the fact that the time dimension has a minus
sign in the metric signature. Thus, if we want the difference in proper
time between a past event and a future event to be positive, we must
cancel that minus sign.}
\[
\d\tau^{2}=-g_{\mu\nu}\thinspace\d x^{\mu}\otimes\d x^{\nu}.
\]
Employing a slight abuse of notation, we may ``divide'' this expression
by $\d\lambda^{2}$ where $\lambda$ is the (arbitrary) curve parameter:
\begin{equation}
\frac{\d\tau^{2}}{\d\lambda^{2}}=\left(\frac{\d\tau}{\d\lambda}\right)^{2}=-g_{\mu\nu}\thinspace\frac{\d x^{\mu}}{\d\lambda}\frac{\d x^{\nu}}{\d\lambda}=-g_{\mu\nu}\thinspace\xd^{\mu}\xd^{\nu}=-\left|\xxd\right|^{2}.\label{eq:dtau-dlambda}
\end{equation}
Thus, we learn that
\[
\frac{\d\tau}{\d\lambda}=\sqrt{-\left|\xxd\right|^{2}}\soosp\d\tau=\sqrt{-\left|\xxd\right|^{2}}\thinspace\d\lambda,
\]
where the expression inside the square root is positive since $\left|\xxd\right|^{2}<0$
for a timelike path. Now we can find the total proper time $\tau$
along a path (that is, from one value of $\lambda$ to a subsequent
value of $\lambda$) by integrating this differential:
\begin{equation}
\tau\equiv\int\d\tau=\int\sqrt{-\left|\xxd\right|^{2}}\thinspace\d\lambda=\int\sqrt{-g_{\mu\nu}\xd^{\mu}\xd^{\nu}}\thinspace\d\lambda.\label{eq:proper-time}
\end{equation}
This allows us, in principle, to calculate $\tau$ as a function of
$\lambda$, and then use the proper time in place of $\lambda$ as
the parameter for our curve. When using the proper time as a parameter
-- and \textbf{only} then! -- the tangent vector $\xd^{\mu}\equiv\d x^{\mu}/\d\tau$
is called the \emph{4-velocity}, and it is automatically normalized\footnote{Indeed, using the chain rule and (\ref{eq:dtau-dlambda}), we have
\[
\left|\xxd\right|^{2}=g_{\mu\nu}\frac{\d x^{\mu}}{\d\tau}\frac{\d x^{\nu}}{\d\tau}=\frac{g_{\mu\nu}\frac{\d x^{\mu}}{\d\lambda}\frac{\d x^{\nu}}{\d\lambda}}{\left(\frac{\d\tau}{\d\lambda}\right)^{2}}=\frac{g_{\mu\nu}\frac{\d x^{\mu}}{\d\lambda}\frac{\d x^{\nu}}{\d\lambda}}{-g_{\mu\nu}\thinspace\frac{\d x^{\mu}}{\d\lambda}\frac{\d x^{\nu}}{\d\lambda}}=-1.
\]
} to $\left|\xxd\right|^{2}=-1$. From now on, unless stated otherwise,
we will \textbf{always} parametrize timelike paths with proper time,
since the normalization $\left|\xxd\right|^{2}=-1$ simplifies many
calculations, and allows us to talk about 4-velocity in a well-defined
way.

\subsubsection{Massless Particles}

So far we have discussed massive particles, whose worldlines are timelike.
The worldline of a massless particle, on the other hand, is a null
path. For a null path we \textbf{cannot }define a proper time, since
by definition its tangent is a null vector, so $\left|\xxd\left(\lambda\right)\right|^{2}=0$
for any choice of parameter $\lambda$ (although, of course, the tangent
vector itself is not the zero vector). Therefore, from (\ref{eq:proper-time})
we have that $\tau=0$ between any two points on the path. This is
why we sometimes say that massless particles, such as photons, ``do
not experience the passage of time''; the proper time along their
worldlines vanishes.

This means that for null paths, there is no preferred parameter; however,
as we will see below, null geodesics possess a family of preferred
parameters called \emph{affine parameters}. Furthermore, note that
massless particles do \textbf{not }have a well-defined 4-velocity,
since the definition of 4-velocity makes use of the proper time.

\subsection{Covariant Derivatives and Connections}

\subsubsection{Defining Tensors}

A function $\T$ defined on a manifold is called a \emph{tensor }of
\emph{rank} $\left(p,q\right)$ if its components have $p$ upper
indices and $q$ lower indices: $T_{\nu_{1}\cdots\nu_{q}}^{\mu_{1}\cdots\mu_{p}}$,
and if, under a transformation from a coordinate system $\{x^{\mu}\}$
to another one $\{x^{\mu'}\}$, each upper index $\mu_{i}$ receives
a factor of $\partial x^{\mu_{i}'}/\partial x^{\mu_{i}}$ and each
lower index $\nu_{i}$ receives a factor of $\partial x^{\nu_{i}}/\partial x^{\nu_{i}'}$
(note that the original and primed coordinates switch places). Vectors
and covectors are specific cases of tensors, with rank $\left(1,0\right)$
and $\left(0,1\right)$ respectively.

For example, the components of a vector $v^{\mu}$, a covector $u_{\mu}$
and a rank $\left(0,2\right)$ tensor $g_{\mu\nu}$ transform as follows\footnote{You don't need to remember where the primes are placed, since they
are located in the only place that makes sense for contracting with
the components of the tensor, taking into account that an \textbf{upper
}index in the denominator, e.g. the $\mu$ on $\partial/\partial x^{\mu}$,
counts as a \textbf{lower }index -- since $\partial_{\mu}\equiv\partial/\partial x^{\mu}$
-- and thus should be contracted with an \textbf{upper }index.}:
\[
v^{\mu'}=\frac{\partial x^{\mu'}}{\partial x^{\mu}}v^{\mu}\sp u_{\mu'}=\frac{\partial x^{\mu}}{\partial x^{\mu'}}u_{\mu}\sp g_{\mu'\nu'}=\frac{\partial x^{\mu}}{\partial x^{\mu'}}\frac{\partial x^{\nu}}{\partial x^{\nu'}}g_{\mu\nu}.
\]
If the tensors transforms this way, we say that they are \emph{covariant
}or \emph{transform covariantly}. This is very important, since the
\textbf{abstract }quantities $\v$, $\u$ and $\g$ are guaranteed
to be \emph{invariant }under the transformation if and only if their
\textbf{components }transform exactly in this way.

\subsubsection{Covariant Derivatives}

The standard way to define differentiation on a manifold is by using
the \emph{covariant derivative} $\nabla_{\mu}$, which generalizes
the usual partial derivative $\partial_{\mu}\equiv\partial/\partial x^{\mu}$
and is defined\footnote{Some sources use a notation where a comma indicates partial derivatives:
$\partial_{\mu}u_{\nu}\equiv u_{\nu,\mu}$, and a semicolon indicates
covariant derivatives: $\nabla_{\mu}u_{\nu}\equiv u_{\nu;\mu}$. We
will not use this notation here.} as follows:
\begin{itemize}
\item On a scalar field $\phi$: $\nabla_{\mu}\phi\equiv\partial_{\mu}\phi$.
\item On a vector field $v^{\nu}$: $\nabla_{\mu}v^{\nu}\equiv\partial_{\mu}v^{\nu}+\Gamma_{\mu\lambda}^{\nu}v^{\lambda}$.
\item On a covector $u_{\nu}$: $\nabla_{\mu}u_{\nu}\equiv\partial_{\mu}u_{\nu}-\Gamma_{\mu\nu}^{\lambda}u_{\lambda}$.
\item On a rank $\left(p,q\right)$ tensor $\T$ with components $T_{\sigma_{1}\cdots\sigma_{q}}^{\nu_{1}\cdots\nu_{p}}$:
The first term of $\nabla_{\mu}T_{\sigma_{1}\cdots\sigma_{q}}^{\nu_{1}\cdots\nu_{p}}$
will be $\partial_{\mu}T_{\sigma_{1}\cdots\sigma_{q}}^{\nu_{1}\cdots\nu_{p}}$,
and to that we \textbf{add} one term $\Gamma_{\mu\lambda}^{\nu_{i}}T_{\sigma_{1}\cdots\sigma_{q}}^{\nu_{1}\cdots\lambda\cdots\nu_{p}}$
for each upper index $\nu_{i}$, and \textbf{subtract }one term $\Gamma_{\mu\sigma_{i}}^{\lambda}T_{\sigma_{1}\cdots\lambda\cdots\sigma_{q}}^{\nu_{1}\cdots\nu_{p}}$
for each lower index $\sigma_{i}$, generalizing the expressions above.
\end{itemize}
$\Gamma_{\mu\nu}^{\lambda}$ are called the \emph{connection coefficients}.
Importantly, $\Gamma_{\mu\nu}^{\lambda}$ are \textbf{not }the components
of a tensor, since they do \textbf{not} transform covariantly under
a change of coordinates. However, the partial derivative itself does
not transform as a tensor either, and it just so happens that the
unwanted terms from each of the transformations exactly cancel each
others, so while $\partial_{\mu}T_{\sigma_{1}\cdots\sigma_{q}}^{\nu_{1}\cdots\nu_{p}}$
is \textbf{not }a tensor, the covariant derivative $\nabla_{\mu}T_{\sigma_{1}\cdots\sigma_{q}}^{\nu_{1}\cdots\nu_{p}}$
of any tensor \textbf{is }itself a tensor, of rank $\left(p,q+1\right)$.
This is exactly why covariant derivatives are so important: partial
derivatives do not transform covariantly in a general curved spacetime,
and thus do not generate tensors, but covariant derivatives do.

\subsubsection{The Levi-Civita Connection}

Just like the partial derivative, the covariant derivative satisfies
linearity and the Leibniz rule. It also commutes with contractions,
meaning that if we contract two indices of a tensor, e.g. $\udi T{\mu\lambda}{\lambda\nu}\equiv g_{\lambda\rho}\udi T{\mu\lambda\rho}{\nu}$,
and then apply the covariant derivative, then in the new tensor (of
one higher rank) that we have formed, the same two indices will still
be contracted: $\nabla_{\sigma}\udi T{\mu\lambda}{\lambda\nu}=g_{\lambda\rho}\nabla_{\sigma}\udi T{\mu\lambda\rho}{\nu}$.
However, the covariant derivative is not unique since it depends on
the choice of $\Gamma_{\mu\nu}^{\lambda}$. A \textbf{unique }choice
of connection coefficients may be obtained by requiring that the connection
is also:
\begin{itemize}
\item \emph{Torsion-free}, meaning that the \emph{torsion tensor} $\udi T{\lambda}{\mu\nu}\equiv\Gamma_{\mu\nu}^{\lambda}-\Gamma_{\nu\mu}^{\lambda}\equiv2\Gamma_{[\mu\nu]}^{\lambda}$
vanishes, or equivalently, the connection coefficients are symmetric
in their lower indices: $\Gamma_{\mu\nu}^{\lambda}=\Gamma_{\nu\mu}^{\lambda}$.
\item Metric-compatible, that is, the covariant derivative of the metric
vanishes: $\nabla_{\lambda}g_{\mu\nu}=0$. Note that this is a stronger
condition than commuting with contractions, since we can also write
$\udi T{\mu\lambda}{\lambda\nu}\equiv\delta_{\lambda}^{\rho}\udi T{\mu\lambda}{\rho\nu}$,
so commuting with contractions merely implies that $\nabla_{\mu}\delta_{\lambda}^{\rho}=0$,
which one would expect given that the Kronecker delta $\delta_{\lambda}^{\rho}$
is just the identity matrix.
\end{itemize}
With these additional constraints, there is one unique connection,
called the \emph{Levi-Civita connection}, whose coefficients are sometimes
called \emph{Christoffel symbols}, and are given as a function of
the metric by:
\begin{equation}
\Gamma_{\mu\nu}^{\lambda}=\hf g^{\lambda\sigma}\left(\partial_{\mu}g_{\nu\sigma}+\partial_{\nu}g_{\mu\sigma}-\partial_{\sigma}g_{\mu\nu}\right).\label{eq:connection-coeff}
\end{equation}
The reader is encouraged to prove this and all other unproven claims
in this section.

\subsection{Parallel Transport and Geodesics}

\subsubsection{\label{subsec:Parallel-Transport}Parallel Transport}

Since each point on the manifold has its own tangent space, it is
unclear how to relate vectors (or more generally, tensors) at different
points, since they belong to different vector spaces. Again, imagine
the sphere, with the tangent spaces being planes which touch it only
at one point. A tangent vector to one plane, if moved to another point,
will not be tangent to the plane at that point.

A unique way of taking a tensor from one point to another point on
the manifold is given by \emph{parallel transport}. Let $x^{\mu}\left(\lambda\right)$
be a curve, and let $v^{\mu}$ be a vector whose value is known on
a particular point on the curve (e.g. at $\lambda=0$). To parallel
transport $v^{\mu}$ to another point along the curve (e.g. $\lambda=1$),
we solve the equation
\[
\xd^{\mu}\nabla_{\mu}v^{\nu}=0,
\]
where $\xd^{\mu}\equiv\d x^{\mu}/\d\lambda$. Using the definition
of the covariant derivative, this may be written explicitly as
\[
\xd^{\mu}\partial_{\mu}v^{\sigma}+\Gamma_{\mu\nu}^{\sigma}\xd^{\mu}v^{\nu}=0.
\]
Furthermore, using the chain rule $\frac{\d x^{\mu}}{\d\lambda}\partial_{\mu}=\frac{\d}{\d\lambda}$,
we may write this as
\[
\frac{\d}{\d\lambda}v^{\sigma}+\Gamma_{\mu\nu}^{\sigma}\xd^{\mu}v^{\nu}=0.
\]
Similar parallel transport equations are easily obtained for tensors
of any rank.

\subsubsection{Geodesics and Affine Parameters}

Let us apply parallel transport to the tangent vector to the same
curve we are parallel-transporting along, that is, take $v^{\nu}\equiv\xd^{\nu}$.
Then we get:
\begin{equation}
\xd^{\mu}\nabla_{\mu}\xd^{\sigma}=0\soosp\xdd^{\sigma}+\Gamma_{\mu\nu}^{\sigma}\xd^{\mu}\xd^{\nu}=0.\label{eq:geodesic-eq}
\end{equation}
This equation is called the \emph{geodesic equation}, and it generalizes
the notion of a ``straight line'' to a curved space. Indeed, for
the flat Minkowski space, we have $\Gamma_{\mu\nu}^{\sigma}=0$ and
thus the geodesics are given by curves satisfying $\xxdd=0$, which
describe straight lines.

Equation (\ref{eq:geodesic-eq}) demands that the vector tangent to
the curve is parallel transported in the direction of the curve. This
means that the resulting vector must have the same direction and magnitude.
We can, in fact, weaken this condition and allow the resulting vector
to have a different magnitude, while still demanding that its direction
remains unchanged. This still captures the idea behind geodesics,
namely that they are a generalization of straight lines in flat space.
The resulting equation takes the form:
\begin{equation}
\xd^{\mu}\nabla_{\mu}\xd^{\sigma}=\alpha\xd^{\sigma},\label{eq:geodesic-general}
\end{equation}
where $\alpha$ is some function on the curve. However, any curve
which satisfies this equation may be reparametrized so that it satisfies
(\ref{eq:geodesic-eq}). Indeed, (\ref{eq:geodesic-general}) may
be written explicitly as follows:
\[
\frac{\d^{2}x^{\sigma}}{\d\lambda^{2}}+\Gamma_{\mu\nu}^{\sigma}\frac{\d x^{\mu}}{\d\lambda}\frac{\d x^{\nu}}{\d\lambda}=\alpha\frac{\d x^{\sigma}}{\d\lambda}.
\]
Now, consider a curve which solves (\ref{eq:geodesic-general}) with
some parameter $\lambda$ and introduce a new parameter $\mu\left(\lambda\right)$.
Then we have
\[
\frac{\d}{\d\lambda}=\frac{\d\mu}{\d\lambda}\frac{\d}{\d\mu}\soosp\frac{\d^{2}}{\d\lambda^{2}}=\frac{\d^{2}\mu}{\d\lambda^{2}}\frac{\d}{\d\mu}+\left(\frac{\d\mu}{\d\lambda}\right)^{2}\frac{\d^{2}}{\d\mu^{2}},
\]
and we may rewrite the equation as follows:
\[
\frac{\d^{2}\mu}{\d\lambda^{2}}\frac{\d x^{\sigma}}{\d\mu}+\left(\frac{\d\mu}{\d\lambda}\right)^{2}\frac{\d^{2}x^{\sigma}}{\d\mu^{2}}+\Gamma_{\mu\nu}^{\sigma}\left(\frac{\d\mu}{\d\lambda}\right)^{2}\frac{\d x^{\mu}}{\d\mu}\frac{\d x^{\nu}}{\d\mu}=\alpha\frac{\d\mu}{\d\lambda}\frac{\d x^{\sigma}}{\d\mu}.
\]
Rearranging, we get
\[
\frac{\d^{2}x^{\sigma}}{\d\mu^{2}}+\Gamma_{\mu\nu}^{\sigma}\frac{\d x^{\mu}}{\d\mu}\frac{\d x^{\nu}}{\d\mu}=\left(\frac{\d\lambda}{\d\mu}\right)^{2}\left(\alpha\frac{\d\mu}{\d\lambda}-\frac{\d^{2}\mu}{\d\lambda^{2}}\right)\frac{\d x^{\sigma}}{\d\mu}.
\]
Therefore the right-hand side will vanish if $\mu\left(\lambda\right)$
is a solution to the differential equation
\begin{equation}
\frac{\d^{2}\mu}{\d\lambda^{2}}=\alpha\left(\lambda\right)\frac{\d\mu}{\d\lambda}.\label{eq:affine-par}
\end{equation}
Such a solution always exists, and thus we have obtained the desired
parameterization. The parameter $\mu$ for which the geodesic equation
reduces to the form (\ref{eq:geodesic-eq}) is called an \emph{affine
parameter}. Note that this is in fact a whole family of parameters,
since any other parameter given by $\nu\equiv A\mu+B$ with $A,B$
real constants is also an affine parameter, as can be easily seen
from (\ref{eq:affine-par}).

The geodesic equation is one of the two most important equations in
general relativity; the other is Einstein's equation, which we will
discuss in Sec. \ref{subsec:Einstein's-Equation}.

\subsubsection{Massive Particles and Geodesics}

A \emph{test particle }is a particle which has a negligible effect
on the curvature of spacetime. Such a particle's path will always
be a \emph{timelike geodesic }if it's a massive particle (such as
an electron), or a \emph{null geodesic }if it's a massless particle
(such as a photon), as long as it is in \emph{free fall} -- meaning
that no forces act on it other than gravity.

For a massive particle with \emph{(rest) mass}\footnote{Some textbooks also define a ``relativistic mass'' which depends
on the velocity or frame of reference. However, this concept is not
useful in general relativity (and even in special relativity it mostly
just causes confusion). In these notes, as in most of the theoretical
physics literature, the rest mass $m$ is assumed to be a constant
which is assigned to the particle once and for all. For example, the
electron has roughly $m=511\thinspace\mathrm{keV}$, or the pure number
$m=4.2\xx10^{-23}$ in Planck units, independently of its velocity
or reference frame.}\emph{ }$m$ and 4-velocity $\xd^{\mu}=\d x^{\mu}/\d\tau$, the \emph{4-momentum}
is given by $p^{\mu}\equiv m\xd^{\mu}$. Recall again that the 4-velocity
is only defined if the curve is parametrized by the proper time $\tau$.
Massless particles have neither mass, nor 4-velocity, since proper
time is undefined for a null geodesic. Therefore the definition $p^{\mu}\equiv m\xd^{\mu}$
would not make sense, and we simply define the 4-momentum as the tangent
vector with respect to some affine parameter: $p^{\mu}\equiv\d x^{\mu}/\d\lambda$.

We will sometimes write the geodesic equation in terms of the particle's
4-momentum as follows:
\[
p^{\nu}\nabla_{\nu}p^{\mu}=0.
\]
In other words, an unaccelerated (free-falling) particle keeps moving
in the direction of its momentum. An observer moving with 4-velocity
$u^{\mu}$ then measures the energy of the particle to be
\[
E\equiv-p_{\mu}u^{\mu}.
\]
As a simple example, consider a flat spacetime with the Minkowski
metric $\eta_{\mu\nu}\equiv\diag\left(-1,1,1,1\right)$. A massive
particle with 4-momentum $p^{\mu}\equiv m\xd^{\mu}$ is measured by
an observer at rest, that is, with 4-velocity $u^{\mu}=\left(1,0,0,0\right)$.
The energy measured will then be
\[
E=-\eta_{\mu\nu}p^{\mu}u^{\nu}=-\eta_{00}p^{0}u^{0}=p^{0}.
\]
In other words, in this case the energy is simply the time component
of the 4-momentum. Motivated by this result, we take the 4-momentum
to be of the form $p^{\mu}\equiv\left(E,\vec{p}\right)$ where $\vec{p}\equiv\left(p^{1},p^{2},p^{3}\right)$
is the (spatial\footnote{We will use bold font, $\v$, for spacetime 4-vectors and an arrow,
$\vec{v},$ for spatial 3-vectors.}) 3-momentum. Since $\left|\xxd\right|^{2}=-1$, the norm of the 4-momentum
is given by
\[
\left|\p\right|^{2}=m^{2}\left|\xxd\right|^{2}=-m^{2}.
\]
On the other hand, by direct calculation we find
\begin{equation}
\left|\p\right|^{2}=-E^{2}+\left|\vec{p}\right|^{2},\label{eq:p-e-p}
\end{equation}
where $\left|\vec{p}\right|^{2}\equiv\left(p^{1}\right)^{2}+\left(p^{2}\right)^{2}+\left(p^{3}\right)^{2}$.
Comparing both expressions, we see that
\begin{equation}
E^{2}=m^{2}+\left|\vec{p}\right|^{2}.\label{eq:mass-energy-equiv}
\end{equation}
In the rest frame of the particle, where $\vec{p}=0$, this reduces
to the familiar \emph{mass-energy equivalence }equation $E=mc^{2}$
(with $c=1$).

\subsubsection{\label{subsec:Massless-Particles-and}Massless Particles and the
Speed of Light}

For massless particles, the situation is simpler. We again take $p^{\mu}\equiv\left(E,\vec{p}\right)$,
so an observer at rest with $u^{\mu}=\left(1,0,0,0\right)$ will measure
the energy $E$. Furthermore, since by definition $p^{\mu}\equiv\xd^{\mu}$
and $\xd^{\mu}$ is null, we have 
\[
\left|\p\right|^{2}=\left|\xxd\right|^{2}=0,
\]
and combining with (\ref{eq:p-e-p}) we find that $E^{2}=\left|\vec{p}\right|^{2}$.
Again, this is the familiar equation $E=pc$, with $c=1$. We conclude
that the relation (\ref{eq:mass-energy-equiv}) applies to all particles,
whether massive or massless.

Even though a massless particle doesn't have a 4-velocity, we know
that it \textbf{locally }(i.e. at the same point as the observer)
always moves at the speed of light, $v=c=1$. This is true both in
special and general relativity. It is easy to see in a flat spacetime.
The line element is
\[
\d s^{2}=-\d t^{2}+\d x^{2}+\d y^{2}+\d z^{2}.
\]
Let us assume the massless particle is moving at this instant in the
$x$ direction (if it isn't, then we can rotate our coordinate system
until it is). Then, since the $y$ and $z$ coordinates remain unchanged,
we have $\d y=\d z=0$. Furthermore, since the particle is moving
along a null path, it has $\d s^{2}=0$ as well. In conclusion, we
have
\[
0=-\d t^{2}+\d x^{2},
\]
which may be rearranged into
\[
\frac{\d x}{\d y}=\pm1.
\]
In other words, the particle is moving at the speed of light 1, either
in the positive or negative $x$ direction.

However, if the particle was massive, we would have $\d s^{2}<0$
since it is moving along a timelike path. Therefore we have
\[
\d s^{2}=-\d t^{2}+\d x^{2}<0,
\]
which may be rearranged into
\[
\left|\frac{\d x}{\d t}\right|<1.
\]
Hence, the particle is necessarily moving locally at strictly slower
than the speed of light.

So far, we have only considered special relativity. In general relativity
we have an arbitrarily curved spacetime, and naively, it seems that
massless particles may move at any speed. This is easy to see by considering,
for example, the following metric:
\[
\d s^{2}=-V^{2}\d t^{2}+\d x^{2},
\]
where $V$ is some real number. Then for a massless particle we have
$\d s^{2}=0$ and thus
\[
\frac{\d x}{\d t}=\pm V,
\]
so the speed of the particle is given by the arbitrary number $V$.
The problem here is that we have calculated the \textbf{coordinate
}speed of the particle, not its the \textbf{local }speed. General
relativity works in any coordinate system -- this is called \emph{general
covariance }or \emph{diffeomorphism invariance} -- and the coordinate
speed will naturally depend on the choice of coordinate system.

The fact that massless particles always \textbf{locally }move at the
speed of light easily follows from the well-known result that at a
particular point $p$ it is always possible to transform to \emph{locally
inertial coordinates}\footnote{See e.g. \cite{Carroll} for the details of how to construct such
coordinates.}, which have the property that $g_{\mu\nu}\left(p\right)=\eta_{\mu\nu}$
and $\partial_{\sigma}g_{\mu\nu}\left(p\right)=0$. Then, as far as
the observer at $p$ is concerned, spacetime is completely flat in
their immediate vicinity, and thus they will see massless particles
passing by at the speed of light.

Finally, it is important to note that if a particle starts moving
along a timelike or null geodesic it can never suddenly switch to
moving along a different type of geodesic or path; this is simply
due to the mathematical fact that the parallel transport preserves
the norm of the tangent vector to the path. Indeed, if the tangent
vector being parallel transported is $v^{\mu}$, then $\xd^{\mu}\nabla_{\mu}v^{\nu}=0$
and thus
\begin{equation}
\xd^{\mu}\nabla_{\mu}\left(\left|\v\right|^{2}\right)=g_{\alpha\beta}\xd^{\mu}\nabla_{\mu}\left(v^{\alpha}v^{\beta}\right)=g_{\alpha\beta}\left(v^{\beta}\left(\xd^{\mu}\nabla_{\mu}v^{\alpha}\right)+v^{\alpha}\left(\xd^{\mu}\nabla_{\mu}v^{\beta}\right)\right)=0,\label{eq:parallel-transport-norm}
\end{equation}
so $\left|\v\right|^{2}$ is constant along the path.

\subsubsection{\label{subsec:Maximization}Inertial Motion, Maximization of Proper
Time, and the Twin ``Paradox''}

It is important to clarify that particles follow geodesics only if
they are in \emph{inertial motion}, or \emph{free-falling}. By this
we mean that the only force acting on the particle is that of gravity,
and there are no other forces which would influence the particle's
trajectory. If non-gravitational forces act on the particle -- or,
for example, a spaceship uses its rockets -- then it will no longer
follow a geodesic.

Now, in Euclidean space, geodesics minimize distance. However, in
a curved space with Lorentzian signature, geodesics instead \textbf{maximize
}proper time -- at least for massive particles, since for massless
particles the proper time vanishes by definition. This can be seen
from the fact that the geodesic equation may be derived by demanding
that the variation of the proper time integral (\ref{eq:proper-time})
vanishes\footnote{See e.g. \cite{Carroll}, chapter 3.3.}.

The two facts we have just mentioned provide an elegant solution to
the famous twin ``paradox'' of special relativity. In this ``paradox''
there are two twins, Alice and Bob. Alice stays on Earth while Bob
goes on a round trip to a nearby planet, traveling at a significant
fraction of the speed of light. When Bob returns to Earth, the twins
discover that Alice is now older than Bob, due to relativistic time
dilation.

The ``paradox'' lies in the naive assumption that, since from Bob's
point of view he was the one who stayed in place while Alice was the
one moving with respect to him, then Bob would expect himself to be
the older twin. However, obviously both twins cannot be older than
each other, which leads to a ``paradox''. The solution to the ``paradox''
lies in the fact that Alice remained in\textbf{ inertial motion} for
the entire time, while Bob was \textbf{accelerating} and thus not
in inertial motion, hence there is an asymmetry between the twins.

Alternatively, if one does not wish to complicate things by involving
acceleration, we may assume that all of the accelerations involved
(speeding up, slowing down, and turning around) were instantaneous.
Then Bob was also in inertial motion for the entire time, except for
the moment of \textbf{turnaround}.

At that moment, the inertial frame for the outbound journey is replaced
with a completely different inertial frame for the return journey.
That one singular moment of turnaround is alone responsible for the
asymmetry between the points of view of the twins. It is easy to see
by drawing spacetime diagrams -- as the reader is encouraged to do
-- that at the moment of turnaround, Bob's notion of simultaneity
(that is, surfaces of constant $t$) changes dramatically. From Bob's
point of view, Alice ages instantaneously at the moment!

This is the standard solution from the special relativistic point
of view. However, now that we know about general relativity and geodesics,
we may provide a much simpler solution. Since Alice remains in inertial
motion, and the only forces that act on her are gravitational forces,
she simply follows a timelike geodesic. On the other hand, since Bob
is using non-gravitational forces (e.g. rockets) to accelerate himself,
he will \textbf{not }follow a timelike geodesic. Of course, he will
still follow a timelike \textbf{path}, but the path will not be a
geodesic.

In other words, both Alice and Bob's paths in spacetime begin and
end at the same point, but Alice follows a timelike geodesic while
Bob follows a timelike path which is not a geodesic. Since timelike
geodesics are exactly the timelike paths which maximize proper time,
the proper time experienced by Alice must be larger than the proper
time experienced by Bob. Therefore, Alice must be the older twin.

\subsection{Curvature}

\subsubsection{The Riemann Curvature Tensor}

The \emph{Riemann curvature tensor} is defined as the commutator\footnote{Note that if there is torsion, that is $\Gamma_{\mu\nu}^{\lambda}\ne\Gamma_{\nu\mu}^{\lambda}$,
then one must add a term $2\Gamma_{\left[\mu\nu\right]}^{\lambda}\nabla_{\lambda}v^{\rho}$
to the right-hand side of this equation, where $2\Gamma_{\left[\mu\nu\right]}^{\lambda}\equiv\Gamma_{\mu\nu}^{\lambda}-\Gamma_{\nu\mu}^{\lambda}$.} of the action of two covariant derivatives on a vector:
\begin{equation}
\udi R{\rho}{\sigma\mu\nu}v^{\sigma}=\left[\nabla_{\mu},\nabla_{\nu}\right]v^{\rho}.\label{eq:Riemann-def}
\end{equation}
Since the covariant derivative facilitates parallel transport, this
can -- roughly speaking -- be understood as taking the vector $v^{\rho}$
from some point $p$ along a path given by $\nabla_{\mu}\nabla_{\nu}v^{\rho}$
to a nearby point $q$, and then taking it back from $q$ along a
path given by $-\nabla_{\nu}\nabla_{\mu}v^{\rho}$ to $p$. In other
words, we take $v^{\rho}$ along a loop. If spacetime was flat, we
would expect that the vector $v^{\rho}$ would remain unchanged after
going around the loop. However, if spacetime is curved, there is a
difference between the initial $v^{\rho}$ and the final $v^{\rho}$
(note that both are in the same tangent space $T_{p}M$, so we may
compare them). This difference is encoded in the Riemann tensor $R_{\ \sigma\mu\nu}^{\rho}$.

The full coordinate expression for the Riemann tensor in terms of
the connection coefficients may be calculated from the definition,
and it is given by:
\[
\udi R{\rho}{\sigma\mu\nu}=\partial_{\mu}\Gamma_{\nu\sigma}^{\rho}-\partial_{\nu}\Gamma_{\mu\sigma}^{\rho}+\Gamma_{\mu\lambda}^{\rho}\Gamma_{\nu\sigma}^{\lambda}-\Gamma_{\nu\lambda}^{\rho}\Gamma_{\mu\sigma}^{\lambda}.
\]
If we lower the first index with the metric ($R_{\rho\sigma\mu\nu}=g_{\lambda\rho}\udi R{\lambda}{\sigma\mu\nu}$),
the resulting tensor satisfies the following identities:
\begin{itemize}
\item Symmetry and anti-symmetry under exchange of indices: $R_{\rho\sigma\mu\nu}=-R_{\sigma\rho\mu\nu}=-R_{\rho\sigma\nu\mu}=R_{\mu\nu\rho\sigma}$.
\item \emph{First Bianchi identity}: $R_{\rho\left[\sigma\mu\nu\right]}=0$
or more explicitly\footnote{Here $6R_{\rho\left[\sigma\mu\nu\right]}\equiv R_{\rho\sigma\mu\nu}+R_{\rho\mu\nu\sigma}+R_{\rho\nu\sigma\mu}-R_{\rho\mu\sigma\nu}-R_{\rho\nu\mu\sigma}-R_{\rho\sigma\nu\mu}$
is the usual anti-symmetrizer, and then we use the anti-symmetry in
the last two indices.} $R_{\rho\sigma\mu\nu}+R_{\rho\mu\nu\sigma}+R_{\rho\nu\sigma\mu}=0$.
\item \emph{Second Bianchi identity}: $\nabla_{[\lambda}R_{\rho\sigma]\mu\nu}=0$.
\end{itemize}

\subsubsection{Related Tensors: Ricci and Weyl}

By contracting the first and third index of the Riemann tensor, we
obtain the \emph{Ricci tensor}:
\[
R_{\mu\nu}\equiv\udi R{\lambda}{\mu\lambda\nu}.
\]
Note that it is symmetric, $R_{\mu\nu}=R_{\nu\mu}$. The trace of
the Ricci tensor is the \emph{Ricci scalar}:
\[
R\equiv\udi R{\mu}{\mu}\equiv g^{\mu\nu}R_{\mu\nu}.
\]
We may also define (in 4 dimensions\footnote{For a spacetime of dimension $d$ we have
\[
C_{\rho\sigma\mu\nu}\equiv R_{\rho\sigma\mu\nu}-\frac{2}{d-2}\left(g_{\rho[\mu}R_{\nu]\sigma}-g_{\sigma[\mu}R_{\nu]\rho}-\frac{1}{d-1}g_{\rho[\mu}g_{\nu]\sigma}R\right).
\]
}) the \emph{Weyl tensor}:
\[
C_{\rho\sigma\mu\nu}\equiv R_{\rho\sigma\mu\nu}-g_{\rho[\mu}R_{\nu]\sigma}+g_{\sigma[\mu}R_{\nu]\rho}+\trd g_{\rho[\mu}g_{\nu]\sigma}R.
\]
The Weyl tensor has all the symmetries of the Riemann tensor (including
the first Bianchi identity), but it is completely traceless: it vanishes
upon contraction of any pair of indices.

\subsection{Extrinsic Curvature}

\subsubsection{Intrinsic and Extrinsic Curvature}

Let us now consider a surface embedded in a higher-dimensional space.
The Riemann tensor describes the \emph{intrinsic curvature }of the
surface; this is the curvature within the surface itself, which exists
intrinsically, regardless of any embedding (as long as the embedding
is isometric, see below).

In contrast, \emph{extrinsic curvature} is the curvature of the surface
which comes from the way in which it is embedded, and from the particular
space in which it is embedded. Intrinsic curvature comes from the
parallel transport of vectors \textbf{tangent }to (a curve on) the
surface, while extrinsic curvature comes from parallel transport of
vectors \textbf{normal }to the surface.

For example, a flat piece of paper has no intrinsic curvature. If
we draw a triangle on it, the angles will sum to $\pi$. However,
if we roll that paper into a cylinder, then as seen from the (flat)
higher-dimensional space in which we live, the angles of the triangle
will no longer sum to $\pi$. Thus, the surface has acquired an extrinsic
curvature.

However, the intrinsic curvature is still flat, since within the surface
itself, the edges of the triangle are still geodesics -- regardless
of its embedding into the higher-dimensional space. Therefore, the
angles still sum to $\pi$ when viewed from inside the surface. In
other words, the intrinsic curvature is completely independent of
any embedding of the surface.

Let us see this with a concrete calculation. We take a cylinder with
circumference $2\pi L$. It can be obtained by ``rolling up'' $\BBR^{2}$,
i.e. by performing the periodic identification
\[
\left(x,y\right)\sim\left(x,y+2\pi L\right).
\]
Topologically, this is homeomorphic\footnote{A \emph{homeomorphism }between topological spaces $X$ and $Y$ is
a continuous function that has a continuous inverse. The importance
of homeomorphisms is that they preserve the topological properties
of the space. If there is a homeomorphism between two spaces, we say
that they are \emph{homeomorphic} to each other.} to $\BBR\xx S^{1}$. The metric is inherited from the original (unrolled)
$\BBR^{2}$, and is thus flat. Now, let us embed the cylinder in $\BBR^{3}$.
Introducing cylindrical coordinates $\left(r,\phi,z\right)$, the
metric on $\BBR^{3}$ takes the form
\begin{equation}
\d s^{2}=\d r^{2}+r^{2}\d\phi^{2}+\d z^{2}.\label{eq:R3-cyl-metric}
\end{equation}
Taking a constant $r=L$ and identifying the points
\[
\left(L,\phi,z\right)\sim\left(L,\phi+2\pi,z\right),
\]
we get the same cylinder with circumference $2\pi L$, and it has
the induced metric (with $\d r=0$ since $r$ is constant)
\[
\d s^{2}\bl_{r=L}=L^{2}\d\phi^{2}+\d z^{2}.
\]
Since the components of this induced metric are constant, the intrinsic
curvature of the cylinder is zero.

\subsubsection{Isometric and Non-Isometric Embeddings}

Let $\left(M,g_{\mu\nu}\right)$ and $\left(N,h_{\mu\nu}\right)$
be two Riemannian manifolds, with metrics $g_{\mu\nu}$ and $h_{\mu\nu}$.
An \emph{immersion }between the manifolds is a differentiable function
$F:M\to N$ with an injective (one-to-one) derivative. An \emph{embedding
}of $M$ into $N$ is an injective immersion such that $M$ is homeomorphic
to its image $f\left(M\right)$. The embedding is called \emph{isometric
}if it also preserves the metric, that is, $g_{\mu\nu}=f^{*}h_{\mu\nu}$
where $f^{*}h_{\mu\nu}$ is the pullback\footnote{The \emph{pullback }$f^{*}\T$ of a tensor $\T$ by the map $f:M\to N$
literally ``pulls back'' $\T$ from $N$ into the source manifold
$M$. In the case of a metric $h_{\mu\nu}$, which is a rank $\left(0,2\right)$
tensor, the pullback acts on its components as follows:
\[
\left(f^{*}h\left(\x\right)\right)_{\mu\nu}=\frac{\partial x^{\alpha}}{\partial y^{\mu}}\frac{\partial x^{\beta}}{\partial y^{\nu}}h_{\alpha\beta}\left(\y\right),
\]
where $\y$ are coordinates on $N$ and $\x$ are coordinates on $M$.} of $h_{\mu\nu}$ by $f$.

The embedding of the cylinder $\BBR\xx S^{1}$ into $\BBR^{3}$ is
isometric, since the induced metric on $\BBR^{3}$ is flat, and thus
equal to the original metric on the cylinder. Therefore, the intrinsic
curvature remains unchanged after the embedding. However, there are
cases when an embedding forces us to change the intrinsic curvature
of a manifold. This happens when we cannot embed our manifold in another
one without stretching or bending it.

An illustrative example is given by the torus, which may be obtained
from $\BBR^{2}$ in a similar manner to the cylinder, except that
now \textbf{both} coordinates are identified, with periods $2\pi L_{1}$
and $2\pi L_{2}$:
\[
\left(x,y\right)\sim\left(x+2\pi L_{1},y+2\pi L_{2}\right).
\]
Topologically, this is homeomorphic to $S^{1}\xx S^{1}$. The metric
inherited from $\BBR^{2}$ is, of course, still flat. However, let
us now embed the torus $\BBR^{3}$, with the same cylindrical coordinates
as before. The surface will be defined as the set of points solving
the equation
\[
z^{2}+\left(r-L_{1}\right)^{2}=L_{2}^{2}.
\]
Indeed, for each value of $\phi$, this equation defines a circle
with radius $L_{2}$ centered at $\left(r,z\right)=\left(L_{1},0\right)$.
The surface of revolution of a circle is a torus; it can also be seen
as a cylinder whose top and bottom have been glued together. Thus
$L_{1}$ is the \emph{major radius}, or the distance from the $z$
axis to the center of the circle, while $L_{2}$ is the \emph{minor
radius}, that of the circle that is being revolved.

Let us now isolate $z$:
\[
z=\pm\sqrt{L_{2}^{2}-\left(r-L_{1}\right)^{2}}.
\]
Then
\[
\d z=-\frac{\left(r-L_{1}\right)\d r}{\sqrt{L_{2}^{2}-\left(r-L_{1}\right)^{2}}},
\]
and by plugging this into the flat metric in cylindrical coordinates
on $\BBR^{3}$, (\ref{eq:R3-cyl-metric}), we obtain the following
induced metric:
\[
\d s^{2}\bl_{z^{2}+\left(r-L_{1}\right)^{2}=L_{2}^{2}}=\frac{L_{2}^{2}}{L_{2}^{2}-\left(r-L_{1}\right)^{2}}\d r^{2}+r^{2}\d\phi^{2},
\]
where $L_{1}-L_{2}\le r\le L_{1}+L_{2}$. This metric is \textbf{not}
flat, as can be checked e.g. by calculating the Ricci scalar, which
is
\[
R=\frac{2\left(r-L_{1}\right)}{L_{2}^{2}r}.
\]
Thus, in this case the intrinsic curvature is not flat after the embedding.

In the case of the cylinder, all we did when we embedded it in $\BBR^{3}$
was to take a flat plane and glue two opposite ends of it together.
This can be easily illustrated by taking a flat piece of paper and
bending it to create a cylinder. However, for the torus, we started
with a cylinder -- which is still intrinsically flat -- and glued
its top and bottom together. This cannot be done without stretching
the paper (try it!), resulting in an intrinsic curvature that is no
longer flat. In other words, this embedding is not isometric\footnote{Note, however, that it is possible to isometrically embed the torus
in $\BBR^{4}$.}.

\subsubsection{Surfaces and Normal Vectors}

Let $\Sigma$ be a surface embedded in a (higher-dimensional) manifold
$M$. This surface may be defined, for example, by an equation of
the form $f\left(\x\right)=0$, as we did for the torus. Then the
vector field $\xi^{\mu}\equiv\nabla^{\mu}f=\partial^{\mu}f$ is everywhere
normal to the surface, meaning that it is orthogonal to every vector
tangent to $\Sigma$ (we will not prove this here). The nature of
the surface will have the opposite sign to that of the normal vector:
if $\xi^{\mu}$ is timelike the surface is spacelike, if $\xi^{\mu}$
is spacelike the surface is timelike, and if $\xi^{\mu}$ is null
the surface is also null.

If $\xi^{\mu}$ is timelike or spacelike, we may normalize it and
define a unit normal vector:
\[
n^{\mu}\equiv\frac{\xi^{\mu}}{\sqrt{\left|\xi_{\lambda}\xi^{\lambda}\right|}}.
\]
If $\xi^{\mu}$ is null, then it is not only normal but also tangent
to $\Sigma$, since it is orthogonal to itself. Thus the \emph{integral
curves }$x^{\mu}\left(\lambda\right)$ to the vector field $\xi^{\mu}$,
which are curves such that their tangent vectors are equal to $\xi^{\mu}$
at every point along the curve:
\begin{equation}
\frac{\d x^{\mu}}{\d\lambda}=\xi^{\mu},\label{eq:null-integral-curves}
\end{equation}
are null curves contained in $\Sigma$. Now, we have
\begin{equation}
\xi^{\mu}\nabla_{\mu}\xi_{\nu}=\xi^{\mu}\nabla_{\mu}\partial_{\nu}f=\xi^{\mu}\left(\partial_{\mu}\partial_{\nu}f+\Gamma_{\mu\nu}^{\lambda}\partial_{\lambda}f\right)=\xi^{\mu}\nabla_{\nu}\partial_{\mu}f=\xi^{\mu}\nabla_{\nu}\xi_{\mu}=\hf\nabla_{\nu}\left(\xi^{\mu}\xi_{\mu}\right),\label{eq:normal-null-geodesic}
\end{equation}
where we used the fact that $\Gamma_{\mu\nu}^{\lambda}$ is symmetric
in its lower indices. Unfortunately, $\xi^{\mu}\xi_{\mu}$ does not
necessarily vanish outside of $\Sigma$, so we cannot conclude that
the last term vanishes. However, if it does not vanish, we simply
redefine the defining equation $f\left(\x\right)=0$ for the surface
using $f\left(\x\right)\equiv\xi^{\mu}\xi_{\mu}$, and then the last
term in (\ref{eq:normal-null-geodesic}) is simply $\hf\nabla_{\nu}f\left(\x\right)$,
which is normal to the surface and thus must be proportional to $\xi^{\mu}$!
If the proportionality function is $\alpha$, then we get
\[
\xi^{\mu}\nabla_{\mu}\xi_{\nu}=\alpha\xi_{\nu},
\]
which is the generalized geodesic equation (\ref{eq:geodesic-general}).
As we have seen when discussing that function, we may reparameterize
the geodesic using an affine parameter $\lambda$ such that $\xi^{\mu}\nabla_{\mu}\xi_{\nu}=0$.
The null geodesics $x^{\mu}\left(\lambda\right)$ defined by (\ref{eq:null-integral-curves})
are called the \emph{generators }of the null surface $\Sigma$, since
the surface is the union of these geodesics.

\subsubsection{The Projector on the Surface}

Let $M$ be a manifold with metric $g_{\mu\nu}$. We define the \emph{projector
}on the surface $\Sigma$ with unit normal $n^{\mu}$ as follows:
\begin{equation}
P_{\mu\nu}\equiv g_{\mu\nu}-\left|\n\right|^{2}n_{\mu}n_{\nu}.\label{eq:projector}
\end{equation}
Note that $\left|\n\right|^{2}\equiv n_{\lambda}n^{\lambda}=\pm1$
with $+$ for timelike and $-$ for spacelike surfaces. This symmetric
tensor projects any vector field $v^{\mu}$ in $M$ to a tangent vector
$P_{\nu}^{\mu}v^{\nu}$ on $\Sigma$. We can see this by showing that
the projected vector is orthogonal to the normal vector:
\[
n^{\mu}\left(P_{\mu\nu}v^{\nu}\right)=n^{\mu}v_{\mu}-\left|\n\right|^{4}n_{\nu}v^{\nu}=0,
\]
since $\left|\n\right|^{4}=1$ whether $n^{\mu}$ is spacelike or
timelike. Furthermore, the projector acts like a metric on vectors
tangent to $\Sigma$:
\[
P_{\mu\nu}v^{\mu}w^{\nu}=g_{\mu\nu}v^{\mu}w^{\nu}=\langle\v,\w\rangle,
\]
since $n_{\mu}v^{\mu}=n_{\mu}w^{\mu}=0$ for tangent vectors. Also,
the projector remains unchanged under its own action:
\[
P_{\lambda}^{\mu}P_{\nu}^{\lambda}=\left(\delta_{\lambda}^{\mu}-\left|\n\right|^{2}n^{\mu}n_{\lambda}\right)\left(\delta_{\nu}^{\lambda}-\left|\n\right|^{2}n^{\lambda}n_{\nu}\right)=\delta_{\nu}^{\mu}-2\left|\n\right|^{2}n^{\mu}n_{\nu}+\left|\n\right|^{6}n^{\mu}n_{\nu}=P_{\nu}^{\mu},
\]
since $\left|\n\right|^{6}=\left|\n\right|^{2}$.

\subsubsection{Definition of Extrinsic Curvature}

The \emph{extrinsic curvature tensor}\footnote{Sometimes called the \emph{second fundamental form }in the differential
geometry literature.} $K_{\mu\nu}$ is a rank $\left(0,2\right)$ symmetric tensor defined
as the Lie derivative\footnote{\label{fn:Lie-derivative}We will not go into the rigorous definition
of the Lie derivative here, but just note that a general formula for
its action on tensors is
\[
\LLL_{\n}\udi T{\mu\cdots}{\nu\cdots}=n^{\lambda}\partial_{\lambda}\udi T{\mu\cdots}{\nu\cdots}-\udi T{\lambda\cdots}{\nu\cdots}\partial_{\lambda}n^{\mu}-\ldots+\udi T{\mu\cdots}{\lambda\cdots}\partial_{\nu}n^{\lambda}+\ldots,
\]
where for each upper index of $\T$ we add a negative term with the
index on $\n$ exchanged with that index, and for each lower index
of $\T$ we add a positive term with the index on the partial derivative
exchanged with that index. Note that the partial derivatives may be
replaced with covariant derivatives, since the extra terms all cancel
(show this).} of the projector $P_{\mu\nu}$ in the direction of the normal vector
$n^{\mu}$:
\begin{equation}
K_{\mu\nu}\equiv\hf\LLL_{\n}P_{\mu\nu}.\label{eq:extrinsic-curvature}
\end{equation}
From the coordinate expression for the Lie derivative (see Footnote
\ref{fn:Lie-derivative}) it is relatively straightforward (try it!)
to show that
\[
K_{\mu\nu}=P_{\mu}^{\alpha}P_{\nu}^{\beta}\nabla_{(\alpha}n_{\beta)}=\hf P_{\mu}^{\alpha}P_{\nu}^{\beta}\LLL_{\n}g_{\alpha\beta},
\]
since by definition $\LLL_{\n}g_{\alpha\beta}=2\nabla_{(\alpha}n_{\beta)}$.

Finally, let us assume that $n^{\mu}$ is tangent to a geodesic. This
is generally possible by extending it off the surface $\Sigma$ using
the geodesic equation, $n^{\lambda}\nabla_{\lambda}n^{\mu}=0$. Then
we have
\[
\LLL_{\n}n_{\nu}=n^{\lambda}\nabla_{\lambda}n_{\nu}+n_{\lambda}\nabla_{\nu}n^{\lambda}=0+\hf\nabla_{\nu}\left(n_{\lambda}n^{\lambda}\right)=0,
\]
since $n_{\lambda}n^{\lambda}$ is a constant. Therefore we find that
\begin{equation}
K_{\mu\nu}=\hf\LLL_{\n}P_{\mu\nu}=\hf\LLL_{\n}\left(g_{\mu\nu}-\left|\n\right|^{2}n_{\mu}n_{\nu}\right)=\hf\LLL_{\n}g_{\mu\nu}=\nabla_{(\mu}n_{\nu)}.\label{eq:extrinsic-curvature-geodesic}
\end{equation}

\subsection{\label{subsec:Einstein's-Equation}Einstein's Equation}

Let us define the \emph{Einstein-Hilbert action}\footnote{The \emph{volume form} $\ep\equiv\frac{1}{4!}\epsilon_{\rho\sigma\mu\nu}\d x^{\rho}\wedge\d x^{\sigma}\wedge\d x^{\mu}\wedge\d x^{\nu}=\sqrt{-g}\thinspace\d^{4}x$
is a 4-form also known as the \emph{Levi-Civita tensor}, and its components
$\epsilon_{\rho\sigma\mu\nu}$ are related to the familiar totally
anti-symmetric \emph{Levi-Civita symbol} $\ept_{\rho\sigma\mu\nu}$
by $\epsilon_{\rho\sigma\mu\nu}=\sqrt{-g}\thinspace\ept_{\rho\sigma\mu\nu}$.
Note that $\ept_{\rho\sigma\mu\nu}$ is not a tensor but a \emph{tensor
density }of weight 1, since it is related to a tensor by a factor
of one power of $\sqrt{-g}$.}:
\begin{equation}
S_{H}\equiv\frac{1}{16\pi}\int R\sqrt{-g}\thinspace\d^{4}x,\label{eq:Einstein-Hilbert}
\end{equation}
where $g\equiv\det\g$. The factor\footnote{If we did not use units where $G\equiv1$, this would instead be $1/16\pi G$.}
of $1/16\pi$ in front of the action is a convention chosen to produce
the correct form of Newton's law of gravitation $F=m_{1}m_{2}/r^{2}$
in the Newtonian limit. By varying this action with respect to the
metric, we obtain the \emph{vacuum Einstein equation}:
\begin{equation}
\frac{1}{\sqrt{-g}}\frac{\delta S_{H}}{\delta g^{\mu\nu}}=\frac{1}{16\pi}\left(R_{\mu\nu}-\hf Rg_{\mu\nu}\right)=0\soosp R_{\mu\nu}-\hf Rg_{\mu\nu}=0.\label{eq:vaccum-Einstein}
\end{equation}
To add matter, we simply add the appropriate action $S_{M}$ for the
type of matter we would like to consider:
\[
S\equiv S_{H}+S_{M}.
\]
Whatever the form of $S_{M}$ is, by varying $S$ with respect to
the metric we will get
\[
\frac{1}{\sqrt{-g}}\frac{\delta S}{\delta g^{\mu\nu}}=\frac{1}{16\pi}\left(R_{\mu\nu}-\hf Rg_{\mu\nu}\right)+\frac{1}{\sqrt{-g}}\frac{\delta S_{M}}{\delta g^{\mu\nu}}=0.
\]
If we now define the \emph{energy-momentum tensor }(or \emph{stress-energy
tensor}) to be the symmetric rank $\left(0,2\right)$ tensor with
components
\begin{equation}
T_{\mu\nu}\equiv-\frac{2}{\sqrt{-g}}\frac{\delta S_{M}}{\delta g^{\mu\nu}},\label{eq:EM-tensor}
\end{equation}
we get the \emph{full Einstein equation }in the presence of matter:
\begin{equation}
R_{\mu\nu}-\hf Rg_{\mu\nu}=8\pi T_{\mu\nu}.\label{eq:Einstein}
\end{equation}
The expression on the left-hand side is also known as the \emph{Einstein
tensor}\footnote{Notice from (\ref{eq:vaccum-Einstein}) and (\ref{eq:EM-tensor})
that the Einstein tensor is, in fact, just the energy-momentum tensor
of the metric field $g_{\mu\nu}$ itself, multiplied by $-8\pi$.
Thus, if we treat the gravitational field as just another matter field,
the Einstein equation reduces to the statement that the total energy-momentum
tensor of all of the fields in the universe is zero: $T_{\mu\nu}=0$!}:
\begin{equation}
G_{\mu\nu}\equiv R_{\mu\nu}-\hf Rg_{\mu\nu}.\label{eq:Einstein-tensor}
\end{equation}
From the second Bianchi identity, one can show that $\nabla^{\mu}G_{\mu\nu}=0$.
This is good, since conservation of energy implies that the energy-momentum
tensor must satisfy $\nabla^{\mu}T_{\mu\nu}=0$.

One may also rewrite the Einstein equation as follows:
\begin{equation}
R_{\mu\nu}=8\pi\left(T_{\mu\nu}-\hf Tg_{\mu\nu}\right),\label{eq:Einstein-T}
\end{equation}
where $T\equiv\udi T{\mu}{\mu}$ is the trace of the energy-momentum
tensor.

\subsection{The Cosmological Constant}

Consider the following action:
\[
S_{\Lambda}\equiv-\frac{\Lambda}{8\pi}\int\sqrt{-g}\thinspace\d^{4}x.
\]
This is simply a constant times the volume of the entire spacetime!
However, since the volume depends on the metric, the variation of
this action with respect to the metric is not trivial; in fact, it
is easy to see that 
\[
\frac{1}{\sqrt{-g}}\frac{\delta S_{\Lambda}}{\delta g^{\mu\nu}}=\frac{\Lambda}{16\pi}g_{\mu\nu}.
\]
If we now take the full action to be
\[
S\equiv S_{H}+S_{\Lambda}+S_{M},
\]
we get the \emph{Einstein equation with a cosmological constant}:
\[
R_{\mu\nu}-\hf Rg_{\mu\nu}+\Lambda g_{\mu\nu}=8\pi T_{\mu\nu}.
\]
The simplest way to interpret $\Lambda$ is as \emph{vacuum energy},
that is, the energy density of empty spacetime. To see this, let us
assume for simplicity there is no matter, so $T_{\mu\nu}=0$:
\[
R_{\mu\nu}-\hf Rg_{\mu\nu}+\Lambda g_{\mu\nu}=0.
\]
Then, we move the cosmological constant to the right-hand side of
the equation:
\[
R_{\mu\nu}-\hf Rg_{\mu\nu}=-\Lambda g_{\mu\nu}.
\]
This may be interpreted as the usual Einstein equation with the energy-momentum
tensor now given by
\[
T_{\mu\nu}=-\frac{\Lambda}{8\pi}g_{\mu\nu}.
\]
Now, let $t^{\mu}$ be a timelike vector normalized such that $\left|\t\right|^{2}=g_{\mu\nu}t^{\mu}t^{\nu}=-1$.
Then the energy density measured by an observer with 4-velocity $t^{\mu}$
is
\begin{equation}
T_{\mu\nu}t^{\mu}t^{\nu}=-\frac{\Lambda}{8\pi}g_{\mu\nu}t^{\mu}t^{\nu}=\frac{\Lambda}{8\pi}.\label{eq:lambda-WEC}
\end{equation}
Hence, a cosmological constant corresponds to a uniform energy density
everywhere in spacetime. Note that the sign of the energy density
is the same as the sign of the cosmological constant; in our universe
we have a positive cosmological constant, and thus positive energy
density.

Finally, note that the trace of the energy-momentum tensor is
\[
T\equiv\udi T{\mu}{\mu}=-\frac{\Lambda}{2\pi}.
\]
Thus
\[
T_{\mu\nu}-\hf Tg_{\mu\nu}=\frac{\Lambda}{8\pi}g_{\mu\nu},
\]
and so
\begin{equation}
\left(T_{\mu\nu}-\hf Tg_{\mu\nu}\right)t^{\mu}t^{\nu}=-\frac{\Lambda}{8\pi}.\label{eq:lambda-SEC-violation}
\end{equation}
We will see the importance of this result in Sec. \ref{subsec:Violations-of-the},
when we will talk about violations of the energy conditions.

\section{Advanced Topics in General Relativity}

\subsection{Causal Structure}

Light cones are sufficient to understand causality in special relativity,
where spacetime is flat. Whatever is inside the past light cone of
a point can affect it, and whatever is in its future light cone can
be affected by it. In a curved spacetime, things are not so simple
anymore. Let us discuss some important concepts now.

\subsubsection{Chronological and Causal Relations}

In general relativity, as we have discussed, it is assumed that massive
particles follow timelike paths, and massless particles follows null
paths. If we want to transmit information, we must use some kind of
particle, either massive or massless. Therefore, in a sense, information
itself must travel along either a timelike or null path. We thus define
a \emph{causal curve }to be a curve whose tangent is non-spacelike
everywhere. Particles following causal curves always travel locally
at or below the speed of light.

Note that the fact that particles must follow causal curves does not
mean they cannot exceed the the speed of light \textbf{globally},
and indeed in these lecture notes we will see some examples where
this in fact happens! However, it is still true that massless and
massive particles particles must \textbf{locally }travel at or below
the speed of light respectively; the faster-than-light travel we will
consider here is a global phenomenon.

At each point $p\in M$ we have a tangent space $T_{p}M$, which is
by definition isomorphic to Minkowski spacetime. Thus, in the tangent
space we may define a light cone passing through the origin, in the
same way that we do in special relativity; it is the region bounded
by null curves emanating from the origin. We may divide this light
cone into a \emph{future light cone} and a \emph{past light cone}.

If there is a choice of past and future that is continuous throughout
spacetime, we say that $M$ is \emph{time orientable}. More precisely,
a spacetime is time orientable if and only if it admits an everywhere
non-vanishing continuous timelike vector field (which points e.g.
to the future). We will assume that all spacetimes are time orientable
from now on.

Let $p$ be a point in our spacetime manifold $M$. We define the
\emph{causal future }$J^{+}\left(p\right)$ to be the set of points
that are connected to $p$ by a future-directed causal curve, and
similarly define the \emph{causal past} $J^{-}\left(p\right)$ to
be the set of points that are connected to $p$ by a past-directed
causal curve, where the future and past directions are determined
by our particular choice of time orientation. The \emph{chronological
future }$I^{+}\left(p\right)$ and \emph{chronological past }$I^{-}\left(p\right)$
are defined similarly, except that we replace causal curves by timelike
curves\footnote{A useful mnemonic is that the extra curve in the bottom of the letter
$J$ compared to $I$ corresponds to the extra type of curves (null
ones) allowed in $J^{\pm}\left(p\right)$ compared to $I^{\pm}\left(p\right)$.}. Finally, if $S\subset M$ is a set of points in $M$, then $J^{+}\left(S\right)$
is defined to be the set of points that are connected by a future-directed
causal curve to any point in $S$, and we similarly define $J^{-}\left(S\right)$
and $I^{\pm}\left(S\right)$.

Note that $I^{+}\left(p\right)$ is an open\footnote{An \emph{open set }in a metric space is a set which contains a neighborhood
of (that is, a ball around) each of its points. For example, $\left(0,1\right)$
is an open set (in $\BBR$) since for each $x\in\left(0,1\right)$
there is a neighborhood $\left(x-\varepsilon,x+\varepsilon\right)\subseteq\left(0,1\right)$
for some $\varepsilon>0$. A \emph{closed set }is a set whose complement
is an open set. For example, $\left[0,1\right]$ is a closed set since
its complement is $\left(-\infty,0\right)\cup\left(1,+\infty\right)$,
which is open.} set; $J^{+}\left(p\right)$ is often a closed set, but not always.
Furthermore, $\overline{I^{+}\left(p\right)}=\overline{J^{+}\left(p\right)}$,
where the bar denotes closure, and $\partial I^{+}\left(p\right)=\partial J^{+}\left(p\right)$.
The proofs of these statements may be found in Hawking and Ellis \cite{HawkingEllis},
pages 182-183.

\subsubsection{\label{subsec:Achronal-Sets-and}Achronal Sets and Cauchy Horizons}

A set $S\subset M$ is called \emph{achronal }if no two points in
it are chronologically connected, that is, no two points can be connected
by a timelike curve. Similarly, a set is \emph{acausal }if no two
points in it are causally connected. We can similarly define achronal
and acausal surfaces. A \emph{spacelike surface} is a surface such
that all tangent vectors to the surface are spacelike.

Acausal sets and spacelike surfaces are completely independent notions.
For example, the surface $t=x$ in flat Minkowski spacetime is achronal,
but it is not spacelike, since its tangent vectors are null. Conversely,
in 2+1-dimensional Minkowski spacetime with polar coordinates $\left(t,r,\phi\right)$,
consider the surface $S$ defined by
\[
\left\{ t=\frac{\phi}{2},r\in\left(1,3\right)\right\} .
\]
This surface is a narrow strip (of width 2) winding around the $t$
axis. Two linearly independent tangent vectors to this 2-dimensional
surface are $\partial_{r}$ (which points across the short side of
the strip) and $\partial_{t}+2\partial_{\phi}$ (which points up the
long side of the strip), and they are both spacelike\footnote{The 2+1D Minkowski metric in polar coordinates is $g_{\mu\nu}=\diag\left(-1,1,r^{2}\right)$.
Thus $\left|\partial_{r}\right|^{2}=1$ and $\left|\partial_{t}+2\partial_{\phi}\right|^{2}=-1+4r^{2}>3$
since $r>1$ inside the surface.}. Thus, the surface is spacelike. However, the point $\left(t=0,r=2,\phi=0\right)$
and the point $\left(t=\pi,r=2,\phi=0\right)$ may be connected by
a timelike curve (with tangent vector $\partial_{t}$), so the surface
is not achronal.

Now, consider a closed achronal set $S$. The \emph{future domain
of dependence} $D^{+}\left(S\right)$ of $S$ is the set of all points
$p$ such that every inextendible\footnote{An \emph{inextendible causal curve} has no ends; it either goes on
forever, or closes in on itself to create a \emph{closed causal curve}.
More precisely, a curve is inextendible if it is not a strict subset
of another curve. For example, consider a curve $C:\left(0,1\right]\to M$,
which has an endpoint: $C\left(1\right)$. Obviously, we may extend
this curve to $\left(0,1+\varepsilon\right)$ for some $\varepsilon>0$,
therefore it is not inextendible.}, past-directed, causal curve through $p$ intersects $S$. In other
words, every point in $D^{+}\left(S\right)$ is causally related to
some point in $S$, so $D^{+}\left(S\right)$ is the region of spacetime
which may be causally influenced by events happening in $S$. The
requirement for the curve to be inextendible is necessary since we
could always have short extendible curves which pass through $p$
and do not intersect $S$, but would intersect it if we extended them.

We similarly define the \emph{past domain of dependence} $D^{-}\left(S\right)$
by replacing ``future'' with ``past'' in the definition above.
The \emph{domain of dependence }is then defined as $D\left(S\right)\equiv D^{+}\left(S\right)\cup D^{-}\left(S\right)$,
and it is the set of all points in spacetime causally connected to
$S$ from either the future or the past. Note that $S$ itself is
inside both $D^{+}\left(S\right)$ and $D^{-}\left(S\right)$.

\subsubsection{Cauchy Surfaces and Globally Hyperbolic Spacetimes}

The \emph{Cauchy horizon }$H\left(S\right)$ separates the points
in $D\left(S\right)$, which are causally connected to $S$, from
the points which are not causally connected to $S$. Thus, the Cauchy
horizon is simply the boundary of $D\left(S\right)$, i.e. $H\left(S\right)=\partial D\left(S\right)$.
In particular, we have the \emph{future Cauchy horizon} $H^{+}\left(S\right)=\partial D^{+}\left(S\right)$
and the \emph{past Cauchy horizon} $H^{-}\left(S\right)=\partial D^{-}\left(S\right)$,
and $H\left(S\right)=H^{+}\left(S\right)\cup H^{-}\left(S\right)$.
Both $H^{+}\left(S\right)$ and $H^{-}\left(S\right)$ are null surfaces.

Note that $D^{+}\left(S\right)\subseteq J^{+}\left(S\right)$, since
by definition $J^{+}\left(S\right)$ is the set of points that are
connected to $S$ by a causal curve. However, there can be points
in $J^{+}\left(S\right)$ which are not in $D^{+}\left(S\right)$,
since a point in $J^{+}\left(S\right)$ only needs to have \textbf{one
}causal curve connected to $S$, while a point in $D^{+}\left(S\right)$
must have \textbf{every }causal curve connected to $S$. Similarly,
$D^{-}\left(S\right)\subseteq J^{-}\left(S\right)$. Also, the closure\footnote{i.e. the set $D^{+}\left(S\right)$ together with all of its limit
points.} $\Db^{+}\left(S\right)$ of $D^{+}\left(S\right)$ is\footnote{The proof is a bit too technical for our purposes, so we will omit
it.} the set of all points $p$ such that every inextendible, past-directed,
\textbf{timelike }curve through $p$ intersects $S$.

We are finally ready to define a \emph{Cauchy surface}. It is a closed
achronal\footnote{Note that some authors require a Cauchy surface to be spacelike, instead
of achronal.} surface $S$ such that its domain of dependence $D\left(S\right)$
is the \textbf{entire spacetime} $M$. Thus, by knowing the information
(e.g. the values of some field) on a Cauchy surface, we are able to
predict (and postdict) what happens throughout all of spacetime. Unfortunately,
not all spacetimes possess Cauchy surfaces; but if a spacetime has
a Cauchy surface, we call it \emph{globally hyperbolic}.

A simple example is any surface of constant $t$ in Minkowski spacetime.
It is the entire universe, ``frozen'' in a single moment in time.
If we know what is happening everywhere in the universe in that particular
moment, then it is straightforward to use the equations of motion
to know what will happen in any other moment.

\subsubsection{Closed Causal Curves and Time Machines}

When a causal (non-spacelike) curve forms a loop, we call it a \emph{closed
causal curve}. If spacetime possesses such a curve, and it passes
through a Cauchy surface, then the information on the surface determines
not only the future, but also -- following the loop -- the information
on the surface itself. A \emph{closed timelike curve }(or \emph{closed
chronological curve}) is defined similarly.

The \emph{causality-violating region }of a spacetime is the set of
all points $p$ which are connected to themselves by a closed causal
curve. Each such point is in its own future (and past), and it is
unclear how causality should work in this situation. We may define
the causality-violating region associated with a point $p$ as:
\[
J^{0}\left(p\right)\equiv J^{+}\left(p\right)\cap J^{-}\left(p\right),
\]
and from this, the causality-violating region for the entire spacetime
manifold $M$:
\[
J^{0}\left(M\right)\equiv\bigcup_{p\in M}J^{0}\left(p\right).
\]
If this region is empty, we say that $M$ is a \emph{causal spacetime}.
Of course, a \emph{chronology-violating region }is defined similarly
with $I$ instead of $J$. If this region is non-empty, we say that
$M$ contains a \emph{time machine}. We will discuss time machines
in detail later in these notes.

\subsection{The Energy Conditions}

\subsubsection{Introduction}

We have seen that the Einstein equation takes the form $G_{\mu\nu}=8\pi T_{\mu\nu}$
where the left-hand side depends on the metric and the right-hand
side depends on the matter fields. Note that if the energy-momentum
tensor is allowed to be arbitrary, then literally any metric trivially
solves this equation! Simply calculate the Einstein tensor $G_{\mu\nu}$
for that metric, and then set $T_{\mu\nu}=G_{\mu\nu}/8\pi$.

Of course, we do know how to write down $T_{\mu\nu}$ for familiar
forms of matter, such as a scalar field or the electromagnetic field.
Using these known expressions, for specific distributions of the matter
fields, definitely limits the metrics which can solve Einstein's equation,
but now it limits them \textbf{too much}. For example, sometimes we
want to derive general results or prove theorems which should be true
in general, and not just for a particular choice of matter fields.

As a compromise, we often do not specify exactly what kind of matter
we have, and only require that it satisfies some conditions that we
consider ``realistic'' or ``physically reasonable''. These are
called the \emph{energy conditions}. Roughly speaking, these conditions
restrict energy density to be non-negative and gravity to be attractive\footnote{Not only as a field of study and research, but also as a force.}.

As we will see later, the strategy taken with the exotic metrics --
which allow faster-than-light travel and/or time travel -- is the
one we described in the beginning; one first writes down the desired
metric, and then calculates the form that $T_{\mu\nu}$ must take
in order to solve the Einstein equation. In almost every single case,
exotic metrics require a type of matter called \emph{exotic matter},
which violates one or more of the energy conditions.

However, this may not necessarily be a problem, since the energy conditions
are just assumptions; so far they have not been proven, in the most
general case, to follow from fundamental principles\footnote{\label{fn:Proof-energy-cond}The energy conditions cannot be proven
from general relativity coupled to classical fields. However, in the
context of quantum field theory (QFT), one may prove that some of
the conditions must hold in specific cases. The averaged null energy
condition (ANEC) -- the weakest (and thus most important) of the
conditions we will present in this chapter -- has been proven in
\cite{Faulkner2016} to be satisfied by a Lorentz-invariant QFT on
a \textbf{flat }spacetime. (Another proof is \cite{Hartman2016},
but since it uses causality as an assumption for the proof, its results
cannot be used to disallow causality violations.) However, quantum
fields on a \textbf{curved }spacetime are known to violate even the
ANEC. Nevertheless, as discussed in \cite{Graham2007}, there is an
even weaker energy condition, known as the \emph{achronal} ANEC, which
is only required to hold on achronal null geodesics (see Subsection
\ref{subsec:Achronal-Sets-and}) and of which there currently exist
no known violations. So far the achronal ANEC has not actually been
proven in the general case, but such a proof could be sufficient to
completely rule out many (although not all) forms of causality violation.
Since QFT is beyond the scope of these notes, we will not go into
any more details here.}. In fact, some known realistic of matter -- in particular, quantum
fields -- are known to violate some or all of the energy conditions
to some extent. We shall see several examples below.

We will begin this section with some important definitions and derivations,
and then proceed to motivate and state the most commonly used energy
conditions.

\subsubsection{Geodesic Deviation}

Consider a \emph{congruence }of timelike geodesics. This is a collection
of geodesics in some (open) region of spacetime such that every point
in that region is on precisely one geodesic. Note that this means
the geodesics do not intersect each other.

More precisely, we consider a 1-parameter family of geodesics $x^{\mu}\left(\sigma,\tau\right)$
such that $\tau$ is the proper time along each geodesic and the parameter
$\sigma$ continuously enumerates the different geodesics. This 1-parameter
family thus defines a 2-dimensional surface with coordinates $\tau,\sigma$.
In addition to the vector $t^{\mu}\left(\sigma,\tau\right)\equiv\partial x^{\mu}/\partial\tau$
tangent to each geodesic, we may also define a \emph{separation vector}:
\[
s^{\mu}\equiv\frac{\partial x^{\mu}}{\partial\sigma}.
\]
The vector with components $V^{\mu}\equiv t^{\nu}\nabla_{\nu}s^{\mu}$,
that is, the covariant derivative of $s^{\mu}$ in the direction of
$t^{\nu}$, describes the \emph{``relative velocity''} of the geodesics,
that is, how the separation between them changes as one moves along
the $\tau$ direction. We can furthermore define a \emph{``relative
acceleration''}:
\[
A^{\mu}\equiv t^{\mu}\nabla_{\mu}V^{\mu},
\]
and one can derive (try it!) the following relation, called the \emph{geodesic
deviation equation}:
\[
A^{\mu}=\udi R{\mu}{\nu\rho\lambda}t^{\nu}t^{\rho}s^{\lambda}.
\]
As expected, we see that the behavior of the separation between our
geodesics is determined by the curvature. Furthermore, one can show
that the ``relative velocity'' satisfies
\[
V^{\mu}=t^{\nu}\nabla_{\nu}s^{\mu}=s^{\nu}\nabla_{\nu}t^{\mu}.
\]
The equation for parallel transporting a vector $t^{\mu}$ along a
curve with tangent vector $s^{\nu}$ is given by $s^{\nu}\nabla_{\nu}t^{\mu}=0$.
Therefore, the amount by which the relative velocity differs from
zero measures how much the tangent vector $t^{\mu}$ fails to be parallel-transported
in the direction of the neighboring geodesics, that is, along $s^{\nu}$.

\subsubsection{\label{subsec:The-Raychaudhuri-Equation}The Raychaudhuri Equation}

In the following we will assume that $t^{\mu}$ is a 4-velocity, so
it is normalized as usual:
\[
\left|\t\right|^{2}=t^{\mu}t_{\mu}=-1.
\]
Let us consider the rank $\left(0,2\right)$ tensor $\nabla_{\mu}t_{\nu}$,
the covariant derivative of the tangent vector to the geodesics. Note
that, if we contract this tensor with $s^{\mu}$, we get $V^{\nu}$.
We may decompose it into a trace, a symmetric traceless component,
and an anti-symmetric component:
\begin{equation}
\nabla_{\mu}t_{\nu}=\trd\theta P_{\mu\nu}+\sigma_{\mu\nu}+\omega_{\mu\nu},\label{eq:Ray-decomp}
\end{equation}
where:
\begin{itemize}
\item $P_{\mu\nu}\equiv g_{\mu\nu}+t_{\mu}t_{\nu}$ is a \emph{projector
}into the subspace (of the tangent space) corresponding to vectors
normal to $t^{\mu}$, defined similarly to (\ref{eq:projector}).
Note that $P^{\mu\nu}P_{\mu\nu}=3$.
\item $\theta\equiv\nabla_{\mu}t^{\mu}$, the trace, is called the \emph{expansion}.
It is the fractional change of volume per unit time, that is, $\theta=\Delta V/\left(V\Delta\tau\right)$
where $V$ is the volume. If $\theta>0$ then the geodesics move away
from each other and the congruence diverges; if $\theta<0$ then the
geodesics get closer together and the congruence converges. Alternatively,
if we start with a sphere of particles, the expansion measures the
change in the volume of the sphere.
\item $\sigma_{\mu\nu}\equiv\nabla_{(\mu}t_{\nu)}-\trd\theta P_{\mu\nu}$,
the symmetric traceless component (such that $P^{\mu\nu}\sigma_{\mu\nu}=0$),
is called the \emph{shear}. If we start out with a sphere of particles,
the shear measures the distortion in its shape into an ellipsoid.
\item $\omega_{\mu\nu}\equiv\nabla_{[\mu}t_{\nu]}$, the anti-symmetric
component, is called the \emph{rotation }or \emph{vorticity}. It measures
how the sphere of particles rotates around itself. If $\omega_{\mu\nu}=0$
then, by \emph{Frobenius' theorem}, the congruence is orthogonal to
a family of hypersurfaces foliating it.
\end{itemize}
Let us now look at the expression $t^{\lambda}\nabla_{\lambda}\nabla_{\mu}t_{\nu}$.
Using the definition of the Riemann tensor as a commutator of covariant
derivatives, (\ref{eq:Riemann-def}), we find that
\[
\nabla_{\lambda}\nabla_{\mu}t_{\nu}=\nabla_{\mu}\nabla_{\lambda}t_{\nu}+R_{\nu\sigma\lambda\mu}t^{\sigma}=\nabla_{\mu}\nabla_{\lambda}t_{\nu}-R_{\sigma\nu\lambda\mu}t^{\sigma},
\]
and thus
\[
t^{\lambda}\nabla_{\lambda}\nabla_{\mu}t_{\nu}=t^{\lambda}\nabla_{\mu}\nabla_{\lambda}t_{\nu}-R_{\sigma\nu\lambda\mu}t^{\sigma}t^{\lambda}.
\]
Taking the trace of this expression (with respect to the indices $\mu,\nu$),
we get after relabeling indices
\[
t^{\lambda}\nabla_{\lambda}\theta=t^{\nu}\nabla_{\mu}\nabla_{\nu}t^{\mu}-R_{\mu\nu}t^{\mu}t^{\nu}.
\]
Next, we note that
\[
t^{\nu}\nabla_{\mu}\nabla_{\nu}t^{\mu}=\nabla_{\mu}\left(t^{\nu}\nabla_{\nu}t^{\mu}\right)-\left(\nabla_{\mu}t_{\nu}\right)\left(\nabla^{\nu}t^{\mu}\right),
\]
and furthermore, using (\ref{eq:Ray-decomp}) we have
\[
\left(\nabla_{\mu}t_{\nu}\right)\left(\nabla^{\nu}t^{\mu}\right)=\trd\theta^{2}+\sigma_{\mu\nu}\sigma^{\mu\nu}-\omega_{\mu\nu}\omega^{\mu\nu}.
\]
Thus, we obtain
\[
t^{\lambda}\nabla_{\lambda}\theta=\nabla_{\mu}\left(t^{\nu}\nabla_{\nu}t^{\mu}\right)+\omega_{\mu\nu}\omega^{\mu\nu}-\sigma_{\mu\nu}\sigma^{\mu\nu}-\trd\theta^{2}-R_{\mu\nu}t^{\mu}t^{\nu}.
\]
Finally, since $t^{\mu}$ is a tangent vector to a timelike geodesic,
it satisfies the geodesic equation, $t^{\nu}\nabla_{\nu}t^{\mu}$,
so the first term on the right-hand side vanishes. Moreover, $t^{\lambda}\nabla_{\lambda}=\d/\d\tau$
by the chain rule. Therefore, our equation becomes
\begin{equation}
\frac{\d\theta}{\d\tau}=\omega_{\mu\nu}\omega^{\mu\nu}-\sigma_{\mu\nu}\sigma^{\mu\nu}-\trd\theta^{2}-R_{\mu\nu}t^{\mu}t^{\nu}.\label{eq:Raychaudhuri}
\end{equation}
This is called the \emph{Raychaudhuri equation}. It has many uses,
among them proving the Penrose-Hawking singularity theorems. Below
we will see that it is used to impose the condition that gravity is
always attractive. Note that this equation applies to a congruence
of timelike geodesics; a similar equation may be derived for a congruence
of null geodesics:
\[
\frac{\d\theta}{\d\lambda}=\omega_{\mu\nu}\omega^{\mu\nu}-\sigma_{\mu\nu}\sigma^{\mu\nu}-\hf\theta^{2}-R_{\mu\nu}n^{\mu}n^{\nu},
\]
where $\lambda$ is an affine parameter and $n^{\mu}$ is a null vector.
The derivation of this equation is a bit more involved, and we will
not do it here.

Finally, recall from (\ref{eq:extrinsic-curvature-geodesic}) that
the extrinsic curvature to a surface $\Sigma$ may be written as $K_{\mu\nu}=\nabla_{(\mu}t_{\nu)}$
where $t^{\mu}$ is a vector tangent to a geodesic passing through
$\Sigma$ and normal to it. Thus, the expansion is none other than
the trace of the extrinsic curvature:
\begin{equation}
\theta=K_{\mu}^{\mu}=\nabla_{\mu}t^{\mu}.\label{eq:expansion-trace}
\end{equation}

\subsubsection{The Strong Energy Condition}

Let us first find how to impose the condition that gravity should
always be attractive. If this condition is satisfied, then a collection
of (free-falling) massive particles should get closer to each other
over time. Since the particles follow timelike geodesics, they may
be collectively described as a congruence of such geodesics, and thus
they satisfy the Raychaudhuri equation, (\ref{eq:Raychaudhuri}).

If gravity is attractive, then we expect $\d\theta/\d\tau$ to be
negative, signifying that the expansion $\theta$ decreases as time
passes, that is, the geodesics get closer together\footnote{If initially $\theta>0$, then $\d\theta/\d\tau<0$ means that the
congruence will diverge less rapidly until it eventually starts converging
when $\theta<0$.}. Let us assume that the congruence has no vorticity, so $\omega_{\mu\nu}=0$.
Since the terms $\sigma_{\mu\nu}\sigma^{\mu\nu}$ and $\theta^{2}$
are always positive, the only term on the right-hand side of (\ref{eq:Raychaudhuri})
which might prevent $\d\theta/\d\tau$ from being negative is the
last term, $R_{\mu\nu}t^{\mu}t^{\nu}$. It is thus reasonable to demand
\[
R_{\mu\nu}t^{\mu}t^{\nu}\ge0\sp\textrm{for any timelike vector }t^{\mu}.
\]
This is a sufficient, but not necessary, condition for gravity to
always be attractive.

Now, using the Einstein equation in the form (\ref{eq:Einstein-T}),
we see that this condition is equivalent\footnote{Note that if the Einstein equation is altered, as in some modified
gravity theories, then this equivalence does not hold anymore.} to
\begin{center}
The Strong Energy Condition (SEC):
\[
\left(T_{\mu\nu}-\hf Tg_{\mu\nu}\right)t^{\mu}t^{\nu}\ge0\sp\textrm{for any timelike vector }t^{\mu}.
\]
\par\end{center}

Imposing the SEC thus guarantees that gravity is attractive for massive
particles. The statement that gravity is attractive if the SEC is
satisfied is called the \emph{focusing theorem}.

\subsubsection{The Null, Weak and Dominant Energy Conditions}

In the SEC, we can take timelike vectors which are arbitrarily close
to being null. Since by definition $g_{\mu\nu}n^{\mu}n^{\nu}=0$ for
a null vector $n^{\mu}$, in the limit of null vectors $t^{\mu}\to n^{\mu}$
we obtain
\begin{center}
The Null Energy Condition (NEC):
\[
T_{\mu\nu}n^{\mu}n^{\nu}\ge0\sp\textrm{for any null vector }n^{\mu}.
\]
\par\end{center}

The NEC thus imposes that gravity is attractive also for massless
particles -- rays of light also converge under the influence of gravity.
If it is violated, then the SEC will be violated as well.

Another reasonable assumption is that the energy density measured
by any observer should be non-negative, which leads us to
\begin{center}
The Weak Energy Condition (WEC):
\[
T_{\mu\nu}t^{\mu}t^{\nu}\ge0\sp\textrm{for any timelike vector }t^{\mu}.
\]
\par\end{center}

Here, $t^{\mu}$ is the 4-velocity of the observer performing the
measurement, and $T_{\mu\nu}t^{\mu}t^{\nu}$ is the energy density.
Despite the naming, the WEC isn't actually a weaker form of the SEC;
they are in fact independent. Note that the NEC may also be derived
from the WEC in the limit of null vectors. Thus, if the NEC is violated,
then so is the WEC.

One may furthermore impose that the energy density of massive particles
propagates \textbf{in a causal way}. To do this, we define the \emph{energy
flux vector} $F^{\mu}\equiv-\udi T{\mu}{\nu}t^{\nu}$, which encodes
the energy and momentum density of the matter as measured by an observer
moving at 4-velocity $t^{\mu}$. We require that it is causal, that
is, $\left|\F\right|^{2}\equiv F^{\mu}F_{\mu}\le0$. Adding this condition
to the WEC, we get
\begin{center}
The Dominant Energy Condition (DEC):
\[
T_{\mu\nu}t^{\mu}t^{\nu}\ge0\textrm{ and }\udi T{\mu}{\nu}t^{\nu}\textrm{ is not spacelike}\sp\textrm{for any timelike vector }t^{\mu}.
\]
\par\end{center}

If the WEC is violated, then the DEC is violated as well. Note that
this means if the NEC is violated, then so will all the other energy
conditions we have defined!

\subsubsection{\label{subsec:Energy-Density-Pressure}Energy Density and Pressure}

Let us now assume that the energy-momentum tensor $T_{\mu\nu}$ is
of the \emph{Hawking-Ellis Type I} form, which means there is some
orthonormal basis\footnote{\label{fn:Frame-Field}An orthonormal basis is a set of four 4-vectors
$\ee_{A}$, referred to as \emph{frame fields}, \emph{vierbeins} or
\emph{tetrads}, with components $e_{A}^{\mu}$ where $\mu$ is the
usual spacetime index and $A\in\left\{ 0,1,2,3\right\} $ is called
an \emph{internal index}, such that $\left\langle \ee_{A},\ee_{B}\right\rangle \equiv g_{\mu\nu}e_{A}^{\mu}e_{B}^{\nu}=\eta_{AB}$.
The energy-momentum tensor is then decomposed as 
\begin{equation}
T^{\mu\nu}=\rho e_{0}^{\mu}e_{0}^{\nu}+\sum_{i=1}^{3}p_{i}e_{i}^{\mu}e_{i}^{\nu}.\label{eq:T-orth-basis}
\end{equation}
Then $\rho$ and $p_{i}$ are the eigenvalues of $T_{\mu\nu}$, while
$\ee_{A}$ are its eigenvectors. The spacetime components $v^{\mu}$
of an abstract vector $\v$ are related to its components $v^{A}$
in the orthonormal basis by $v^{\mu}=v^{A}e_{A}^{\mu}$.} such that its components take the diagonal form
\begin{equation}
T^{\mu\nu}=\diag\left(\rho,p_{1},p_{2},p_{3}\right),\label{eq:T-diag}
\end{equation}
where $\rho$ is the \emph{energy density }and $p_{i}$ are the three
\emph{principle pressures}. An example of a type of matter which has
an energy-momentum tensor of this form is a \emph{perfect fluid}\footnote{This is a special case of (\ref{eq:T-orth-basis}) where we identify
$\ee_{0}$ with the 4-velocity $\u$ and take $p_{i}=p$. Using the
fact that $g^{\mu\nu}=\eta^{AB}e_{A}^{\mu}e_{B}^{\nu}=-u^{\mu}u^{\nu}+\sum_{i=1}^{3}e_{i}^{\mu}e_{i}^{\nu}$,
we can then write $\sum_{i=1}^{3}p_{i}e_{i}^{\mu}e_{i}^{\nu}=p\left(u^{\mu}u^{\nu}+g^{\mu\nu}\right)$.}, which has
\[
T^{\mu\nu}=\left(\rho+p\right)u^{\mu}u^{\nu}+pg^{\mu\nu},
\]
where $u^{\mu}$ is the 4-velocity of the fluid.

In the orthonormal frame where the energy-momentum tensor is of the
form (\ref{eq:T-diag}), we find that the four energy conditions we
have considered above imply the following:
\begin{itemize}
\item Null Energy Condition: $\rho+p_{i}\ge0$ for all $i$.
\item Weak Energy Condition: $\rho+p_{i}\ge0$ for all $i$ and $\rho\ge0$.
\item Strong Energy Condition: $\rho+p_{i}\ge0$ for all $i$ and $\rho+\sum_{i}p_{i}\ge0$.
\item Dominant Energy Condition: $\left|p_{i}\right|\le\rho$ for all $i$
and $\rho\ge0$.
\end{itemize}

\subsubsection{The Averaged Energy Conditions}

If we integrate the Raychaudhuri equation (\ref{eq:Raychaudhuri})
along a timelike geodesic $\Gamma$ with tangent vector $t^{\mu}$,
we get
\[
\Delta\theta=\int_{\Gamma}\left(\omega_{\mu\nu}\omega^{\mu\nu}-\sigma_{\mu\nu}\sigma^{\mu\nu}-\trd\theta^{2}-R_{\mu\nu}t^{\mu}t^{\nu}\right)\d\tau,
\]
where $\Delta\theta$ is the difference in the expansion between the
initial and final points of the geodesic. If we assume $\omega_{\mu\nu}=0$
as before, we get that
\[
\Delta\theta\le-\int_{\Gamma}R_{\mu\nu}t^{\mu}t^{\nu}\d\tau.
\]
Therefore, for the geodesics to converge (i.e. for $\theta$ to decrease)
it is in fact enough to require that
\[
\int_{\Gamma}R_{\mu\nu}t^{\mu}t^{\nu}\d\tau\ge0\sp\textrm{for any timelike geodesic }\Gamma\textrm{ with tangent }t^{\mu}.
\]
As before, using the Einstein equation, we obtain
\[
\textrm{The Averaged Strong Energy Condition (ASEC):}
\]
\[
\int_{\Gamma}\left(T_{\mu\nu}-\hf Tg_{\mu\nu}\right)t^{\mu}t^{\nu}\d\tau\ge0\sp\textrm{for any timelike geodesic }\Gamma\textrm{ with tangent }t^{\mu}.
\]
In the null limit we have
\[
\textrm{The Averaged Null Energy Condition (ANEC):}
\]
\[
\int_{\Gamma}T_{\mu\nu}n^{\mu}n^{\nu}\d\lambda\ge0\sp\textrm{for any null geodesic }\Gamma\textrm{ with tangent }n^{\mu}.
\]
Here $\lambda$ is an affine parameter\footnote{If we allow the parameter $\lambda$ to be arbitrary, the ANEC will
be equivalent to the (non-averaged) NEC.} along the null geodesic. The ANEC is the weakest energy condition
we have -- but unfortunately, even this condition is violated in
some cases.

Finally, we also define
\[
\textrm{The Averaged Weak Energy Condition (AWEC):}
\]
\[
\int_{\Gamma}T_{\mu\nu}t^{\mu}t^{\nu}\d\tau\ge0\sp\textrm{for any timelike geodesic }\Gamma\textrm{ with tangent }t^{\mu}.
\]
The idea behind the averaged energy conditions is that the (non-averaged)
energy conditions are allowed to be violated in some regions along
the geodesics, as long as they are balanced by positive contributions
from other regions; they are thus satisfied ``on average''.

\subsection{\label{subsec:Violations-of-the}Violations of the Energy Conditions}

Despite the fact that the energy conditions are supposed to ``reasonable'',
some types of matter are known to violate some or all of these conditions.
These include both classical and quantum matter. A comprehensive list
(with references) may be found in \cite{Curiel}. We will not discuss
examples from quantum field theory here, since they are beyond the
scope of these notes; instead, we will focus on classical examples.

A particularly simple example is the cosmological constant; in (\ref{eq:lambda-SEC-violation})
we saw that
\[
\left(T_{\mu\nu}-\hf Tg_{\mu\nu}\right)t^{\mu}t^{\nu}=-\frac{\Lambda}{8\pi}.
\]
Thus, a positive\footnote{In (\ref{eq:lambda-WEC}) we found that $T_{\mu\nu}t^{\mu}t^{\nu}=\Lambda/8\pi$
and thus a negative cosmological constant would violate the WEC (and,
trivially, also the AWEC) since it corresponds to a negative energy
density everywhere in spacetime. However, our universe happens to
have a positive $\Lambda$, and therefore we already know that the
cosmological constant in our universe in fact does satisfy the WEC.} cosmological constant (as is the case in our universe) violates the
SEC. Since the SEC was derived by requiring gravity to be attractive,
a positive cosmological constant may potentially make gravity repulsive
instead. This makes sense, given that the cosmological constant accelerates
the expansion of the universe, so in a way it makes galaxies ``repel''
each other .

\subsubsection{Basic Facts about Scalar Fields}

We now move to a more complicated example, that of a classical scalar
field. The discussion is based largely on \cite{VisserBarcelo}.

The action for a classical scalar field $\phi$ coupled to gravity
is
\begin{equation}
S=-\int\d^{4}x\sqrt{-g}\left(\hf\left(g^{\mu\nu}\nabla_{\mu}\phi\nabla_{\nu}\phi+\xi R\phi^{2}\right)+V\left(\phi\right)\right),\label{eq:scalar-action}
\end{equation}
where $R$ is the Ricci scalar, $V\left(\phi\right)$ is the field's
potential, and $\xi$ is a coupling constant. If $\xi=0$, the field
is referred to as \emph{minimally coupled }to gravity. Scalar field
theories describe, for example, the Higgs field, scalar mesons (pions,
etc.), and some hypothesized fields such as axions and inflatons.

The equation of motion may be found by varying the action with respect
to $\phi$:
\[
\delta S=-\int\d^{4}x\sqrt{-g}\left(g^{\mu\nu}\nabla_{\mu}\phi\nabla_{\nu}\delta\phi+\xi R\phi\delta\phi+\frac{\d V}{\d\phi}\delta\phi\right).
\]
To isolate $\delta\phi$, we need to integrate the first term by parts.
This is easy because the connection is metric-compatible, $\nabla_{\lambda}g_{\mu\nu}=0$,
so
\[
\sqrt{-g}\thinspace g^{\mu\nu}\nabla_{\mu}\phi\nabla_{\nu}\delta\phi=-\sqrt{-g}\left(g^{\mu\nu}\nabla_{\nu}\nabla_{\mu}\phi\right)\delta\phi,
\]
where we ignored boundary terms since we assume the variation vanishes
at infinity. It is customary to define the \emph{covariant d'Alembertian}
or \emph{``box'' operator}:
\[
\sq\equiv g^{\mu\nu}\nabla_{\nu}\nabla_{\mu}.
\]
Then we get
\[
\delta S=\int\d^{4}x\sqrt{-g}\left(\sq\phi-\xi R\phi-\frac{\d V}{\d\phi}\right)\delta\phi,
\]
so the equation of motion is
\[
\sq\phi-\xi R\phi-\frac{\d V}{\d\phi}=0.
\]
The corresponding energy-momentum tensor may be calculated using (\ref{eq:EM-tensor}).
For this we must vary the action with respect to the inverse metric
$g^{\mu\nu}$. For this we will use the relation\footnote{To derive this, recall the familiar relation $\ln\left(\det\M\right)=\tr\left(\ln\M\right)$,
where $\M$ is any matrix. Taking the variation, we find
\[
\frac{\delta\left(\det\M\right)}{\det\M}=\tr\left(\delta\M\M^{-1}\right).
\]
Note that the order of $\delta\M$ and $\M^{-1}$ inside the trace
does not matter, since it is cyclic. Applying this relation to the
metric $\g$, we find
\[
\frac{\delta g}{g}=\tr\left(\delta\g\g^{-1}\right)=\delta g_{\mu\nu}g^{\mu\nu}.
\]
Now, since $g_{\mu\nu}g^{\mu\nu}=4$, we have
\[
\delta\left(g_{\mu\nu}g^{\mu\nu}\right)=\delta g_{\mu\nu}g^{\mu\nu}+g_{\mu\nu}\delta g^{\mu\nu}=0,
\]
so $\delta g_{\mu\nu}g^{\mu\nu}=-g_{\mu\nu}\delta g^{\mu\nu}$, and
we can write $\delta g=-gg_{\mu\nu}\delta g^{\mu\nu}$. Finally, we
have
\[
\delta\sqrt{-g}=-\frac{\delta g}{2\sqrt{-g}}=-\frac{\left(-g\right)g_{\mu\nu}\delta g^{\mu\nu}}{2\sqrt{-g}}=-\hf\sqrt{-g}g_{\mu\nu}\delta g^{\mu\nu}.
\]
}
\[
\delta\sqrt{-g}=-\hf\sqrt{-g}g_{\mu\nu}\delta g^{\mu\nu}.
\]
We will also need the variation of $R$, which turns out to be (prove
this!)
\[
\delta R=R_{\mu\nu}\delta g^{\mu\nu}+g_{\mu\nu}\sq\delta g^{\mu\nu}-\nabla_{\mu}\nabla_{\nu}\delta g^{\mu\nu}.
\]
Taking the variation of the action with respect to $g^{\mu\nu}$,
we get after some manipulations, using (\ref{eq:EM-tensor}):
\[
T_{\mu\nu}=\nabla_{\mu}\phi\nabla_{\nu}\phi-g_{\mu\nu}\left(\hf g^{\rho\sigma}\nabla_{\rho}\phi\nabla_{\sigma}\phi+V\left(\phi\right)\right)+\xi\left(g_{\mu\nu}\sq+G_{\mu\nu}-\nabla_{\mu}\nabla_{\nu}\right)\phi^{2},
\]
where $G_{\mu\nu}\equiv R_{\mu\nu}-\hf Rg_{\mu\nu}$ is the Einstein
tensor (\ref{eq:Einstein-tensor}). Now, let us plug this into the
Einstein equation (\ref{eq:Einstein}):
\[
G_{\mu\nu}=8\pi\left(\nabla_{\mu}\phi\nabla_{\nu}\phi-g_{\mu\nu}\left(\hf g^{\rho\sigma}\nabla_{\rho}\phi\nabla_{\sigma}\phi+V\left(\phi\right)\right)+\xi\left(g_{\mu\nu}\sq+G_{\mu\nu}-\nabla_{\mu}\nabla_{\nu}\right)\phi^{2}\right).
\]
Note that the Einstein tensor $G_{\mu\nu}$ appears on both sides.
If we isolate it, we get
\[
G_{\mu\nu}=\frac{8\pi}{1-8\pi\xi\phi^{2}}\left(\nabla_{\mu}\phi\nabla_{\nu}\phi-g_{\mu\nu}\left(\hf g^{\rho\sigma}\nabla_{\rho}\phi\nabla_{\sigma}\phi+V\left(\phi\right)\right)+\xi\left(g_{\mu\nu}\sq\phi^{2}-\nabla_{\mu}\nabla_{\nu}\phi^{2}\right)\right).
\]

Therefore we may define a simpler \emph{effective energy-momentum
tensor}:
\begin{equation}
\Tt_{\mu\nu}=\frac{1}{1-8\pi\xi\phi^{2}}\left(\nabla_{\mu}\phi\nabla_{\nu}\phi-g_{\mu\nu}\left(\hf g^{\rho\sigma}\nabla_{\rho}\phi\nabla_{\sigma}\phi+V\left(\phi\right)\right)+\xi\left(g_{\mu\nu}\sq\phi^{2}-\nabla_{\mu}\nabla_{\nu}\phi^{2}\right)\right).\label{eq:effective-T}
\end{equation}
This tensor then appears on the right-hand side of the Einstein equation,
$R_{\mu\nu}-\hf Rg_{\mu\nu}=8\pi\Tt_{\mu\nu}$, such that now the
right-hand side does \textbf{not }depend directly on the Ricci tensor
$R_{\mu\nu}$. Therefore it is the energy-momentum tensor of interest
with respect to the energy conditions. Indeed, recall that we derived
the SEC by first demanding that gravity attracts, thus finding a condition
on $R_{\mu\nu}$, and then used Einstein's equation to convert it
into a condition on $T_{\mu\nu}$. Hence, in that derivation we must
use $\Tt_{\mu\nu}$ and not the original $T_{\mu\nu}$. From now on
we will just write $T_{\mu\nu}$ for the effective energy-momentum
tensor, for brevity.

\subsubsection{\label{subsec:The-Scalar-Field}The Scalar Field and the Energy Conditions}

We are now ready to test whether the scalar field satisfies the energy
conditions. Already from the overall coefficient $1/\left(1-8\pi\xi\phi^{2}\right)$
in (\ref{eq:effective-T}) we can see that something fishy is going
on, since for $\xi\ne0$ the sign of $\Tt_{\mu\nu}$ depends on $\phi$!

Let us begin by testing the SEC:
\[
\left(T_{\mu\nu}-\hf Tg_{\mu\nu}\right)t^{\mu}t^{\nu}\ge0,
\]
where $t^{\mu}$ is a timelike vector. In fact, this energy condition
is already violated for a minimally-coupled scalar field, that is,
with $\xi=0$, where the energy-momentum tensor (\ref{eq:effective-T})
simplifies to
\[
T_{\mu\nu}=\nabla_{\mu}\phi\nabla_{\nu}\phi-g_{\mu\nu}\left(\hf g^{\rho\sigma}\nabla_{\rho}\phi\nabla_{\sigma}\phi+V\left(\phi\right)\right),
\]
and its trace is
\[
T=g^{\mu\nu}T_{\mu\nu}=-g^{\rho\sigma}\nabla_{\rho}\phi\nabla_{\sigma}\phi-4V\left(\phi\right).
\]
For a timelike vector $t^{\mu}$ normalized as usual such that $\left|\t\right|^{2}=-1$,
we get
\[
\left(T_{\mu\nu}-\hf Tg_{\mu\nu}\right)t^{\mu}t^{\nu}=\left(t^{\mu}\nabla_{\mu}\phi\right)^{2}-V\left(\phi\right).
\]
If $t^{\mu}\nabla_{\mu}\phi$ is small (the field changes slowly in
the direction of $t^{\mu}$) and the potential is positive (e.g. $V\left(\phi\right)=\hf m^{2}\phi^{2}$
for a free massive field), we see that the SEC is easily violated!

Next we move to the NEC:
\[
T_{\mu\nu}n^{\mu}n^{\nu}\ge0,
\]
where $n^{\mu}$ is a null vector. Since $g_{\mu\nu}n^{\mu}n^{\nu}=0$,
we get
\begin{equation}
T_{\mu\nu}n^{\mu}n^{\nu}=\frac{1}{1-8\pi\xi\phi^{2}}\left(\left(n^{\mu}\nabla_{\mu}\phi\right)^{2}-\xi n^{\mu}n^{\nu}\nabla_{\mu}\nabla_{\nu}\phi^{2}\right).\label{eq:Tnn-xi}
\end{equation}
For minimal coupling, $\xi=0$, this simplifies to
\[
T_{\mu\nu}n^{\mu}n^{\nu}=\left(n^{\mu}\nabla_{\mu}\phi\right)^{2},
\]
which is obviously non-negative. So a minimally-coupled scalar field
satisfies the NEC. However, for $\xi\ne0$, let us assume that $n^{\mu}$
is tangent to a null geodesic, so that $n^{\mu}\nabla_{\mu}n^{\nu}=0$.
If $\lambda$ is an affine parameter, such that $n^{\mu}\nabla_{\mu}=\d/\d\lambda$,
then we may write (\ref{eq:Tnn-xi}) as
\[
T_{\mu\nu}n^{\mu}n^{\nu}=\frac{1}{1-8\pi\xi\phi^{2}}\left(\left(\frac{\d\phi}{\d\lambda}\right)^{2}-\xi\frac{\d^{2}\left(\phi^{2}\right)}{\d\lambda^{2}}\right).
\]
At a local extremum of $\phi^{2}$ along the null geodesic we have
\[
\frac{\d\left(\phi^{2}\right)}{\d\lambda}=2\phi\frac{\d\phi}{\d\lambda}=0,
\]
and thus as long as $\phi\ne0$, we have $\d\phi/\d\lambda=0$ there.
Hence
\[
T_{\mu\nu}n^{\mu}n^{\nu}=\frac{1}{8\pi\phi^{2}-1/\xi}\frac{\d^{2}\left(\phi^{2}\right)}{\d\lambda^{2}}.
\]
Now, if $\xi<0$, or $\xi>0$ and $\phi^{2}>1/8\pi\xi$ at the extremum,
then the denominator will be positive and thus any local maximum of
$\phi^{2}$, which has $\d^{2}\left(\phi^{2}\right)/\d\lambda^{2}<1$,
will violate the NEC. Similarly, if $\xi>0$ and $\phi^{2}<1/8\pi\xi$
then the denominator will be negative and thus any local minimum of
$\phi^{2}$ will violate the NEC. Recall that the NEC is the weakest
energy condition, in the sense that if the NEC is violated, then so
will all the other energy conditions. Therefore, a classical scalar
field which is not minimally coupled violates all of the energy conditions!

In fact, we can go even weaker and show the the ANEC is violated,
which means all of the averaged energy conditions are violated. We
have
\[
\int T_{\mu\nu}n^{\mu}n^{\nu}\d\lambda=\int\frac{1}{1-8\pi\xi\phi^{2}}\left(\left(\frac{\d\phi}{\d\lambda}\right)^{2}-\xi\frac{\d^{2}\left(\phi^{2}\right)}{\d\lambda^{2}}\right)\d\lambda.
\]
Integrating the second term by parts, discarding the boundary term,
and collecting terms, we get
\[
\int T_{\mu\nu}n^{\mu}n^{\nu}\d\lambda=\int\left(\frac{1-8\pi\xi\left(1-4\xi\right)\phi^{2}}{\left(1-8\pi\xi\phi^{2}\right)^{2}}\right)\left(\frac{\d\phi}{\d\lambda}\right)^{2}\d\lambda.
\]
Thus, the ANEC is violated when
\[
\xi\left(1-4\xi\right)>\frac{1}{8\pi\phi^{2}}.
\]
This is only possible for $\xi\in\left(0,1/4\right)$, since otherwise
the left-hand side will be negative. In fact, the value $\xi=1/6$,
which is within this range, corresponds to \emph{conformal coupling
}which is thought to be the most natural way in which quantum scalar
fields should be coupled to gravity. In terms of $\phi$ we may write
this as
\begin{equation}
\phi^{2}>\frac{1}{8\pi\xi\left(1-4\xi\right)}>\frac{2}{\pi},\label{eq:phi-greater}
\end{equation}
where in the last step we used the fact that $\xi\left(1-4\xi\right)$
has a local maximum at $\xi=1/8$, where $\xi\left(1-4\xi\right)=1/16$.
Since we are working in Planck units, this means that $\phi$ must
take a value at the Planck scale in order for the ANEC to be violated.

\subsubsection{The Jordan and Einstein Frames}

Let us combine the scalar field action (\ref{eq:scalar-action}) with
the Einstein-Hilbert action (\ref{eq:Einstein-Hilbert}):
\[
S=-\int\d^{4}x\sqrt{-g}\left(\hf\left(g^{\mu\nu}\nabla_{\mu}\phi\nabla_{\nu}\phi+\left(\xi\phi^{2}-\frac{1}{8\pi}\right)R\right)+V\left(\phi\right)\right).
\]

If our theory is described by such an action, where there is a term
in which the Ricci scalar $R$ is multiplied by the scalar field $\phi$,
we say that it is in the \emph{Jordan frame}. If we take a conformal
transformation \cite{Faraoni1998} 
\[
g_{\mu\nu}\mt\Omega^{2}g_{\mu\nu}\sp\Omega^{2}=1-8\pi\xi\phi^{2},
\]
and redefine the scalar field
\[
\d\phi\mt\frac{\sqrt{1-8\pi\xi\left(1-6\xi\right)\phi^{2}}}{1-8\pi\xi\phi^{2}}\d\phi,
\]
we obtain gravity with a \textbf{minimally-coupled} scalar field:
\[
S=-\int\d^{4}x\sqrt{-g}\left(\hf\left(g^{\mu\nu}\nabla_{\mu}\phi\nabla_{\nu}\phi-\frac{1}{8\pi}R\right)+V\left(\phi\right)\right).
\]
We call this the \emph{Einstein frame}. This is still the same action
as before, only written in different variables. Does this mean that
a minimally-coupled scalar field may also violate the ANEC, as we
have shown above for the action (\ref{eq:scalar-action}) with $\xi\in\left(0,1/4\right)$?
In fact, that is not the case. Note that this transformation is only
possible if 
\[
\Omega^{2}=1-8\pi\xi\phi^{2}>0\sp1-8\pi\xi\left(1-6\xi\right)\phi^{2}>0.
\]
Let us only consider conformal coupling, $\xi=1/6$, for simplicity.
Then the second condition is always satisfied, and the first condition
reduces to
\begin{equation}
\phi^{2}<\frac{3}{4\pi}.\label{eq:phi-less}
\end{equation}
However, in order to violate the ANEC, (\ref{eq:phi-greater}) must
also be satisfied. For $\xi=1/6$, (\ref{eq:phi-greater}) becomes
\begin{equation}
\phi^{2}>\frac{9}{4\pi}.\label{eq:phi-greater-conf}
\end{equation}
Clearly, we cannot have both (\ref{eq:phi-less}) and (\ref{eq:phi-greater-conf})
at the same time. Thus, we cannot use this trick to make a minimally-coupled
scalar field violate the ANEC.

\subsubsection{Conclusions}

To summarize, we have found that a positive cosmological constant
and a minimally-coupled scalar field violate the SEC, while a conformally-coupled
scalar field violates the NEC and thus all the other energy conditions,
and can even violate the ANEC, which is the weakest energy condition
we have defined.

As we commented in the beginning, many other examples exist, some
classical and some quantum. However, the few simple examples we have
considered here should already make it clear that the energy conditions
are definitely not fundamental axioms or results of general relativity,
but rather additional assumptions that do not necessarily apply to
all conceivable forms of matter.

As we will see below, it is possible to define spacetime geometries
in general relativity which potentially allow for faster-than-light
travel and/or time travel. These spacetimes almost universally require,
through the Einstein equation, matter which violates the energy conditions,
also known as \emph{exotic matter. }It is currently unknown whether
exotic matter may realistically be used to facilitate these exotic
spacetime geometries. We will discuss this in more detail below.

\section{Faster-than-Light Travel and Time Travel in Special Relativity}

\subsection{The Nature of Velocity}

In this section we will work in special relativity, in which spacetime
is flat and described by the Minkowski metric $\eta_{\mu\nu}\equiv\diag\left(-1,1,1,1\right)$.
These results also work in general relativity, but only locally, if
one uses locally inertial coordinates as described in Sec. \ref{subsec:Massless-Particles-and}.
Indeed, one of the major differences between special and general relativity
is that in the former the inertial coordinates are global, while in
the latter one can generally only define them locally, that is, at
one particular point in spacetime.

\subsubsection{Movement in the Time Direction}

Recall that massive particles move along timelike paths. Such paths,
when parametrized as usual by the proper time $\tau$, have tangent
vectors $\xd^{\mu}$ with $\left|\xxd\right|^{2}=-1$. The tangent
vector is the 4-velocity of the particle. If the particle is at rest,
then it has
\[
\xd^{\mu}=\left(1,0,0,0\right).
\]
In other words, it has no velocity along any of the spatial coordinates,
but it ``moves at the speed of light\footnote{Recall that we are using units where $c\equiv1$.}''
along the time coordinate. Indeed, a massive particle must \textbf{always
}move in time, since if we take $\td=0$ then there is no way to satisfy
the condition
\[
\left|\xxd\right|^{2}=-\td^{2}+\xd^{2}+\yd^{2}+\zd^{2}<0.
\]
Thus, we learn that the 4-velocity for a massive particle must have
a non-zero value in the time component.

Note that we said ``non-zero'', not ``positive''; even if $\td<0$,
we still have $\td^{2}>0$. In other words, $\xd^{\mu}=\left(-1,0,0,0\right)$
is also a perfectly good 4-velocity, except that it is past-directed
instead of future-directed. Unfortunately, time travel is not as easy
as taking $\td<0$; in practice we always assume\footnote{Another way to phrase this assumption is that the \emph{``arrow of
time'' }points to the future. This result \textbf{cannot }be derived
from relativity; in fact, relativity is completely invariant under
\emph{time reversal}, $t\mt-t$. However, we nonetheless know that
physics is in general not time-reversal-invariant; entropy increases
when going forward in time. Understanding this discrepancy is a major
open problem in physics.} that the 4-velocity is future-directed, $\td>0$.

\subsubsection{The Lorentz Factor}

Let us now imagine a particle moving at some constant spatial 3-velocity
$v$ along, say, the $x^{1}$ direction:
\[
v\equiv\frac{\d x^{1}}{\d t}=\frac{\d x^{1}}{\d x^{0}}.
\]
Since the 4-velocity is defined as the derivative with respect to
the proper time $\tau$, \textbf{not} the coordinate time $t=x^{0}$,
the corresponding spatial component $\xd^{1}$ of the 4-velocity will
be
\[
\xd^{1}\equiv\frac{\d x^{1}}{\d\tau}=\frac{\d t}{\d\tau}\frac{\d x^{1}}{\d t}=\gamma v,
\]
where $\gamma\equiv\d t/\d\tau$ is called the \emph{Lorentz factor},
and it measures the relation between coordinate time and proper time.
In other words, $\gamma$ measures the amount of \emph{time dilation},
since $\d t=\gamma\thinspace\d\tau$.

Similarly, the time component $\xd^{0}=\td$ of the 4-velocity will
be
\[
\td\equiv\frac{\d t}{\d\tau}=\gamma.
\]
In conclusion, the 4-velocity is
\begin{equation}
\xd^{\mu}=\gamma\left(1,v,0,0\right).\label{eq:4-velocity}
\end{equation}
To derive an explicit value for $\gamma$, we use the normalization
condition:
\begin{equation}
\left|\xxd\right|^{2}=\gamma^{2}\left(-1+v^{2}\right)=-1\soosp\gamma=\frac{1}{\sqrt{1-v^{2}}}.\label{eq:norm-timelike-v}
\end{equation}
We now see that there is nothing mysterious about the Lorentz factor
in special relativity; it is just a normalization factor!

In (\ref{eq:parallel-transport-norm}) we showed that parallel transport
preserves the norm of the tangent vector to the path. Thus, a particle
which starts moving along a timelike path will forever stay along
a timelike path, and so a massive particle can never reach or exceed
the speed of light (locally), since that would mean changing to a
null path. Another way to see this is to note that $\gamma\to\infty$
as $v\to1$. However, the energy of a massive particle with mass $m$
moving at speed $v$ is $E=\gamma m$. Therefore, it would require
infinite energy to accelerate the particle to the speed of light.

\subsubsection{The Speed of Light}

In Sec. \ref{subsec:Massless-Particles-and} we proved that particles
moving along null paths always locally move at the speed of light.
Let us now show the opposite -- that a particle moving locally at
the speed of light always moves along a null path. Notice that in
the discussion of the previous section we did not actually use the
fact that the path is timelike until (\ref{eq:norm-timelike-v}).
If $v=1$, then (\ref{eq:norm-timelike-v}) becomes
\[
\left|\xxd\right|^{2}=\gamma^{2}\left(-1+v^{2}\right)=0,
\]
and thus the path must be null. In that case $\gamma$ can take any
non-zero value, and it is conventional to set it to 1, so the 4-velocity
will be
\[
\xd^{\mu}=\left(1,1,0,0\right).
\]
From (\ref{eq:parallel-transport-norm}), we see that a particle which
starts on the null path must stay on a null path. Thus, a particle
moving at the speed of light can never decelerate or accelerate to
a different speed.

Since the norm $\left|\xxd\right|^{2}$ is by definition invariant
under a Lorentz transformation, the path will remain null when transforming
to any other frame of reference. We deduce that the speed of light
is the same for all observers, which is of course one of the two fundamental
postulates of relativity.

\subsubsection{Tachyons}

If $v>1$, that is, the particle is moving faster than the speed of
light, then the norm of the 4-velocity (\ref{eq:4-velocity}) will
necessarily be $\left|\xxd\right|^{2}>0$ and thus the path must be
spacelike. In that case we can normalize to $\left|\xxd\right|^{2}=1$,
which gives\footnote{Note that the Lorentz factor is real, as it should be. Naively, if
one used the definition of $\gamma$ which was derived by assuming
$v<1$, then $\gamma$ will be imaginary for $v>1$; but that is simply
not the right normalization factor to use. Furthermore, since $\gamma$
is real, so is the energy $E=\gamma m$. A tachyon need not have imaginary
energy or mass, as is sometimes naively claimed.}
\[
\gamma=\frac{1}{\sqrt{v^{2}-1}}.
\]
Particles moving faster than light, and along spacelike paths, are
called \emph{tachyons}. Just as normal massive particles moving along
timelike paths can never accelerate to the speed of light or beyond,
so can tachyons never decelerate to the speed of light or below. This
follows from the fact that a particle which starts on a spacelike
path will stay on a spacelike path.

Another way to see this is to note that, just for normal particles,
we have $\gamma\to\infty$ as $v\to1$ (except now $v$ approaches
1 from above). Therefore, it would require infinite energy $E=\gamma m$
to decelerate a tachyon to the speed of light. Paradoxically, a tachyon
in fact has \textbf{less }energy at higher velocities, since $\gamma\to0$
as $v\to\infty$! Indeed, a tachyon is at rest when its velocity is
\textbf{infinite}, $v\to\infty$, since that's when its energy is
minimized (and \emph{proper distance} $\d s^{2}=g_{\mu\nu}\thinspace\d x^{\mu}\otimes\d x^{\nu}$
is maximized). Taking $\gamma=1/\sqrt{v^{2}-1}$, we see that $\lim_{v\to\infty}\gamma=0$,
but $\lim_{v\to\infty}\gamma v=1$. Therefore a tachyon at rest has
\[
\xd^{\mu}=\lim_{v\to\infty}\gamma\left(1,v,0,0\right)=\left(0,1,0,0\right),
\]
so it is moving only along a spacelike direction and not along the
timelike direction, just as a normal massive particle at rest moves
only along a timelike direction and not along a spacelike direction.

\subsection{Why is There a Universal Speed Limit?}

A question which often arises when one learns that particles cannot
move faster than light is: Why is there a speed limit in the first
place? It seems quite arbitrary. One possible, and perhaps unexpected,
answer to this question is that we are using the \textbf{wrong variable
}to measure velocity. The correct variable is called \emph{rapidity}.

\subsubsection{The Velocity Addition Formula}

Let us work in 1+1 spacetime dimensions for simplicity. A Lorentz
transformation to a frame with relative velocity $v$ takes the form:
\[
\left(\begin{array}{c}
t'\\
x'
\end{array}\right)=\gamma\left(\begin{array}{cc}
1 & -v\\
-v & 1
\end{array}\right)\left(\begin{array}{c}
t\\
x
\end{array}\right).
\]
Combining it with another transformation of velocity $v'$, we get
\[
\left(\begin{array}{c}
t''\\
x''
\end{array}\right)=\gamma'\left(\begin{array}{cc}
1 & -v'\\
-v' & 1
\end{array}\right)\left(\begin{array}{c}
t'\\
x'
\end{array}\right)=\gamma\gamma'\left(\begin{array}{cc}
1+vv' & -v-v'\\
-v-v' & 1+vv'
\end{array}\right)\left(\begin{array}{c}
t\\
x
\end{array}\right).
\]
We may write the product of Lorentz factors as follows:
\[
\gamma\gamma'=\frac{1}{\sqrt{1-v^{2}}}\frac{1}{\sqrt{1-v^{\prime2}}}=\frac{1}{\left(1+vv'\right)\sqrt{1-\left(\frac{v+v'}{1+vv'}\right)^{2}}},
\]
which then allows us to simplify the combined transformation to:
\[
\left(\begin{array}{c}
t''\\
x''
\end{array}\right)=\frac{1}{\sqrt{1-\left(\frac{v+v'}{1+vv'}\right)^{2}}}\left(\begin{array}{cc}
1 & -\frac{v+v'}{1+vv'}\\
-\frac{v+v'}{1+vv'} & 1
\end{array}\right)\left(\begin{array}{c}
t\\
x
\end{array}\right).
\]
Clearly, the product of two Lorentz transformations with velocities
$v$ and $v'$ is another Lorentz transformation with velocity
\[
v\oplus v'\equiv\frac{v+v'}{1+vv'}.
\]
This is the familiar \emph{velocity addition formula} from special
relativity. For small velocities, $v,v'\ll1$, the denominator is
approximately 1 and thus velocities add normally. However, for relativistic
velocities, it turns out that we must use this counter-intuitive formula.

\subsubsection{Rapidity}

The velocity addition formula comes from the fact that a Lorentz transformation
is actually a type of rotation -- more precisely, a \emph{hyperbolic
rotation}. This is simply a version of rotation where one of the dimensions
involved in the rotation has a negative signature in the metric. Like
any rotation, a hyperbolic rotation also has an angle -- which is
appropriately called a \emph{hyperbolic angle}. The correct variable
to parametrize a Lorentz transformation is \textbf{not} velocity,
but rather this hyperbolic angle -- which is sometimes called \emph{rapidity}.

A usual rotation in Euclidean space with angle $\theta$ is given
by
\[
\left(\begin{array}{c}
x'\\
y'
\end{array}\right)=\left(\begin{array}{cc}
\cos\theta & -\sin\theta\\
\sin\theta & \cos\theta
\end{array}\right)\left(\begin{array}{c}
x\\
y
\end{array}\right).
\]
The determinant of the matrix is $\cos^{2}\theta+\sin^{2}\theta=1$,
so we know that it only rotates vectors and does not change their
magnitude. Also, it is easy to calculate that the product of two rotations
with angles $\theta$ and $\theta'$ is a rotation with angle $\theta+\theta'$.

Analogously, a hyperbolic rotation in Minkowski space with hyperbolic
angle $\psi$ is given by
\[
\left(\begin{array}{c}
t'\\
x'
\end{array}\right)=\left(\begin{array}{cc}
\cosh\psi & -\sinh\psi\\
-\sinh\psi & \cosh\psi
\end{array}\right)\left(\begin{array}{c}
t\\
x
\end{array}\right).
\]
The determinant of the matrix is $\cosh^{2}\psi-\sinh^{2}\psi=1$,
and it is easy to calculate that the product of two rotations with
angles $\psi$ and $\psi'$ is a rotation with angle $\psi+\psi'$.
Hence, we learn that we may simply replace the Lorentz transformation
matrix with a hyperbolic rotation matrix of hyperbolic angle (or rapidity)
\[
\psi=\sinh^{-1}\left(\gamma v\right)=\tanh^{-1}v.
\]
Indeed, we then have
\[
\cosh\psi=\cosh\left(\tanh^{-1}v\right)=\frac{1}{\sqrt{1-v^{2}}}=\gamma,
\]
and thus
\[
\left(\begin{array}{cc}
\cosh\psi & -\sinh\psi\\
-\sinh\psi & \cosh\psi
\end{array}\right)=\gamma\left(\begin{array}{cc}
1 & -v\\
-v & 1
\end{array}\right).
\]
Now, since $v=\tanh\psi$, and using the fact that $\tanh0=0$ and
$\lim_{\psi\to\pm\infty}\tanh\psi=\pm1$, we see that $\psi$ can
take any real value, but $v$ will be limited to $\left|v\right|\le1$.
We conclude that there is no ``universal rapidity limit''! If rapidity
is the variable with which you measure your movement, then you can
accelerate as much as you want, never reaching any upper bound. Perhaps,
in a relativistic setting, rapidity is the natural variable to use,
and velocity is an unnatural variable, which only makes sense in the
non-relativistic limit $\psi\to0$\footnote{A similar case is encountered with temperature. The variable $T$
that we use to measure temperature is unnatural, since negative temperatures
(i.e. below absolute zero) exist, and they are in fact \textbf{hotter
}than any positive temperature! A variable which makes more sense
as a fundamental variable is the reciprocal of the temperature, $\beta\equiv1/T$
(in units where $k_{B}\equiv1$), also known as the \emph{coldness}.
One finds that $\beta\to+\infty$ corresponds to absolute zero, $T=0$,
and as $\beta$ decreases towards zero (coldness increases, so things
get hotter), $T$ approaches $+\infty$. As $\beta$ crosses zero,
$T$ suddenly jumps down to $-\infty$, and as $\beta$ decreases
towards $-\infty$, $T$ increases to zero from below. Thus negative
values of $T$ are indeed hotter than any positive value of $T$,
since they correspond to lower values of coldness!}.

\subsection{\label{subsec:Tachyons-and-Time}Tachyons and Time Travel}

In the next section, we will present exotic spacetime geometries in
general relativity which allow faster-than-light travel globally while
lawfully maintaining the speed of light limit locally -- by which
we mean that massless or massive particles always locally follow null
or timelike paths respectively, and no particles \textbf{locally }follow
spacelike paths, but there is nonetheless some notion in which the
particles may \textbf{globally }follow spacelike paths.

However, even in the much simpler setting of special relativity, we
have seen that it is possible to mathematically describe particles
called tachyons, which follow spacelike paths even \textbf{locally},
and always travel faster than light. If tachyons exist, then they
also allow time travel, or at the very least, communication to the
past. Let us see how this works.

Consider a spacetime which is flat and completely empty aside from
from two space stations, $S$ and $S'$. Both stations are in inertial
motion, and they are moving with respect to each other with constant
relative speed $u<1$. The rest frame of station $S$ has coordinates
$\left(t,x\right)$ and the rest frame of station $S'$ has coordinates
$\left(t',x'\right)$. Hence, the worldline of station $S$ is along
the $t$ axis and the worldline of station $S'$ is along the $t'$
axis.

When station $S$ reached the origin of its coordinate system, $\left(t,x\right)=\left(0,0\right)$,
it sends a tachyon to station $S'$ at a speed $v>1$. The tachyon
arrives at station $S'$ whenever its worldline intersects the $t'$
axis. This will necessarily be in the future, since while the tachyon
is superluminal, it is still going forward in time. So no time travel
has occurred yet. For simplicity, we assume that this point of contact
is the origin of the $\left(t',x'\right)$ coordinate system\footnote{This means the coordinate systems are related by both a Lorentz transformation
and a spacetime translation, or in other words, a Poincaré transformation.}.

\begin{figure}[h]
\begin{centering}
\includegraphics[width=1\textwidth]{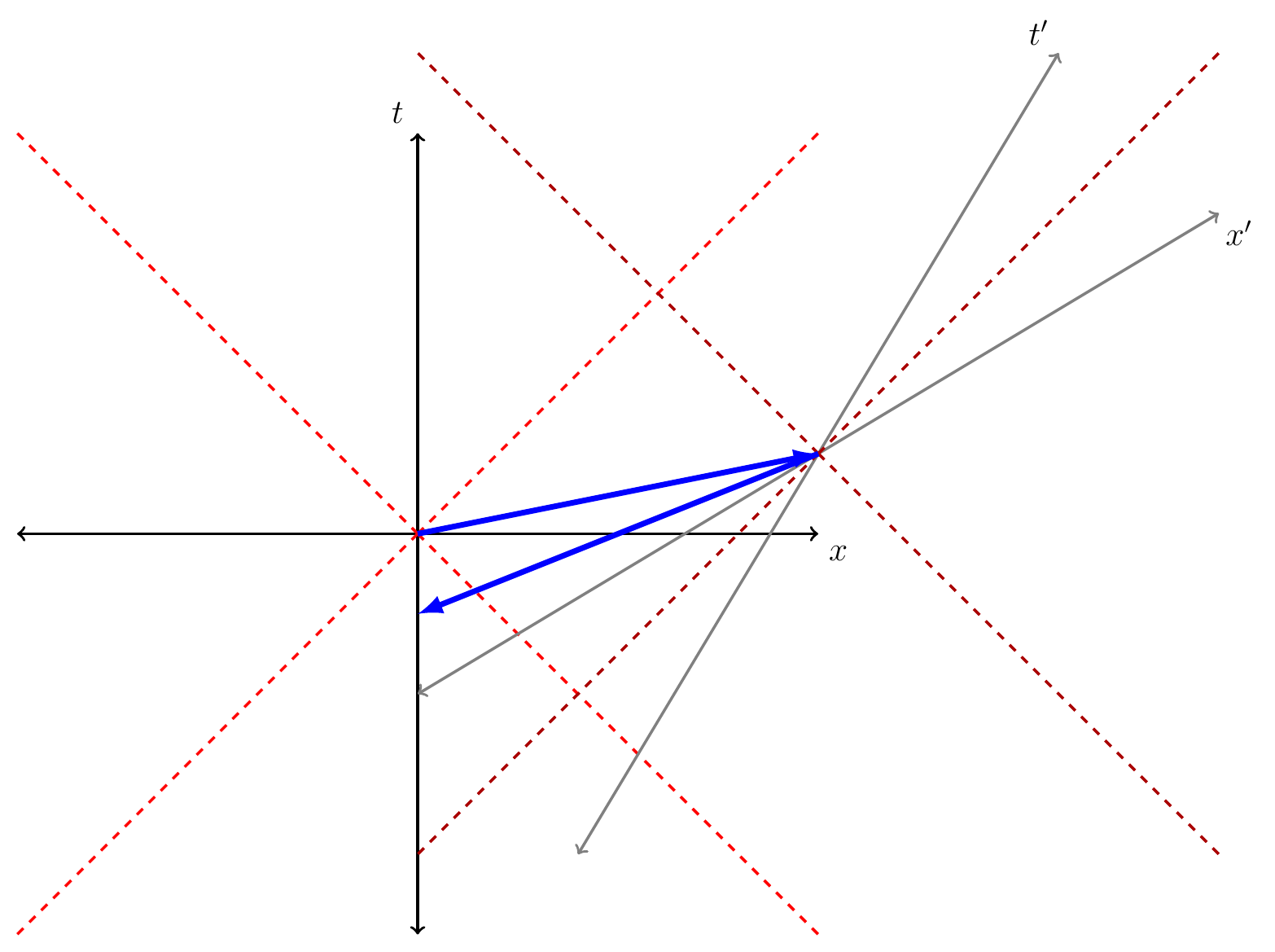}
\par\end{centering}
\caption{\label{fig:Tachyon}Using tachyons to send a message back in time.
The black lines are the axes for the $\left(t,x\right)$ system, the
gray lines are the axes for the $\left(t',x'\right)$ system, the
bright red lines are the light cones for the $\left(t,x\right)$ system,
the dark red lines are the light cones for the $\left(t',x'\right)$
system, and the blue lines are the (spacelike) worldlines of the tachyons.}
\end{figure}

When station $S'$ receives the tachyon, which is by assumption at
time $t'=0$, it sends another tachyon back to station $S$ at speed
$v'>1$. Now, from the point of view of $S'$, the tachyon it emits
is again going forward in time. This means that its worldline must
be above the $x'$ axis. However, by drawing a spacetime diagram as
in Fig. \ref{fig:Tachyon}, one can easily see that, for sufficiently
large $u$, the $x'$ axis intersects the $t$ axis at a negative
value. Therefore, all $S'$ has to do in order to send a message to
the past of $S$ is to make the velocity $v'$ fast enough so that
the tachyon ends up at a point on the $t$ axis corresponding to a
negative $t$ value (while still having a positive $t'$ value).

Now, if the tachyon is used to encode a message\footnote{The message could even simply be the detection of the tachyon itself.},
a person at space station $S$ may use the tachyon to send a message
to their past selves. This will inevitably create a paradox! This
kind of paradox is common to all forms of time travel and/or communication
to the past, and it resembles the \emph{grandfather paradox}, which
we will discuss in detail later. For the tachyon, the paradox may
be formulated as follows.

Let us assume that station $S$ sends a tachyon at time $t=0$ only
if it did \textbf{not }receive a tachyon at any time $t<0$. Furthermore,
station $S'$ sends a tachyon at time $t'=0$ only if it \textbf{did
}receive a tachyon at that time; in other words, it simply acts as
a tachyon ``mirror''. So if $S$ sent a tachyon at $t=0$, this
means that it did \textbf{not }receive a tachyon at an earlier time;
but in that case, since the sent tachyon was relayed by $S'$, it
in fact must have received a tachyon at an earlier time, which means
it couldn't have sent the initial tachyon in the first place.

In other words, station $S$ sends a tachyon at $t=0$ \textbf{if
and only if }it does not send a tachyon at $t=0$! This is a paradox,
since an event cannot both happen and not happen at the same time.

\section{Warp Drives}

Let us now discuss how, in general relativity, it is in principle
possible to travel faster than light globally while still not breaking
the speed of light barrier locally. We begin with the warp drive.
To gain an intuitive understanding of how it works, we first present
the expanding universe as motivation.

\subsection{Motivation: The Expanding Universe}

Our expanding universe is described by the Friedmann--Lemaitre--Robertson--Walker
(FLRW) metric, which has the line element:
\[
\d s^{2}=-\d t^{2}+a\left(t\right)^{2}\d\Sigma^{2},
\]
where $\d\Sigma^{2}$ is the line element for the 3-dimensional spatial
hypersurfaces of constant $t$, which are assumed to have uniform
curvature. Usually we take them to be flat, so 
\[
\d\Sigma^{2}=\d x^{2}+\d y^{2}+\d z^{2}=\d r^{2}+r^{2}\left(\d\theta^{2}+\sin^{2}\theta\thinspace\d\phi^{2}\right),
\]
depending on the coordinate system we use. The function $a\left(t\right)$
is called the \emph{scale factor}, and it simply scales the spatial
distances measured within the spatial hypersurfaces. At the present
time, $a\left(t\right)$ is defined to be equal to 1 so that the whole
spacetime metric is flat.

We define \emph{proper distance }to be the spatial distance measured
with this metric, which is modified by the scale factor. In an expanding
universe like ours, $a\left(t\right)$ increases with time, and thus
if two galaxies are at rest, the proper distance between them nevertheless
increases with time. However, we also define \emph{comoving distance},
which factors out the scale factor so that the comoving distance between
two galaxies at rest is constant.

A galaxy that is currently a proper distance $D_{0}$ from us will
be a distance
\[
D\left(t\right)=a\left(t\right)D_{0}
\]
from us at a later time $t$. From this we can calculate the \emph{recession
velocity }of the galaxy with respect to us:
\[
\Dd\left(t\right)=\adt\left(t\right)D_{0}=\frac{\adt\left(t\right)}{a\left(t\right)}D\left(t\right)\equiv H\left(t\right)D\left(t\right),
\]
where $H\left(t\right)\equiv\adt\left(t\right)/a\left(t\right)$ is
called the \emph{Hubble parameter}. At the present time we have
\[
\Dd=H_{0}D\sp H_{0}\ap70\thinspace\mathrm{(km/s)/Mpc}.
\]
This is \emph{Hubble's law}: the recession velocity of a galaxy from
us is proportional to its distance from us.

As a numerical example, a galaxy at a distance $1\thinspace\mathrm{Mpc\ap3\thinspace Mly\ap3\xx10^{19}\thinspace km}$
away from us has a recession velocity of $\ap70\thinspace\mathrm{km/s}$.
However, it's easy to see that a galaxy further away than approximately
$4.5\thinspace\mathrm{Gpc}\ap14\thinspace\mathrm{Gly}$ is receding
faster than light!

Importantly, the recession velocity $\Dd$ is only a \textbf{global
}velocity due to space itself expanding. The galaxy's \textbf{local
}velocity (sometimes called \emph{peculiar velocity}) in space, relative
to nearby galaxies, is independent from $\Dd$ and always slower than
light, since it locally follows a timelike path. Nonetheless, it seems
that we've found a \textbf{loophole} in the universal speed limit:
While the local velocity of an object \textbf{within space }cannot
exceed the speed of light, the global velocity due to the expansion
of \textbf{space itself }is unlimited! We will now see how this loophole
is exploited by the warp drive.

\subsection{The Warp Drive Metric}

\begin{figure}
\begin{centering}
\includegraphics[width=0.75\textwidth]{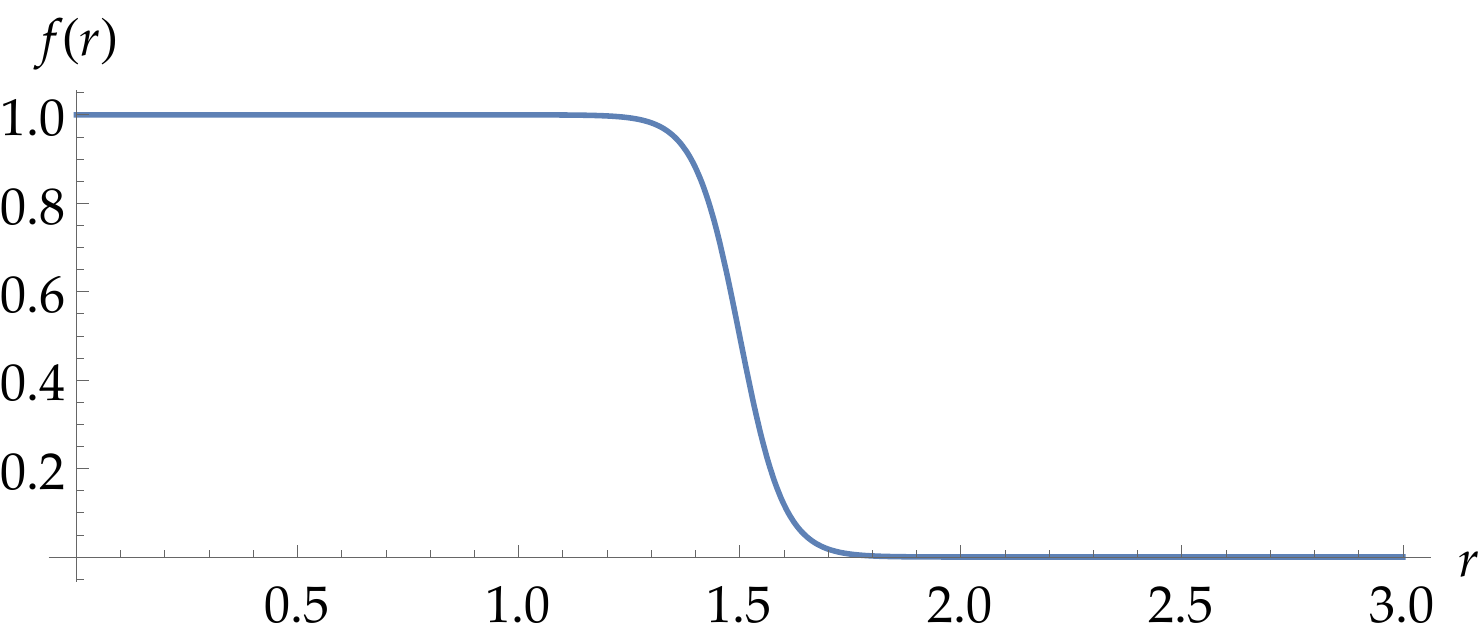}
\par\end{centering}
\caption{\label{fig:FormFunction}The function $f\left(r\right)$ with $R=1.5$
and $\varepsilon=0.1$.}

\end{figure}

The Alcubierre warp drive metric, presented by Alcubierre in 1994
\cite{Alcubierre}, is given by the following line element:
\begin{equation}
\d s^{2}=-\d t^{2}+\d x^{2}+\d y^{2}+\left(\d z-v\left(t\right)f\left(r\left(t\right)\right)\d t\right)^{2},\label{eq:warp-metric}
\end{equation}
corresponding to the metric
\[
g_{\mu\nu}=\left(\begin{array}{cccc}
-1+v^{2}f^{2} & 0 & 0 & -vf\\
0 & 1 & 0 & 0\\
0 & 0 & 1 & 0\\
-vf & 0 & 0 & 1
\end{array}\right),
\]
where:
\begin{itemize}
\item $r\left(t\right)\equiv\sqrt{x^{2}+y^{2}+\left(z-\zeta\left(t\right)\right)^{2}}$
is the spatial distance at time $t$ to the center of the spherical
\emph{warp bubble}, located at position $\zeta\left(t\right)$ along
the $z$ axis at time $t$ and moving at (global) velocity $v\left(t\right)\equiv\d\zeta\left(t\right)/\d t$.
The radius of the bubble is $R$ and the thickness of its wall is
$\varepsilon$. The center of the bubble is where our spaceship will
be located.
\item $f\left(r\left(t\right)\right)$ is the \emph{form function}; its
exact form does not matter, as long as it has the value $f\left(0\right)=1$
at the center of the bubble and $f\to0$ as $r\to\infty$ outside
the bubble, and drops sharply from $\ap1$ to $\ap0$ at the wall
of the bubble, at $r=R$. How sharp the transition is depends on the
wall thickness $\varepsilon$. This means spacetime is approximately
flat everywhere outside the bubble. An example of a possible form
function is
\begin{equation}
f\left(r\right)\equiv\frac{\tanh\frac{r+R}{\varepsilon}-\tanh\frac{r-R}{\varepsilon}}{2\tanh\frac{R}{\varepsilon}}.\label{eq:form-function}
\end{equation}
In the limit $\varepsilon\to0$, this function approaches a step function:
\[
\lim_{\varepsilon\to0}f\left(r\right)=\begin{cases}
1 & \textrm{if }r\in\left[0,R\right],\\
0 & \textrm{if }r\in\left(R,\infty\right).
\end{cases}
\]
However, for positive $\varepsilon$ there is a smooth transition,
as can be seen in Fig. \ref{fig:FormFunction}.
\item For constant $t$ hypersurfaces, we have $\d t=0$ and thus the induced
metric is simply the flat Euclidean metric.
\end{itemize}

The crucial fact about this metric is that the velocity $v$ of the
bubble can be faster than light! We will now study how this is possible.

\subsection{Properties of the Metric}

\subsubsection{(No) Time Dilation}

For any $v$, even $v>1$, if we substitute the location of the center
of the bubble $\left(x,y,z\right)=\left(0,0,\zeta\right)$ into the
metric (\ref{eq:warp-metric}) we get
\[
\d s^{2}=-\d t^{2},
\]
since $f\left(0\right)=1$, $\d x=\d y=0$ and 
\[
\d z=\d\left(\zeta\left(t\right)\right)=\frac{\d\zeta}{\d t}\d t=v\left(t\right)\d t.
\]
Since the proper time is given by $\d t=\sqrt{-\d s^{2}}$, we see
that
\[
\d\tau=\d t.
\]
Hence, the proper time of the spaceship inside the bubble is the same
as the coordinate time. The spaceship will experience no time dilation
during the journey.

This is very different from the case of a spaceship traveling normally
(e.g. using rockets) at close to the speed of light, using just the
usual laws of special relativity. The journey to a distant star at
close to light speed might take a very short proper time, but will
correspond to a very long coordinate time.

As a concrete example, let us assume the ship travels at $v=0.999999$,
which corresponds to a time dilation factor $\gamma=1000$. The crew
travels to a star 100 light years away and back, and the entire round
trip journey only lasts about 10 weeks. However, when they arrive
back on Earth, they see that 200 years have passed, and everyone they
know is long dead. The warp drive solves this problem; since there
is no time dilation, only 10 weeks will have passed on Earth during
the journey\footnote{We have, of course, neglected issues of acceleration in this example.}.

\subsubsection{Timelike Paths and Tilted Light Cones}

Globally, the path taken by the bubble is going to be spacelike if
$v>1$. This does not violate anything in relativity, since the bubble
is not a massive object obeying the geodesic equation; it is just
a specific choice of the curvature of spacetime itself. Furthermore,
from the fact that $\d s^{2}<0$ we learn that, even for $v>1$, the
spaceship travels along a \textbf{timelike }path \textbf{inside} the
bubble, as expected from a massive object. (Below we will show it
is, in fact, at rest!)

\begin{figure}
\begin{centering}
\includegraphics[width=1\textwidth]{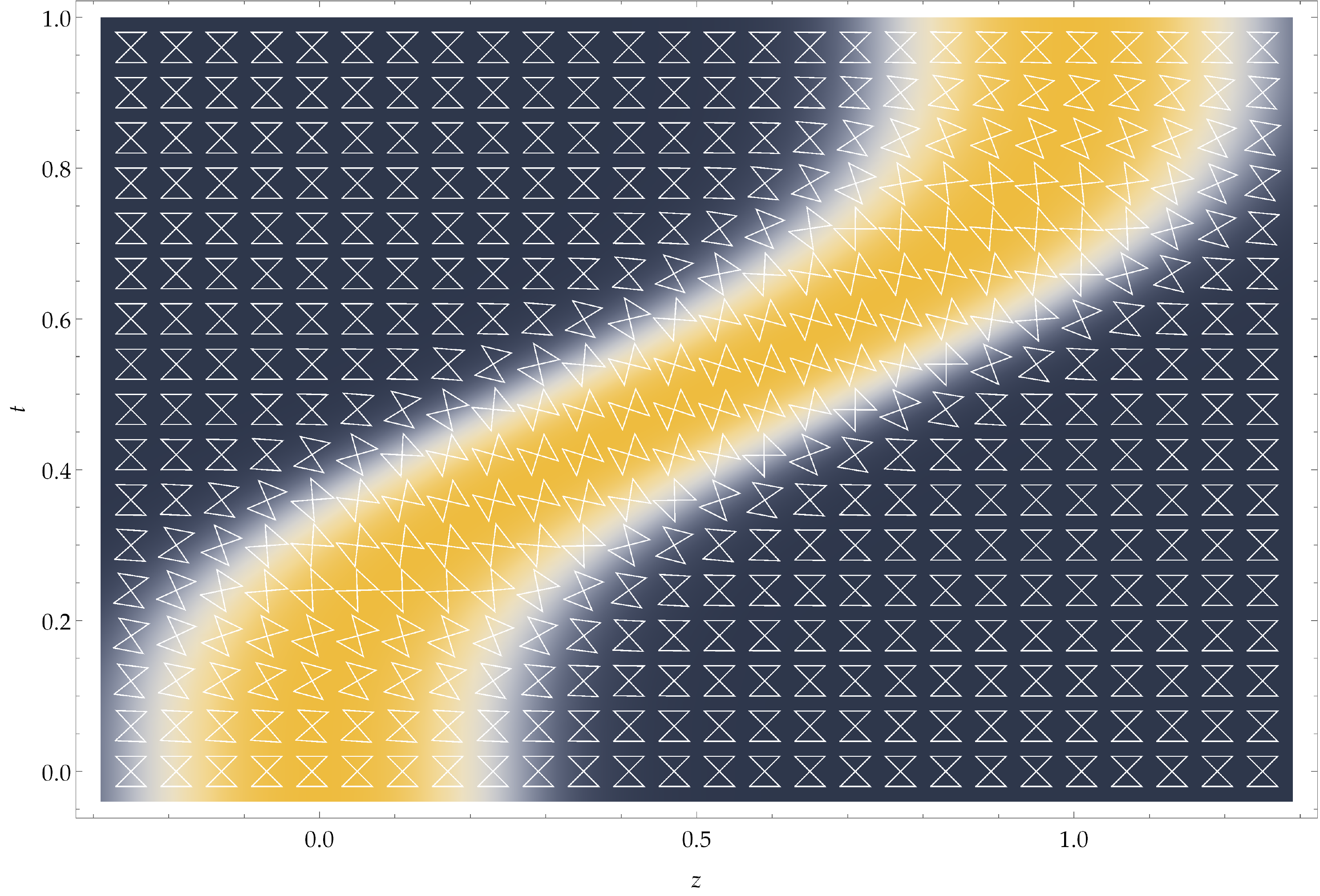}
\par\end{centering}
\caption{\label{fig:WarpLightCones}A spacetime diagram for a warp bubble (in
yellow) with $R=0.25$, $\varepsilon=0.1$, and $v\left(t\right)=2\sin^{2}\left(\pi t\right)$.
It starts at rest at $t=0$, accelerates up to a maximum velocity
of $v=2$ at $t=0.5$, and then decelerates to rest at $t=1$. The
light cones have slopes $\pm1+vf$, so when $v>0$ they are ``tipped
over'' inside the bubble such that the spaceship locally remains
within its own light cone at all times.}

\end{figure}

To see this more clearly, let us calculate the light cones of this
metric, which correspond to $\d s^{2}=0$. We will only look at the
light cones in the $z$ direction, so $\d x=\d y=0$. We find
\[
\frac{\d z}{\d t}=\pm1+v\left(t\right)f\left(r\left(t\right)\right).
\]
Outside the bubble, where $f\ap0$, we have $\d z/\d t\ap\pm1$ which
is the usual light cone for a flat spacetime. However, inside the
bubble, where $f\ap1$, we have a ``tilt'' which is proportional
to $v$. This is illustrated in Fig. \ref{fig:WarpLightCones}. It
is now clear that, even though the spaceship travels faster than light
globally, it is locally still within its own light cone along the
path, so no relativistic traffic laws are broken.

\subsubsection{Geodesics and Free Fall}

Let us calculate the geodesics of the warp drive metric. Recall that
a geodesic is calculated by finding an extremum (which will, in fact,
be a maximum) of the proper time, given by (\ref{eq:proper-time}):
\[
\tau=\int\sqrt{-g_{\mu\nu}\xd^{\mu}\xd^{\nu}}\thinspace\d\lambda.
\]
In our case, for the metric (\ref{eq:warp-metric}), we have
\[
\tau=\int\sqrt{-\td^{2}+\xd^{2}+\yd^{2}+\left(\zd-v\left(t\right)f\left(r\left(t\right)\right)\td\right)^{2}}\thinspace\d\lambda.
\]
Since the square root is a monotonically increasing function, we may
instead maximize the following Lagrangian:
\[
L=\hf\left(-\td^{2}+\xd^{2}+\yd^{2}+\left(\zd-V\left(\x\right)\td\right)^{2}\right),
\]
where we defined $V\left(\x\right)\equiv v\left(t\right)f\left(r\left(t\right)\right)$
for simplicity; we treat it as a general function which depends on
all 4 coordinates (but not their derivatives). Furthermore, we parameterize
the geodesic using the proper time $\tau$ itself, which automatically
normalizes it as a 4-velocity, $\left|\xxd\right|^{2}=-1$. Thus,
the dot is now a derivative with respect to $\tau$.

To maximize this Lagrangian, we use the Euler-Lagrange equations:
\[
\frac{\d}{\d\tau}\left(\frac{\partial L}{\partial\xd^{\mu}}\right)-\frac{\partial L}{\partial x^{\mu}}=0.
\]
In total we have 4 equations, one for each value of $\mu$. We leave
the $\tau$ derivatives implicit\footnote{Leaving the derivative implicit simplifies things, since we always
end up taking the derivative of a constant (or zero) anyway.}:
\[
\frac{\d}{\d\tau}\left(-\td-V\left(\x\right)\left(\zd-V\left(\x\right)\td\right)\right)+\frac{\partial V\left(\x\right)}{\partial t}\td\left(\zd-V\left(\x\right)\td\right)=0,
\]
\[
\frac{\d\xd}{\d\tau}+\frac{\partial V\left(\x\right)}{\partial x}\td\left(\zd-V\left(\x\right)\td\right)=0\sp\frac{\d\yd}{\d\tau}+\frac{\partial V\left(\x\right)}{\partial y}\td\left(\zd-V\left(\x\right)\td\right)=0,
\]
\[
\frac{\d}{\d\tau}\left(\zd-V\left(\x\right)\td\right)+\frac{\partial V\left(\x\right)}{\partial z}\td\left(\zd-V\left(\x\right)\td\right)=0.
\]
It's easy to see that
\begin{equation}
\xd^{\mu}=\left(1,0,0,v\left(t\right)f\left(r\left(t\right)\right)\right)\label{eq:warp-geodesic}
\end{equation}
solves all 4 equations, since then we have $\zd-V\left(\x\right)\td=0$.

Observers with the 4-velocity (\ref{eq:warp-geodesic}) are called
\emph{Eulerian observers}. They are following timelike geodesics,
even though, at $r=0$ where $f=1$, their spatial velocity is $v$
which may be faster than light! Note that
\[
\xd_{\mu}=g_{\mu\nu}\xd^{\mu}=\left(-1,0,0,0\right),
\]
which is the covector normal to the hypersurfaces given by $\d t=0$,
which as we said above, have a flat Euclidean metric. (Of course,
$\xd_{\mu}\xd^{\mu}=-1$ as expected.)

\subsubsection{Expansion and Contraction of Space}

How exactly does the warp bubble move? Just like with the expansion
of the universe, where the galaxies move faster than light with respect
to each other due to the expansion of space itself, here also the
spaceship moves faster than light with respect to its point of origin
(or the coordinate system) due to the expansion and contraction of
space around the warp bubble.

We are essentially using the same ``loophole''. Space expands behind
the warp bubble and contracts in front of it, thus pushing the bubble
forward at velocity $v$. The ship, which is \textbf{at rest }inside
the bubble, moves along with the bubble at an arbitrarily large \textbf{global
}velocity.

To see how this works, recall that if the unit normal vector $n^{\mu}$
is tangent to a geodesic, then from (\ref{eq:extrinsic-curvature-geodesic})
the expansion $\theta$ is given by the trace of the extrinsic curvature:
\[
\theta=K_{\mu}^{\mu}=\nabla_{\mu}n^{\mu}.
\]
In this case, as we have seen above, the unit normal covector is $n_{\mu}=\left(-1,0,0,0\right)$,
which is dual to the vector $n^{\mu}=\left(1,0,0,vf\right)$, which
is indeed tangent to a geodesic. Thus
\[
\theta=\nabla_{z}n^{z}=\partial_{z}\left(v\left(t\right)f\left(r\left(t\right)\right)\right)=\frac{z-\zeta}{r\left(t\right)}v\left(t\right)f'\left(r\left(t\right)\right),
\]
where we replaced the covariant derivative with a partial derivative
since the metric on the surface is flat.

\begin{figure}
\begin{centering}
\includegraphics[width=1\textwidth]{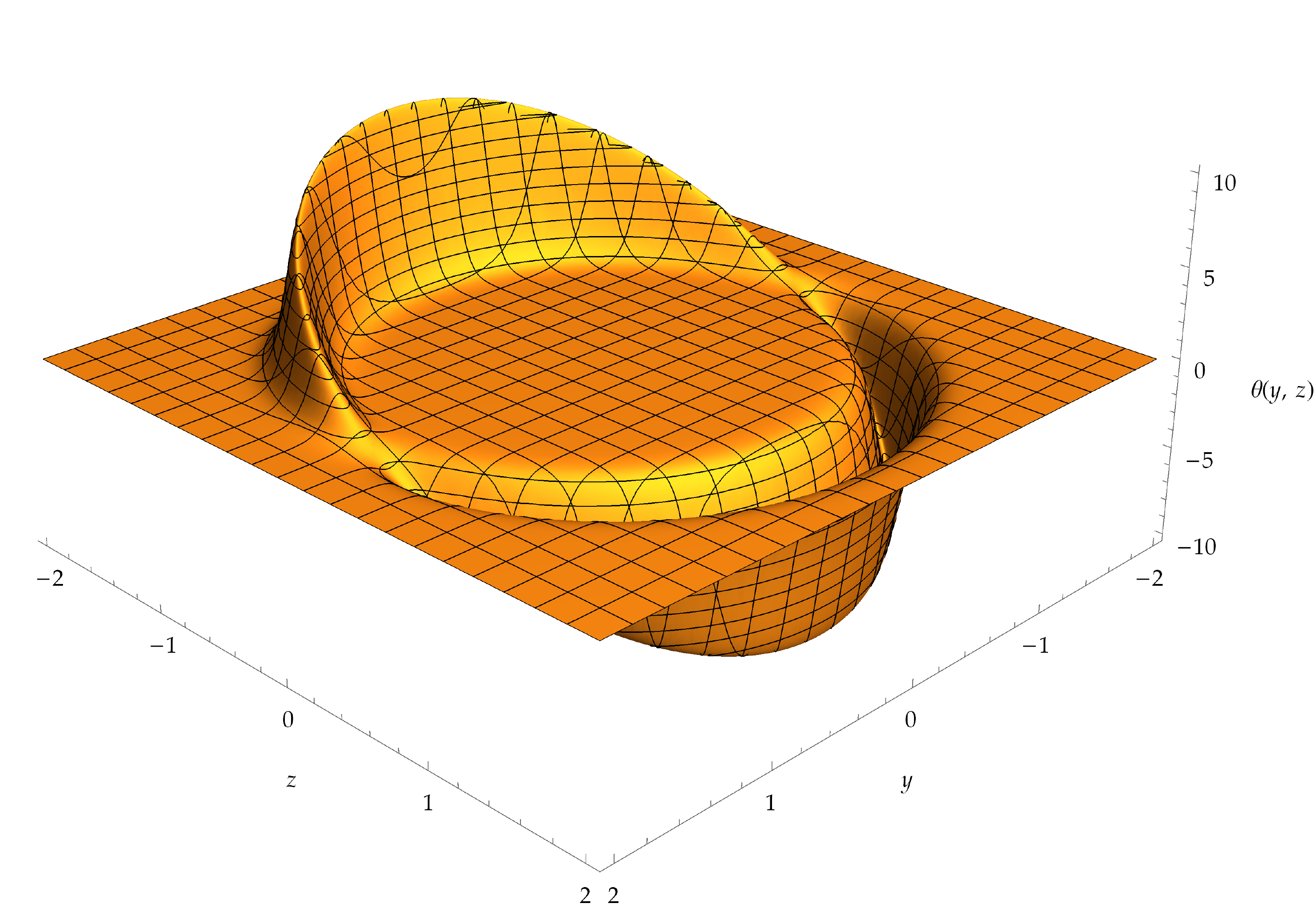}
\par\end{centering}
\caption{\label{fig:Expansion}A plot of the expansion $\theta$ of a warp
drive bubble with $v=2$, $R=1.5$, and $\varepsilon=0.1$ moving
along the positive $z$ axis.}
\end{figure}

Now, since $f\ap1$ inside the bubble and $f\ap0$ outside the bubble,
we have that $\theta\ap0$ in those regions. The only place where
we would have $\theta\ne0$ is on the wall of the bubble, where, as
$r$ (the distance from the center of the bubble) increases, $f$
sharply changes from $1$ to $0$, and thus has a negative derivative,
$f'\left(r\left(t\right)\right)<0$. Behind the bubble we have $z<\zeta$
and thus $\theta>0$, so space is expanding; in front of the bubble
we have $z>\zeta$ and thus $\theta<0$, so space is contracting\footnote{Interestingly, Natario \cite{Natario} has generalized the warp drive
metric and shown that it is in fact possible to construct it in such
a way that $\theta=0$ everywhere, so space does not expand nor contract!
Instead, his interpretation is that the warp bubble simply ``slides''
through space, taking the spaceship with it, and the expansion and
contraction are not necessary for the warp drive to function.}. This is illustrated in Fig. \ref{fig:Expansion}.

\subsection{Violations of the Energy Conditions}

Let us plug the warp drive metric into the left-hand side of the Einstein
equation (\ref{eq:Einstein}):
\[
R_{\mu\nu}-\hf Rg_{\mu\nu}=8\pi T_{\mu\nu}.
\]
Let the timelike vector $t^{\mu}$, normalized such that $\left|\t\right|^{2}=-1$,
be the 4-velocity of an Eulerian observer. We have
\[
T_{\mu\nu}t^{\mu}t^{\nu}=\frac{1}{8\pi}\left(R_{\mu\nu}-\hf Rg_{\mu\nu}\right)t^{\mu}t^{\nu}.
\]
The calculation of the right-hand side is tedious, but it has been
done (see e.g. \cite{LoboVisser}), and the result is
\[
T_{\mu\nu}t^{\mu}t^{\nu}=-\frac{v^{2}}{32\pi}\left(\left(\frac{\partial f}{\partial x}\right)^{2}+\left(\frac{\partial f}{\partial y}\right)^{2}\right)\le0,
\]
which is always non-positive, and zero only if $v=0$ and/or $f=0$.
Thus, the weak energy condition is violated!

\begin{figure}
\begin{centering}
\includegraphics[width=1\textwidth]{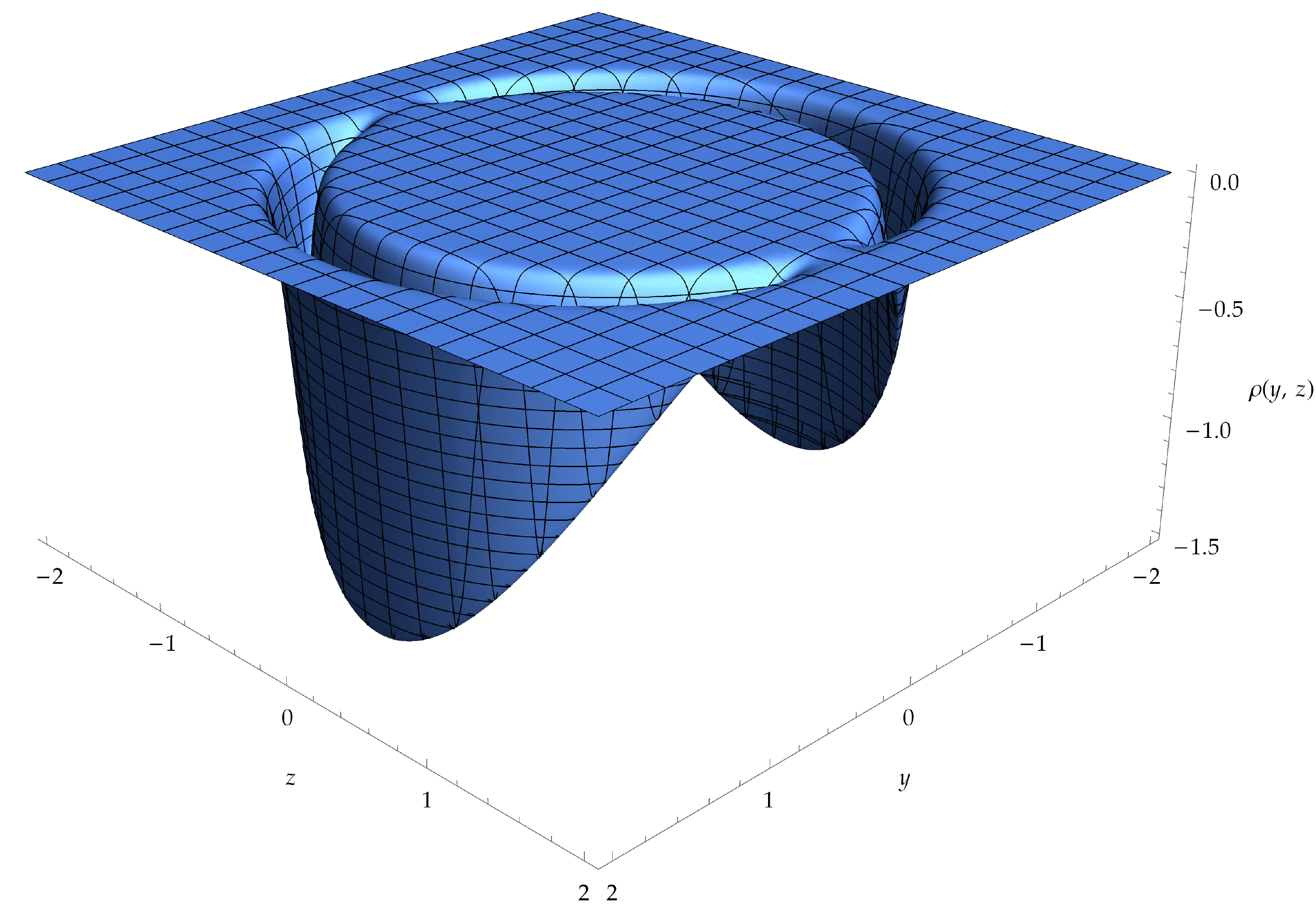}
\par\end{centering}
\caption{\label{fig:EnergyDensity}The distribution of energy density around
a warp drive bubble with $v=2$, $R=1.5$, and $\varepsilon=0.1$
moving along the positive $z$ axis.}
\end{figure}

In particular, since $f$ is a function of $r\left(t\right)\equiv\sqrt{x^{2}+y^{2}+\left(z-\zeta\left(t\right)\right)^{2}}$,
we may write this as
\[
T_{\mu\nu}t^{\mu}t^{\nu}=-\frac{v^{2}}{32\pi}\frac{x^{2}+y^{2}}{r^{2}}\left(\frac{\d f}{\d r}\right)^{2}\le0.
\]
Just like with the expansion, the dependence on the derivative $f'$
means that the violation of the energy conditions is negligible inside
and outside the bubble, and becomes noticeable only at the wall of
the bubble, where $f'$ sharply changes from $1$ to $0$. This is
illustrated in Fig. \ref{fig:EnergyDensity}.

The expression $T_{\mu\nu}t^{\mu}t^{\nu}$ is the energy density measured
by an observer with 4-velocity $t^{\mu}$, and we may integrate it
over a spatial slice to find the total energy:
\begin{equation}
E=\int T_{\mu\nu}t^{\mu}t^{\nu}\d^{3}x=-\frac{v^{2}}{32\pi}\int\frac{x^{2}+y^{2}}{r^{2}}\left(\frac{\d f}{\d r}\right)^{2}r^{2}\thinspace\d r\thinspace\d^{2}\Omega=-\frac{v^{2}}{12}\int\left(\frac{\d f}{\d r}\right)^{2}r^{2}\thinspace\d r.\label{eq:warp-energy}
\end{equation}
Given (\ref{eq:form-function}), we may estimate
\[
E\ap-\frac{v^{2}R^{2}}{\varepsilon}.
\]
Thus, we need a larger amount of negative energy for higher bubble
velocities, larger bubble sizes, and thinner bubble walls, which makes
sense.

A slightly better estimate is given by Pfenning and Ford \cite{PfenningFord}.
They use a piecewise-continuous function instead of (\ref{eq:form-function}):
\[
f\left(r\right)=\begin{cases}
1 & r\le R-\frac{\varepsilon}{2},\\
\hf+\frac{R-r}{\varepsilon} & r\in\left(R-\frac{\varepsilon}{2},R+\frac{\varepsilon}{2}\right),\\
0 & r\ge R+\frac{\varepsilon}{2},
\end{cases}
\]
where again $R$ is the radius of the bubble and $\varepsilon$ is
its wall thickness. Plugging this into (\ref{eq:warp-energy}), we
get
\[
E=-\frac{v^{2}}{12}\int_{R-\varepsilon/2}^{R+\varepsilon/2}\frac{1}{\varepsilon^{2}}r^{2}\d r=-\frac{v^{2}}{12}\left(\frac{R^{2}}{\varepsilon}+\frac{\varepsilon}{12}\right).
\]
Note that this is assuming $v$ is constant. Taking $R=100\thinspace\mathrm{m}$
and assuming $\varepsilon\le10^{2}\thinspace v$ in Planck units we
get
\[
E\le-6.2\xx10^{70}v\ap1.3\xx10^{63}v\thinspace\mathrm{kg}.
\]
This is 10 orders of magnitude larger than the mass of the observable
universe, which is roughly estimated at $10^{53}\thinspace\mathrm{kg}$.
This by itself is already bad news, but on top of that, the energy
must also be negative!

However, it is unclear how meaningful this result is. One issue comes
from the requirement that $\varepsilon\le10^{2}\thinspace v$, which
is only relevant if our source of exotic matter is a quantum field
(and even then, relies on some assumptions which may or may not be
applicable). The reason is that quantum fields must obey the \emph{quantum
energy inequality} \cite{Ford:1994bj}, which is a limitation on how
much negative energy may be contributed by the field over some period
of time -- and it is a pretty small amount\footnote{We will not go into the details of the quantum energy inequality here,
since that would require knowledge of quantum field theory.}.

If the negative energy instead comes from a classical field, then
this restriction does not apply, and we may allow a thicker wall,
for example with $\varepsilon\ap1\thinspace\mathrm{m}$. In this case,
we would only need
\[
E\le-5.0\xx10^{29}\thinspace\mathrm{kg}
\]
of exotic matter, which is roughly of the order of the mass of the
sun, and thus perhaps a bit more realistic -- provided, of course,
that a suitable form of exotic matter exists in this amount.

\subsection{The Event Horizon}

Let us write down the warp drive metric in the reference frame of
an observer on the spaceship. Their $z$ coordinate becomes $z'\equiv z-\zeta\left(t\right)$,
so (\ref{eq:warp-metric}) becomes
\[
\d s^{2}=-\d t^{2}+\d x^{2}+\d y^{2}+\left(\d z'+v\left(t\right)\left(1-f\left(r\left(t\right)\right)\right)\d t\right)^{2}.
\]
Now, suppose the engineers on the spaceship want to modify the geometry
of the warp bubble -- for example, in order to accelerate or to stop
moving. In order to do that, they must be able to causally influence
the bubble wall in front of them. Let us assume they try to do so
by sending a photon forward towards the bubble wall. The photon has
$\d s^{2}=0$, and it is going in the $z'$ direction, so $\d x=\d y=0$.
Thus, we obtain from the metric that
\[
\frac{\d z'}{\d t}=1-v\left(1-f\right),
\]
where we chose the positive square root so that the photons moves
along the positive direction of the $z'$ axis. At the center of the
bubble we have $f=1$, so $\d z'/\d t=1$ and the photon moves at
the speed of light with respect to the spaceship (in the original
frame, where $\d z/\d t=1+vf$, it moves at speed $1+v$).

Now, if $v<1$, then since $f\in\left[0,1\right]$, $\d z'/\d t$
will always remain positive. However, if $v>1$ then, at some point
$z=z_{\mathrm{h}}$ where $f=1-1/v$, the photon's speed will become
zero and it will therefore remain forever at that point. Thus, there
is a \emph{horizon }at $z_{\mathrm{h}}$.

Since $f=0$ outside the bubble, this means the photon will be stuck
somewhere inside the bubble wall. Therefore, the outer edge of the
bubble wall is outside the forward light cone of the spaceship, or
more precisely, outside of its causal future $J^{+}$. From this,
we conclude that the crew on the spaceship cannot modify the bubble,
and thus cannot control their journey, which severely limits the feasibility
of the warp drive! However, perhaps it could be possible, in principle,
that the entire trajectory of the bubble may be constructed beforehand.

Let us see the existence of the horizon more precisely. We will follow
a paper by Hiscock \cite{Hiscock}. For simplicity, we work in 2 dimensions
$\left(t,x\right)$, so the Alcubierre metric is
\[
\d s^{2}=-\left(1-v^{2}f^{2}\right)\d t^{2}-2vf\thinspace\d t\otimes\d x+\d x^{2},
\]
and we will assume that $v$ is constant. Let us now transform to
coordinates $\left(t,r\right)$ by taking $\d x=\d r+v\thinspace\d t$:
\[
\d s^{2}=-A\left(r\right)\left(\d t-\frac{v\left(1-f\right)}{A\left(r\right)}\d r\right)^{2}+\frac{\d r^{2}}{A\left(r\right)},
\]
where
\[
A\left(r\right)\equiv1-v^{2}\left(1-f\right)^{2}.
\]
We can then make the obvious transformation, $\d\tau=\d t-\frac{v\left(1-f\right)}{A\left(r\right)}\d r$,
to diagonalize the metric:
\[
\d s^{2}=-A\left(r\right)\d\tau^{2}+\frac{\d r^{2}}{A\left(r\right)}.
\]
Note that this metric is static, since $A$ does not depend on $\tau$.
Looking at this metric, we see that we run into a problem. We have
$\left(1-f\right)^{2}\in\left[0,1\right]$, and therefore, if $v<1$
we have $A>0$. However, for $v>1$ there is a point $r=r_{\mathrm{h}}$
where $A=0$, given implicitly by $f\left(r_{\mathrm{h}}\right)=1-1/v.$
At this point, there is a coordinate singularity, and an event horizon
forms\footnote{\label{fn:Schwarzschild}This situation is very similar to the event
horizon of a Schwarzschild black hole, which has the metric:
\[
\d s^{2}=-A\left(r\right)\d t^{2}+\frac{\d r^{2}}{A\left(r\right)}+r^{2}\left(\d\theta^{2}+\sin^{2}\theta\thinspace\d\phi^{2}\right)\sp A\left(r\right)\equiv1-\frac{2M}{r},
\]
where $M$ is the mass of the black hole. Since $A=0$ when $r=2M$,
there is a coordinate singularity at that point. This is where the
event horizon forms. However, it is not an actual singularity --
only an artifact of the particular system of coordinates we chose;
the curvature is in fact finite there. The point at $r=0$, where
$A$ diverges, is an actual singularity since there is a scalar quantity,
called the \emph{Kretschmann scalar}, which diverges there: $R^{\mu\nu\alpha\beta}R_{\mu\nu\alpha\beta}=48M^{2}/r^{6}\to\infty$.
Since this is a scalar, it is independent of any choice of coordinates.

Intuitively, when $A$ changes sign at the event horizon $r=2M$,
the $t$ and $r$ directions switch places: $t$ becomes spacelike
and $r$ becomes timelike. In a typical spacetime, where $t$ is the
timelike coordinate, we must always travel along the $t$ direction
whether we want to or not. Inside the event horizon, where $r<2M$,
we must always travel along the negative $r$ direction -- and thus
into the singularity at $r=0$ -- whether we want to or not.}.

\subsection{Using a Warp Drive for Time Travel}

Since a warp drive allows one to travel along paths which globally
seem spacelike, the discussion of tachyons in Sec. \ref{subsec:Tachyons-and-Time}
applies to a warp drive as well. Of course, here the spacetime is
not entirely flat, but it is essentially flat everywhere except inside
the bubble, so this should not invalidate the results of the tachyon
discussion.

However, it is important to clarify several key differences between
a warp drive and a tachyon:
\begin{itemize}
\item As we have seen, a tachyon must always move faster than light; it
can never slow down to the speed of light or slower. A warp drive,
in principle, can travel at any speed, and should be able to accelerate
all the way from zero to arbitrarily high speeds, with the only difference
between $v<1$ and $v>1$ being the appearance of a horizon.
\item A tachyon does not depend on the spacetime geometry to move faster
than light. A warp drive requires a very specific spacetime geometry.
\item A particle inside a warp drive always moves along a locally timelike
path; in fact, it's at rest. A tachyon always moves along a locally
(and globally) spacelike path. Usually the expression ``closed timelike
curve'' is synonymous with ``time machine'', but we see that while
warp drives can be used to create a closed timelike curve, tachyons
allow a different form time travel, which does not involve closed
timelike curves.
\end{itemize}
There is one important similarity, though. We have never seen any
tachyons in reality; and warp drives may be impossible to construct,
due to the absence of exotic matter. Both concepts seem more like
science fiction than science; but unlike science function, they may
be represented using concrete mathematical models, within the context
of physical theories that are known to describe the real universe.
Therefore, whether or not tachyons or warp drives can actually exist
in reality, studying them might teach us a lot about how our universe
works.

\section{Wormholes}

\subsection{The Traversable Wormhole Metric}

\begin{figure}
\begin{centering}
\includegraphics[width=1\textwidth]{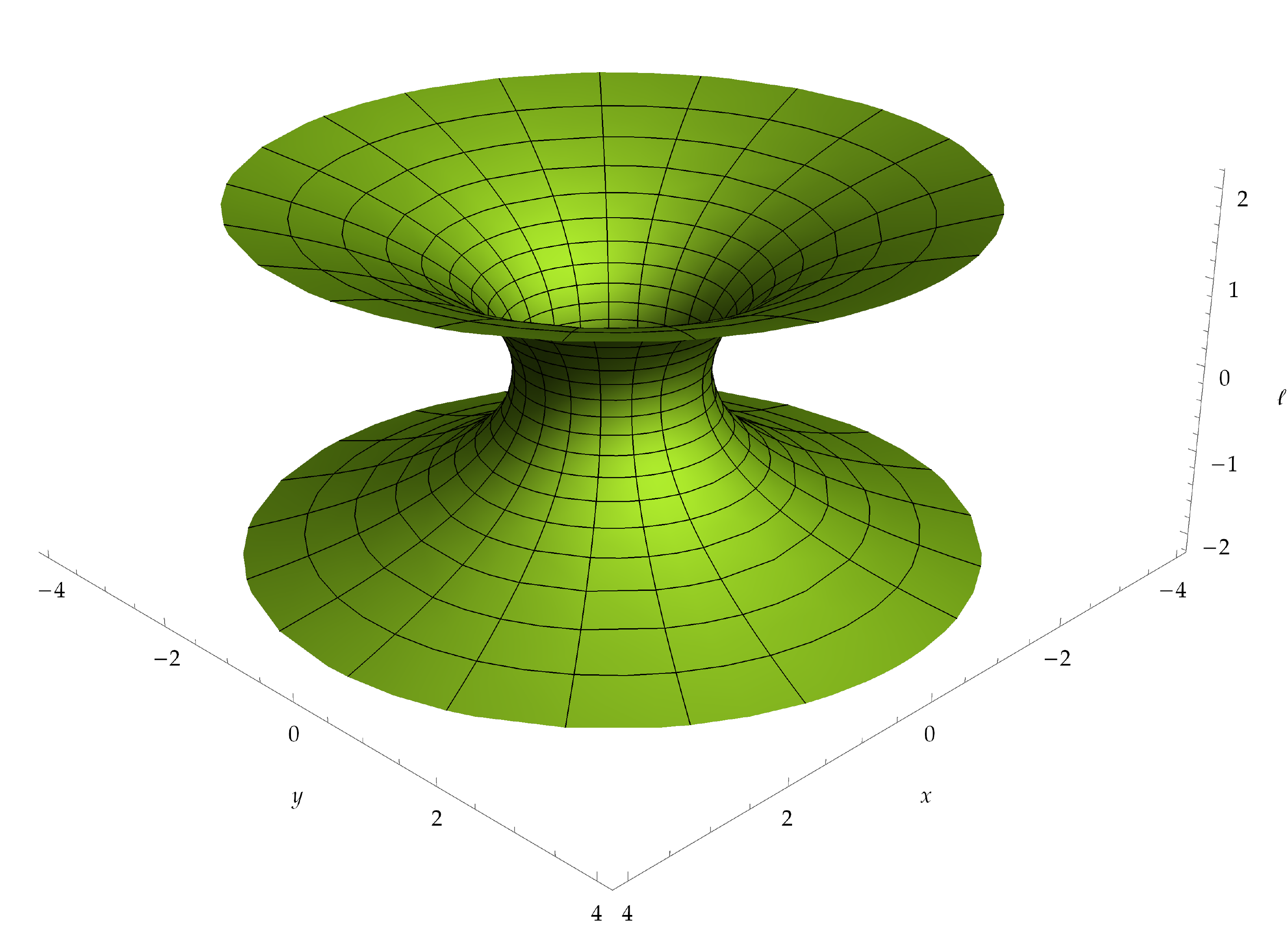}
\par\end{centering}
\caption{\label{fig:Wormhole}A slice of the wormhole geometry at the equator,
$\theta=\pi/2$, and at a constant value of $t$. The upper universe
corresponds to $\ell\protect\to+\infty$, while the lower universe
(which we will assume is another part of the same universe) corresponds
to $\ell\protect\to-\infty$. Note that each circle is actually a
slice of a sphere; we are simply suppressing other values of $\theta$
on the sphere so that we can plot the wormhole in three dimensions.}

\end{figure}

The \emph{Morris-Thorne wormhole} \cite{MorrisThorne}, a traversable,
static, spherically symmetric wormhole\footnote{There are many other types of wormholes studied in the literature,
although many of them are only of historical significance. Visser's
book \cite{Visser} provides a very thorough introduction to virtually
all known types of wormholes.} may be described, in spherical coordinates $\left(t,r,\theta,\phi\right)$,
by the following line element:
\[
\d s^{2}=-\e^{2\Phi\left(r\right)}\d t^{2}+\frac{\d r^{2}}{1-b\left(r\right)/r}+r^{2}\left(\d\theta^{2}+\sin^{2}\theta\thinspace\d\phi^{2}\right),
\]
where:
\begin{itemize}
\item There are two coordinate patches, both covering the range $r\in\left[r_{0},\infty\right)$.
\begin{itemize}
\item The two patches corresponds to two different universes, or two different
locations in the same universe, each on one side of the wormhole.
\item The value $r_{0}$ satisfies $b\left(r_{0}\right)=r_{0}$, and it
is where the wormhole's ``throat'' is located; it is the passage
between the two patches.
\item Note that there is a coordinate singularity at $r_{0}$, but it turns
out to not be a problem; the Riemann tensor is continuous across the
throat.
\end{itemize}
\item $\Phi\left(r\right)$ is called the \emph{redshift function}, since
it is related to gravitational redshift.
\begin{itemize}
\item It must be finite everywhere, including at $r\to\infty$, since if
we have $g_{tt}=0$, an event horizon will form and the wormhole won't
be traversable.
\item For simplicity we assume that $\Phi\left(r\right)$ is the same function
in both coordinate patches; if it's not, then time flows at different
rates on each side.
\end{itemize}
\item $b\left(r\right)$ is called the \emph{shape function}, and it determines
the wormhole's shape\footnote{By comparison with the Schwarzschild metric (see Footnote \ref{fn:Schwarzschild}),
we see that the mass of the wormhole is given by $M=\lim_{r\to\infty}b\left(r\right)/2$.}.
\begin{itemize}
\item For simplicity we assume that $b\left(r\right)$ is the same function
in both coordinate patches; if not, then the wormhole has a different
shape on each side.
\item Away from the throat, we have $b\left(r\right)<r$, so that $1-\frac{b\left(r\right)}{r}>0$.
\end{itemize}
\item The \emph{proper radial distance }$\ell$ is given by:
\[
\frac{\d\ell}{\d r}=\pm\frac{1}{\sqrt{1-\frac{b\left(r\right)}{r}}}\soosp\ell\left(r\right)\equiv\pm\int_{r_{0}}^{r}\frac{1}{\sqrt{1-\frac{b\left(r\right)}{r}}}\d r,
\]
and it is required to be finite everywhere\footnote{Recall that an integral can converge even if the integrand diverges
at the boundary of the integration region}.
\begin{itemize}
\item The two signs correspond to the two separate coordinate patches. The
range $\ell\in\left(-\infty,0\right)$ corresponds to one coordinate
patch and one universe, while the range $\ell\in\left(0,+\infty\right)$
corresponds to the other coordinate patch and either another universe,
or another location in the same universe.
\item The throat is located at $\ell=0$, where $r\left(\ell=0\right)=r_{0}=\min\left[r\left(\ell\right)\right]$.
\item The two regions at the limits $\ell\to\pm\infty$ are asymptotically
flat, so we must have $\lim_{\ell\to\pm\infty}\Phi=0$ and $\lim_{\ell\to\pm\infty}b/r=0$
in order to get the Minkowski metric there.
\end{itemize}
\end{itemize}
The coordinate $r$ is a bit confusing. Let us embed the wormhole
in a 3-dimensional space, as in Fig. \ref{fig:Wormhole}. Then $r$
is the radius of each circle that we see in the figure. At the top
universe, at $\ell\to+\infty$, we have $r\to+\infty$. Then, as we
go south towards the throat, $r$ decreases monotonically until it
reaches the minimum value $r_{0}$, which is the radius of the throat
-- the smallest circle in the figure. As we keep going south into
the bottom universe, $r$ now increases monotonically until, at $\ell\to-\infty$,
we have $r\to+\infty$ again. Note that $r$ is always positive (and
larger than $r_{0}$), even though $\ell$ can take any real value.

\subsection{Violations of the Energy Conditions}

Now, in an orthonormal basis (find it!), the non-zero components of
the Einstein tensor are:
\[
G_{tt}=\frac{b'}{r^{2}}\sp G_{rr}=2\left(1-\frac{b}{r}\right)\frac{\Phi'}{r}-\frac{b}{r^{3}},
\]
\[
G_{\theta\theta}=G_{\phi\phi}=\left(1-\frac{b}{r}\right)\left(\Phi''+\Phi'\left(\Phi'+\frac{1}{r}\right)\right)-\frac{b'r-b}{2r^{2}}\left(\Phi'+\frac{1}{r}\right).
\]
Using the Einstein equation, $G_{\mu\nu}=8\pi T_{\mu\nu}$, we see
that the energy-momentum tensor must have the same components (divided
by $8\pi$). In particular, the energy density is
\[
T_{tt}=\rho\left(r\right)=\frac{1}{8\pi}\frac{b'}{r^{2}},
\]
and it is negative whenever $b'<0$. We also find that the pressure
along the radial direction is
\[
T_{rr}=p\left(r\right)=\frac{1}{8\pi}\left(2\left(1-\frac{b}{r}\right)\frac{\Phi'}{r}-\frac{b}{r^{3}}\right).
\]
Recalling from Sec. \ref{subsec:Energy-Density-Pressure} that the
null energy condition is equivalent to $\rho+p\ge0$, we calculate
this quantity and find:
\begin{align*}
\rho\left(r\right)+p\left(r\right) & =-\frac{1}{8\pi}\frac{1}{r}\left(\frac{1}{r}\left(\frac{b}{r}-b'\right)-2\Phi'\left(1-\frac{b}{r}\right)\right)\\
 & =-\frac{1}{8\pi}\frac{1}{r}\left(\frac{\d}{\d r}\left(1-\frac{b}{r}\right)-2\Phi'\left(1-\frac{b}{r}\right)\right)\\
 & =-\frac{1}{8\pi}\frac{\e^{2\Phi}}{r}\frac{\d}{\d r}\left(\e^{-2\Phi}\left(1-\frac{b}{r}\right)\right).
\end{align*}
Now, at the throat, $r=r_{0}$, we have $b\left(r_{0}\right)=r_{0}$,
so
\[
\e^{-2\Phi}\left(1-\frac{b}{r}\right)\blll_{r=r_{0}}=0.
\]
Away from the throat we have $b\left(r\right)<r$, so
\[
\e^{-2\Phi}\left(1-\frac{b\left(r\right)}{r}\right)\blll_{r>r_{0}}>0.
\]
Thus, there must be some point $r_{*}$ such that
\[
\frac{\d}{\d r}\left(\e^{-2\Phi}\left(1-\frac{b\left(r\right)}{r}\right)\right)\blll_{r\in\left(r_{0},r_{*}\right)}>0.
\]
In conclusion, we have
\[
\rho\left(r\right)+p\left(r\right)\blll_{r\in\left(r_{0},r_{*}\right)}<0,
\]
and thus there is a non-zero region where the null energy condition
is violated. This implies a violation of the weak, strong and dominant
energy conditions.

Exactly how much exotic matter is needed depends on many different
factors, and we won't go into that here. However, there are some indications
that the required amount of negative energy can be made arbitrarily
small; see, for example, the paper by Visser, Kar, and Dadhich \cite{VisserKarDadhich}.
Furthermore, Barceló and Visser \cite{barcelo1999traversable} managed
to create a traversable wormhole from a classical conformally-coupled
scalar field by employing the violations of the energy conditions
we discussed in Subsection \ref{subsec:The-Scalar-Field}.

There are many more things one can study about wormholes, including
questions of traversability, horizons, singularities, tidal forces,
geodesics, and more. These discussions are beyond the scope of our
lecture notes; here we just wanted to give the basic details. The
reader is referred to the vast literature on wormholes for more information;
\cite{Visser} and \cite{Lobo} are good places to start.

\subsection{Using a Wormhole for Time Travel}

Creating a time machine using a wormhole is fairly easy. Consider
a wormhole with its two mouths located in the same universe, a distance
$L$ from each other. Normally, if you enter one mouth, you exit through
the other mouth at the same value of the time coordinate. Now, assuming
it is possible to move one of the mouths (presumably by moving the
exotic matter used to construct it?), we may use special- or general-relativistic
time dilation in order to introduce a relative time shift $T$ between
the two mouths.

For special-relativistic time dilation, we move one of the mouths
at close to light speed; this will be a version of the twin ``paradox'',
where the mouths play the roles of the twins. For general-relativistic
time dilation, we place the mouth near a strong gravitational source.
Finally, we move the mouths closer together.

Once the distance $L$ is less than the time shift $T$, a closed
timelike curve forms. Indeed, using coordinates $\left(t,x\right)$,
the first wormhole mouth follows the path $\left(t-T,0\right)$, while
the other one follows the path $\left(t,L\right)$, both paths parameterized
by $t$. The existence of the wormhole means that the following points
are identified along the paths, for any $t$:
\[
\left(t-T,0\right)\sim\left(t,L\right).
\]
Now, consider a spaceship at the first mouth, at $\left(0,0\right)$.
The ship moves at close to the speed of light until it covers a distance
$L$, which would roughly take time $L$. Therefore, its location
is now $\left(L,L\right)$. At this location, it enters the second
mouth, and arrives almost instantaneously back at the first mouth,
at $\left(L-T,0\right)$. Thus, if $L<T$, the ship has followed a
closed timelike curve and arrived at the first wormhole before it
left!

\section{Time Travel Paradoxes and Their Resolutions}

\subsection{The Paradoxes}

We have seen that it is theoretically possible to create time machines,
within the framework of general relativity, using special spacetime
geometries such as warp drives and wormholes. It is unclear if these
exotic geometries can exist in reality, especially due to the requirement
of exotic matter. However, let us now assume that some sufficiently
advanced civilization has succeeded in building some kind of time
machine. This seems to inevitably give rise to paradoxes, which we
will now describe\footnote{For an extensive and detailed analysis of time travel paradoxes from
the philosophical point of view, the reader is referred to the book
by Wasserman \cite{wasserman2018paradoxes}.}.

\subsubsection{Consistency Paradoxes}

A \emph{consistency paradox }is any situation where changing the past
removes the conditions which allowed (or required) changing the past
in the first place, or when an event happens if and only if it does
not happen. The most familiar example is called the \emph{grandfather
paradox}. In this paradox, the time traveler goes into a time machine
in 2019 and travels back to a time before his or her grandparents
met, let's say 1939. The time traveler then prevents this meeting,
thus ensuring that one of his or her parents will never be born\footnote{In most versions of the paradox the time traveler kills his or her
grandfather, but that is not necessary in order to create a paradox.
In fact, even if the meeting was not prevented but just delayed a
bit, the Butterfly Effect still guarantees that even if the prospective
grandparents do have a child, it will not actually be the time traveler's
parent, due to different initial conditions.}. But if the time traveler was never born, then he or she could not
have gone back in time and prevent his or her grandparents from meeting
when they did... In other words, the meeting happened \textbf{if and
only if }it did not happen, which doesn't make any sense.

Here is a more concrete example of a consistency paradox. The advantage
of this example is that it uses inanimate objects instead of people,
and thus leaves ``free will''\footnote{Whatever that means...}
out of the equation. In the grandfather paradox as we formulated it
above, one could always suggest that the time traveler simply changes
his or her mind, but it is impossible for an inanimate object to ``change
its mind''.

Consider a uranium lump with mass $m=\frac{2}{3}m_{\mathrm{crit}}$,
where $m_{\mathrm{crit}}$ is the critical mass -- which, if exceeded,
will cause a nuclear explosion. The lump follows a closed timelike
curve: it goes into a time machine, comes out a short time earlier,
and then collides with its younger itself. At that instant the total
mass of both lumps together will exceed the critical mass, creating
an explosion. This, of course, means that the younger lump will not
reach the time machine (or perhaps even the time machine itself will
be annihilated by the explosion), and thus the collision will, in
fact, not take place. Again, this is a situation where the explosion
happens \textbf{if and only if }it does not happen.

\subsubsection{Bootstrap Paradoxes}

A \emph{bootstrap paradox }is any situation where something (information,
an object, etc.) is ``created from nothing'', or when an effect
is its own cause. For example, just as I began writing these lecture
notes, a future version of myself appeared and gave me a USB stick
with the LaTeX file for the final version of the lecture notes. In
the future, when I invent my time machine, I will make sure to go
back to that same day and give my past self the same USB drive with
the same lecture notes. Everything in this scenario is perfectly consistent\footnote{It should be noted, though, that avoiding a consistency paradox in
this situation is in fact basically impossible. No matter how hard
I try to look and act the same as my future self did when I met him,
there is no way I could remember all the details perfectly; and even
if I did, there is absolutely no way I can guarantee that \textbf{every
single atom }in my body is exactly the same as it is supposed to be.
Furthermore, the USB drive will inevitably wear down; even if the
data stored on it has not changed, it cannot possibly be true that
every single atom in the USB drive is exactly the same as it was when
I received it. Consistency paradoxes seem to be completely inevitable.
Still, for the purpose of this example we may assume that the Novikov
conjecture (see below) holds true and automatically guarantees that
everything will be consistent. This conjecture cannot, however, prevent
a bootstrap paradox; the only known way of avoiding both types of
paradoxes, while still allowing time travel, is using multiple timelines
(see below).}. However, one might still wonder: who actually wrote the notes? It
seems that both information (the lecture notes) and a physical object
(the USB drive) were created from nothing.

As another example \cite{Lossev_1992}, consider a wormhole time machine.
A billiard ball comes out of the past mouth and travels directly into
the future mouth. From the point of view of an external observer,
the billiard ball only exists for the short period of time where it
travels between the two mouths. It never existed before that, and
will never exist after that; it was effectively created from nothing.
From the point of view of the ball, it keeps moving in an endless
loop between the two mouths. Neither point of view is compatible with
what we are used to in the absence of time travel\footnote{Again, it seems impossible to avoid a consistency paradox even in
this case. Much like the USB drive wearing down, the ball also changes
between cycles -- for example, it radiates heat, so it will have
less energy in each cycle.}.

Bootstrap paradoxes are not necessarily ``true'' paradoxes; Krasnikov
\cite{Krasnikov} refers to them as ``pseudo-paradoxes''. They are
not as disturbing as consistency paradoxes, since they do not seem
to pose logical inconsistencies, only mild discomfort. Furthermore,
bootstrap paradoxes seem to require assumptions which may not themselves
make sense. In both examples provided here, we never specified how
such a loop would exist in the first place, so perhaps there is no
reason to assume these scenarios are actually possible to create.

If time travel is possible, the paradoxes described above must not
be true paradoxes, since the universe, after all, has to make sense.
Let us discuss some options suggested in the literature in attempt
to resolve these paradoxes.

\subsection{The Hawking Chronology Protection Conjecture}

The \emph{Hawking chronology protection conjecture }\cite{Hawking}
suggests that time travel is simply impossible. For example, while
wormholes might in principle exist, perhaps they cannot be used to
create a time machine. As we get closer to building a time machine
by creating a time shift between the mouths, something will inevitably
happen that will cause everything to break down before the time machine
is constructed. Note that this also means we shouldn't be able to
(globally) travel faster than light using wormholes, since then we
could potentially exploit the same scenario described in Fig. \ref{fig:Tachyon}
in order to time travel\footnote{It is interesting to note that recently, Maldacena, Milekhin, and
Popov \cite{MaldacenaMilekhinPopov} found a traversable wormhole
solution which seems realistic -- it does not require more negative
energy than is possible to realistically generate using quantum fields
-- but it is a ``long'' wormhole, which means that its throat is
at least as long as the distance between the mouths in the space in
which the wormhole is embedded. Therefore it does not provide a shortcut
through space, and cannot lead to violations of causality. Perhaps
it is a general result that only ``long'' wormholes are realistic,
while ``short'' wormholes, which may allow time travel, are forbidden.
However, at the moment there is no proof of this. The interested reader
is referred to \cite{Gao2016,Maldacena2017,Maldacena2018,Fu2018,Horowitz2019}
for further information.}.

In quantum field theory, divergences appear everywhere, but usually
it is possible to ``fix'' them through a method called \emph{renormalization}.
However, if a quantity still diverges even after renormalization,
there is no known way to get rid of that divergence. Kim and Thorne
\cite{KimThorne} have shown that the renormalized energy-momentum
tensor diverges when approaching the Cauchy horizon. This could prevent
time machines from forming.

However, Kim and Thorne also conjectured that such divergences should
get cut off by quantum gravity effects. Furthermore, other authors,
such as Krasnikov \cite{Krasnikov1996}, have found spacetimes containing
time machines where the energy-momentum tensor is in fact bounded
near the Cauchy horizon. Therefore, Kim and Thorne's result is not
universal.

Grant \cite{grant1993cosmic} studied a particular causality-violating
spacetime, originally by Gott \cite{gott1991closed}, and found a
similar divergence. Furthermore, Kay, Radzikowski, and Wald \cite{kay1997quantum}
proved theorems which show that the renormalized expectation value
of a scalar field and its energy-momentum tensor are ill-defined or
singular in the presence of time machines; however, Krasnikov \cite{krasnikov1998quantum}
showed that this can be avoided by replacing some assumptions.

The chronology protection conjecture definitely seems plausible. However,
even if it is true, we are nowhere near close to proving it, in part
because we do not yet have the right tools to do so. A full theory
of quantum gravity, which will tell us exactly how spacetime behaves
at the quantum level, would most likely be necessary to prove this
conjecture. Currently, the most we can do is semi-classical gravity
-- which involves quantum matter, but classical gravity.

Many papers have been published on Hawking's conjecture. However,
they all make use of quantum field theory, which is beyond the scope
of these notes. Therefore, we will not discuss them further here.
For more information, the reader is referred to the reviews by Visser
\cite{visser2003quantum} and by Earman, Smeenk, and Wüthrich \cite{earman2009laws}.

\subsection{Multiple Timelines}

\subsubsection{Introduction}

Another possible solution -- which is perhaps the most exciting one
-- is to assume that the universe may have more than one history
or timeline. Whenever time travel occurs, the universe \emph{branches
}into two independent timelines. The time traveler leaves from one
timeline and arrives at a different timeline. Both timelines have
the same past, up to that point in time, but their futures may differ\footnote{If this reminds you of the Everett (``many-worlds'') interpretation
of quantum mechanics, you're in good company; see below how this interpretation
was used by Deutsch as a potential solution to time travel paradoxes.}.

If the traveler prevents his or her grandparents from meeting, this
will happen in the \textbf{new }timeline -- which does \textbf{not
}causally influence the original timeline. It is perfectly consistent
for the time traveler to not be born in this new timeline, since he
or she entered the time machine in the \textbf{original }timeline,
not in the new one. Hence, there are no paradoxes. Similarly, when
the lump of uranium travels into the time machine, it emerges at an
earlier point in time in a new timeline. In this new timeline, it
meets a copy of itself and explodes. This again creates no paradox,
since the explosion in the new timeline does not prevent the lump
from going into the time machine in the original timeline. In the
same way, multiple timelines can solve all known consistency paradoxes.

In the case of the bootstrap paradox, in timeline 1 I sat down for
weeks and worked day and night on these lecture notes. Exhausted but
satisfied, I copied them onto a USB drive and got into my time machine.
I arrived in timeline 2, at the exact moment when I began writing
the notes, and gave my past self (actually, a \textbf{copy} of myself)
the finished notes. It is clear, in this case, that information did
\textbf{not }come from nothing; it was created by the original me,
who lived in timeline 1. In other words, while information is not
conserved in each timeline individually, it is conserved in the union
of both timelines\footnote{Furthermore, my copy in timeline 2 doesn't have to go back to the
past and continue the cycle in order to avoid a consistency paradox.}.

In order to avoid any possibility of a paradox, a new timeline must
be created whenever a time machine is used, and one may never return
to their original timeline, unless they return to a moment after they
left. This can be understood by noticing that the way multiple timelines
solve paradoxes is by ``opening up'' closed timelike curves; a curve
which starts at $t=0$, goes to $t=1$ and then loops back to $t=0$
will inevitably create a paradox, but if it goes back to $t=0$ in
a different timeline (and hence a different point on the manifold),
no loop will be formed. Therefore, to avoid paradoxes there should
be at least enough timelines to ``open up'' every possible CTC.

\subsubsection{Non-Hausdorff Manifolds}

It is obvious that models of multiple timelines would solve time travel
paradoxes. However, it's one thing to describe this solution in words;
it's another thing entirely to actually write down a well-defined
mathematical model.

In the literature, multiple timelines are often said to be possible
if one relaxes the \emph{Hausdorff condition }in the definition of
the spacetime manifold. A topology satisfies the Hausdorff condition
(or ``is Hausdorff'') if and only if for any two distinct points
$x_{1}\ne x_{2}$ there exist two open neighborhoods $\OOO_{1}\ni x_{1}$
and $\OOO_{2}\ni x_{2}$ such that $\OOO_{1}\cap\OOO_{2}=\emptyset$.
This property is useful since it allows us to separate any pair of
points by finding neighborhoods that are sufficiently small to not
intersect each other. In a non-Hausdorff manifold, it is possible
that no two such neighborhoods exist for particular pairs of points.

\begin{figure}
\begin{centering}
\includegraphics[width=0.4\paperwidth]{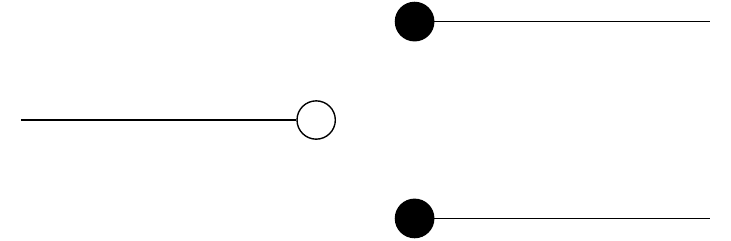}
\par\end{centering}
\caption{\label{fig:Non-Hausdorff}A 1-dimensional manifold which is locally
Euclidean, but not Hausdorff. The line on the left is $\left(-\infty,0\right)$,
and it does not include the point $0$ (the white disk). The two lines
on the right are $\left[0_{1},+\infty_{1}\right)$ and $\left[0_{2},+\infty_{2}\right)$,
and they do include the points $0_{1}$ and $0_{2}$ respectively
(the black disks). Both of these points are in the closure of $\left(-\infty,0\right)$.}
\end{figure}

A well-known example is the \emph{branching line}, which is obtained
by taking the real line $\BBR$, removing the interval $\left[0,+\infty\right)$,
and replacing it with two copies of itself, $\left[0_{1},+\infty_{1}\right)$
and $\left[0_{2},+\infty_{2}\right)$. The topology of this manifold
is Hausdorff everywhere, except for the pair of points $0_{1},0_{2}$.
Indeed, since any possible neighborhood of each of these two points
must contain at least some portion of the interval $\left(-\infty,0\right)$,
one cannot find two neighborhoods which do not intersect. This example
is illustrated in Fig. \ref{fig:Non-Hausdorff}.

\begin{figure}
\begin{centering}
\includegraphics[width=0.4\paperwidth]{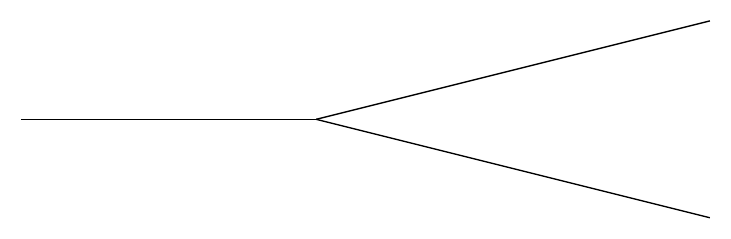}
\par\end{centering}
\caption{\label{fig:Non-Locally-Euclidean}A 1-dimensional manifold which is
Hausdorff, but not locally Euclidean at the point of branching.}
\end{figure}

Another option\textbf{, }discussed by McCabe \cite{McCabe}, is to
use a manifold which is Hausdorff, but not homeomorphic to Euclidean
space at the point of branching. In other words, the manifold is not
\emph{locally Euclidean} at that point. A manifold in the shape of
a ``Y'' is Hausdorff but not locally Euclidean at the point of branching,
since there is no neighborhood of that point which is homeomorphic
to $\BBR$. This example is illustrated in Fig. \ref{fig:Non-Locally-Euclidean}.

It is unclear which option is better for defining branching timelines,
a non-Hausdorff manifold or a non-locally-Euclidean manifold; it might
also turn out that a completely different mathematical structure is
required. In fact, at the time of writing there do not exist any actual
models in the literature which use either type of manifold in order
to resolve paradoxes in any well-defined mathematical way.

Recently, Placek \cite{placek2014branching} developed a detailed
theory of branching spacetime which employs non-Hausdorff (but locally
Euclidean) manifolds in order to take multiple histories into account,
although he did not discuss time travel paradoxes in this context.
Luc \cite{luc2019generalised} (see also \cite{pittphilsci16172}
with Placek) discussed the issue of non-Hausdorff manifolds and concluded
that they can, in fact, be used as fundamental mathematical objects
describing spacetime in general relativity.

The basic notion of using such manifolds to describe branching universes
with multiple timelines seems reasonable, but there are many conceptual
and mathematical issues which need to be resolved first. For example:
\begin{enumerate}
\item What exactly is the physical mechanism which causes the branching?
\item Realistic time machines (unlike those we usually see in science fiction)
don't allow one to jump discretely from one point in spacetime to
another. Instead, as we have seen, the time traveler moves continuously
along a closed timelike curve which doesn't generally have a well-defined
beginning and end. At which point along this curve does the branching
actually happen?
\end{enumerate}
Clearly, much more work is needed in order to construct a well-defined
multiple-timeline solution to the paradoxes, and it seems that this
will inevitably require some major modifications to our current theories
of physics.

\subsubsection{Deutsch's Quantum Time Travel Model}

A consistency paradox occurs when something has to both happen and
not happen at the same time. Classically, this is impossible; however,
in quantum mechanics, we know perfectly well that a system can be
in a superposition of two (or more) different states at once. Perhaps,
then, we could solve paradoxes by invoking this and/or other properties
of quantum mechanics?

It turns out that it's not as easy as just using superposition. However,
Deutsch \cite{Deutsch1991} famously showed that consistency paradoxes
can be avoided if one uses mixed states instead of pure states. Let
us recall what that means. A quantum system is said to be in a \emph{pure
state }if its state, $\left|\psi\right\rangle $, is known exactly.
However, if we only know that it could be in the states $\left|\psi_{i}\right\rangle $
with probabilities $p_{i}$ (where $i$ indexes the possible states),
then the system is said to be in a \emph{mixed state}, which is given
by a \emph{density operator} (or \emph{density matrix}):
\[
\rho\equiv\sum_{i}p_{i}\left|\psi_{i}\right\rangle \left\langle \psi_{i}\right|.
\]
Note that we always have $\tr\left(\rho\right)=1$, since probabilities
sum to 1. However, one can check that a pure state satisfies $\tr\left(\rho^{2}\right)=1$
while a mixed state has $\tr\left(\rho^{2}\right)<1$.

We also need to define the partial trace. Given two physical systems
1 and 2, with density operators $\rho_{1}$ and $\rho_{2}$, we can
describe their \emph{joint state} as another density operator $\rho_{12}$.
Then we define the \emph{reduced density operator }for system 1 by
\[
\rho_{1}\equiv\tr_{2}\left(\rho_{12}\right),
\]
where $\tr_{2}$ is the \emph{partial trace }with respect to system
2. In other words, we ``trace out'' the information about system
2, which means we are left with system 1 alone. Let $\left|A_{1}\right\rangle $,
$\left|B_{1}\right\rangle $ be states of system 1 and $\left|A_{2}\right\rangle $,
$\left|B_{2}\right\rangle $ states of system 2. Then we define the
partial trace by
\begin{equation}
\tr_{2}\left(\left|A_{1}\right\rangle \left\langle B_{1}\right|\otimes\left|A_{2}\right\rangle \left\langle B_{2}\right|\right)=\left|A_{1}\right\rangle \left\langle B_{1}\right|\langle B_{2}|A_{2}\rangle.\label{eq:partial-trace}
\end{equation}
Since $\langle B_{2}|A_{2}\rangle=\tr\left(\left|A_{2}\right\rangle \left\langle B_{2}\right|\right)$
is just a number, we are left with a density matrix for system 1 alone.
By demanding that $\tr_{2}$ is linear, we are able to calculate the
partial trace of any joint mixed state by looking at it term by term.

Now, consider the following paradox given by Deutsch. First of all,
a \emph{qubit} is a generalization of a classical bit, which is described
by a superposition of two basis states: $\left|0\right\rangle $ and
$\left|1\right\rangle $. This paradox involves a gate such that two
qubits enter it and two leave it. Since this system involves time
travel, this might get a bit confusing, so let us label the two input
qubits and two output qubits as follows:
\begin{itemize}
\item $I_{1}$ = first input qubit,
\item $I_{2}$ = second input qubit,
\item $O_{1}$ = first output qubit ($I_{1}$ after undergoing a transformation),
\item $O_{2}$ = second output qubit ($I_{2}$ after undergoing a transformation).
\end{itemize}
$I_{1}$ is the only part we have control of -- it is the qubit sent
by the experimenter into the gate, and we assume it's in a pure state
$\left|x\right\rangle $, where $x\in\left\{ 0,1\right\} $. $I_{2}$
is another qubit, in the pure state $\left|y\right\rangle $, where
$y\in\left\{ 0,1\right\} $. Both $I_{1}$ and $I_{2}$ pass through
the gate, which imposes the following interaction:
\[
I_{1}\otimes I_{2}\soo O_{1}\otimes O_{2},
\]
given by
\[
\left|x\right\rangle \otimes\left|y\right\rangle \soo\left|1-y\right\rangle \otimes\left|x\right\rangle .
\]
Now, to create a paradox, we assume that after passing through the
gate, $O_{1}$ goes through a time machine -- and \textbf{becomes
}$I_{2}$:
\[
O_{1}=I_{2}.
\]
In other words, the experimenter's qubit $I_{1}$ goes through the
gate, leaves as $O_{1}$, and then travels back in time as $I_{2}$
and interacts with itself. After going through the gate once more,
it becomes $O_{2}$. Exactly one qubit, $I_{1}$, enters the lab and
exactly one qubit, $O_{2}$, leaves the lab. By looking at the interaction,
we see that this means we must have
\[
\left|y\right\rangle =\left|1-y\right\rangle .
\]
This is obviously inconsistent, and it is something we have no control
of, since it does not depend on the initial condition $\left|x\right\rangle $.
Therefore, we have created a consistency paradox.

To resolve this paradox, let us now generalize the system by allowing
the qubits to be described by mixed states. We assume that $I_{1}$
is initially in a pure state $\left|\psi\right\rangle $, which is
some superposition\footnote{That is, $\left|\psi\right\rangle =\alpha\left|0\right\rangle +\beta\left|1\right\rangle $
where $\left|\alpha\right|^{2}+\left|\beta\right|^{2}=1$, but the
values of $\alpha$ and $\beta$ are irrelevant to the discussion.} of $\left|0\right\rangle $ and $\left|1\right\rangle $. Then its
density operator is $\left|\psi\right\rangle \left\langle \psi\right|$.
Let $I_{2}$ be described by a mixed state with some density operator
$\rho$. Then, at the gate entrance, the two qubits $I_{1}$ and $I_{2}$
have the joint density operator
\begin{equation}
\rho_{\textrm{input}}=\left|\psi\right\rangle \left\langle \psi\right|\otimes\rho.\label{eq:rho-in}
\end{equation}
The gate is described by the following unitary operator (check this!):
\[
U\equiv\sum_{x,y\in\left\{ 0,1\right\} }\left|1-y\right\rangle \otimes\left|x\right\rangle \left\langle x\right|\otimes\left\langle y\right|,
\]
such that the two qubits $O_{1}$ and $O_{2}$ leave the gate with
the density operator
\[
\rho_{\textrm{output}}=U\left(\left|\psi\right\rangle \left\langle \psi\right|\otimes\rho\right)U^{\dagger}.
\]
Now, we want the density operator of $O_{1}$ to be the same as that
of $I_{2}$. Let us therefore take the partial trace of $\rho_{\textrm{output}}$
with respect to $O_{2}$, which then leaves us with the density operator
of $O_{1}$ alone, and demand that it is the same as $\rho$, the
density operator of $I_{2}$:
\begin{equation}
\rho=\tr_{2}\left(U\left(\left|\psi\right\rangle \left\langle \psi\right|\otimes\rho\right)U^{\dagger}\right).\label{eq:eom}
\end{equation}
The reader should use (\ref{eq:partial-trace}) to calculate the partial
trace. It turns out that there is a family of solutions to this equation
(i.e. fixed points of the operator on the right), given by:
\[
\rho=\hf\left(1+\lambda\left(\left|0\right\rangle \left\langle 1\right|+\left|1\right\rangle \left\langle 0\right|\right)\right)\sp\lambda\in\left[0,1\right].
\]
Therefore, regardless of the initial state $\left|\psi\right\rangle $,
there is always a consistent solution to the evolution of the system,
and we have avoided a paradox. Furthermore, Deutsch proves that there
is always a solution for \textbf{any }$U$, which makes this solution
apply to any other paradoxes one could define in this particular setting.

How is all this related to multiple timelines? For that, we need to
employ the \emph{Everett }or \emph{``many-worlds'' interpretation
}of quantum mechanics \cite{saunders2010many}. Consider a qubit in
the state $\alpha\left|0\right\rangle +\beta\left|1\right\rangle $
and an observer. After the observer measures the qubit, the joint
state of the qubit and the observer is
\[
\alpha\left|0\right\rangle \otimes\left|\textrm{Observer measured "0"}\right\rangle +\beta\left|1\right\rangle \otimes\left|\textrm{Observer measured "1"}\right\rangle .
\]
Without going into details, we can interpret this as if the observer
(and eventually, after interacting with the environment, the entire
universe) ``branched'' into two independent (and non-interacting)
histories or timelines, one in which the measurement yielded 0 and
another in which it yielded 1. Usually, it is impossible to travel
between these timelines. However, in Deutsch's model, one can interpret
the CTC as connecting one timeline to another and allowing the qubit
to travel between them. This provides a natural mechanism for creating
multiple timelines \cite{deutsch1994quantum}\footnote{Note that we secretly assumed the initial state (\ref{eq:rho-in})
is a tensor product of the states of $I_{1}$ and $I_{2}$. This means
that those two states are not correlated, and this is only possible
if $I_{1}$ and $I_{2}$ are completely independent of each other.
Hence, $I_{2}$ must have come from a different timeline.}.

There are several problems with Deutsch's solution, as pointed out
by Deutsch himself. One of the main problems is that, while unitarity
and linearity are two crucial properties of quantum mechanics, the
equation of motion (\ref{eq:eom}) is neither unitary (due to the
partial trace) nor linear (since we are only allowing the specific
states who solve this equation to exist). This has some strange effects,
such as allowing both classical and quantum computers to do incredibly
powerful computations very easily (as shown by Aaronson and Watrous
\cite{aaronson2008closed}), allowing the violation of the no-cloning
theorem, and so on. Accepting this model as a solution to time travel
paradoxes thus requires modifying quantum mechanics in a very significant
way, at least in the vicinity of CTCs\footnote{Note that, as Deutsch himself mentions in his paper, using this model
does not actually allow us to avoid the mathematical issues regarding
branching spacetime manifolds discussed in the previous subsection.}.

Moreover, while (\ref{eq:eom}) always has a solution, it does not,
in general, have a \textbf{unique }solution, and it is thus unclear
which one of the solutions will be realized in nature. Indeed, in
the case described above there is a solution for any value of $\lambda\in\left[0,1\right]$.
This is a version of the bootstrap paradox, since the information
about the value of $\lambda$ seems to be created out of nothing and
is completely independent of anything else in the system. Deutsch
suggests that the state with the maximal entropy should be the unique
solution (in this case it is the maximally mixed state, corresponding
to $\lambda=0$) but it is unclear why this must be so.

In addition, one may wonder if Deutsch's solution works for general
objects that are not test particles, that is, objects which have enough
mass or energy to noticeably influence the curvature of spacetime.
This would allow, for example, ``turning off'' the time machine
in one of the timelines, which seems to break the model. It seems
that a theory of quantum gravity, which should allow the geometry
of spacetime itself to be in a superposition, would be necessary for
Deutsch's model to work in this case.

\subsection{The Novikov Self-Consistency Conjecture}

\subsubsection{Classical Treatment}

The \emph{Novikov self-consistency conjecture} \cite{Novikov} states
that ``the only solutions to the laws of physics that can occur locally
in the real universe are those which are globally self-consistent''.
Alternatively, it may be understood as the statement that the universe
has only one history (or timeline), and it must be consistent, no
matter what.

For example, the time traveler who tries to prevent his or her grandparents
from meeting will simply find that the universe has conspired to make
him or her automatically fail in this task, since anything else would
lead to an inconsistency. In fact, the time traveler \textbf{already
}tried it and \textbf{already} failed... Perhaps even his or her attempt
at meddling is what \textbf{caused }their grandparents to meet in
the first place\footnote{This is a common trope in science fiction.}!

A simple toy model, known as \emph{Polchinski's paradox}, is given
by a billiard ball which goes into a time machine, travels to the
past, and knocks its past self out of its original path -- thus preventing
itself from going into the time machine in the first place. Although
this seems like a paradox, Echeverria, Klinkhammer, and Thorne \cite{EcheverriaKlinkhammerThorne}
found that it has self-consistent solutions, where the ball collides
with itself and knocks itself into the time machine in just the right
angle to product the former collision\footnote{In fact, there is an \textbf{infinite }number of self-consistent solutions
for each initial condition.}.

Note that such solutions cannot exist for the uranium lump paradox,
since it will always explode independently of the trajectories or
angles of collision. Nevertheless, Novikov \cite{novikov1992time}
described a similar problem with an exploding ball and showed that
there could still be a self-consistent evolution by keeping track
of the fragments of the ball after the explosion. It was also conjectured
by Frolov, Novikov, and others \cite{Carlini:1995st} that the principle
of self-consistency could be a consequence of the \emph{principle
of least action}, and this was demonstrated using the billiard ball
scenario. Rama and Sen \cite{KalyanaRama:1994ag} attempted to generalize
the billiard ball model in order to create a paradox that cannot be
solved; however, Krasnikov \cite{krasnikov1997causality} showed that
their model does in fact have self-consistent solutions.

From these results, one may deduce that the there might be self-consistent
solution for any possible system. However, this turns out not to be
the case; it is possible to construct paradoxical scenarios for which
there are no known self-consistent solutions. One such example is
given by Krasnikov \cite{krasnikov2002time}. It generalizes the billiard
ball scenario, in 1+1 dimensions, by considering the trajectories
of particles which have specific properties and follow specific rules
of interactions. This is an example of a \textbf{true }paradox, to
which the Novikov conjecture does not seem to apply, as far as we
know.

Still, the Novikov conjecture seems to make sense. Indeed, if the
Hawking conjecture is incorrect and time travel exists, and if there
is only one possible timeline, then it's hard to see any other way
to avoid time travel paradoxes. However, if it is true, it may lead
to very peculiar scenarios.

For example, let us assume that the Novikov conjecture is correct.
Then, if I travel 10 years into the past and try really hard to create
a paradox by killing my past self, I will find that my murder attempts
always fail for some reason. For instance, if I use a gun, then I
will inevitably find that it gets jammed, or that the bullet misses
the target. At first, I might consider this a temporary fluke. However,
over the next 10 years, every single day, I try to kill my past self
100 times per day. If every single one of those 365,000 assassination
attempts fails, no matter what the circumstances are (weapon used,
location, time of day, etc.), this might start to seem just a tiny
bit ridiculous!

In fact, in this class of thought experiments, the number of failed
attempts may be made arbitrarily large. Consider some measurement
which may be performed in a very short time and has two outcomes of
equal probability -- e.g., the measurement of a qubit prepared in
the quantum\footnote{A quantum measurement is preferred for this thought experiment, since
unlike a classical coin toss, which is deterministic, the result of
a measurement of spin is (as far as we know) truly random.} state $\left(\left|0\right\rangle +\left|1\right\rangle \right)/\sqrt{2}$.
If we measure $\left|0\right\rangle $ nothing happens, but if we
measure $\left|1\right\rangle $, something happens which creates
a paradox. If the Novikov conjecture is true, then we will find that
the outcome of the measurement is \textbf{always }$\left|0\right\rangle $.
We repeat this process over and over, producing $N$ outcomes of $\left|0\right\rangle $
and exactly zero outcomes of $\left|1\right\rangle $. As $N\to\infty$,
the probability for this sequence of outcomes approaches zero.

The fact that the Novikov conjecture implies bizarre situations such
as the ones we described here does not mean that it is wrong; indeed,
we have seen above that Deutsch's multiple-timeline model has very
weird implications of its own. However, our discussion does imply
that the Novikov conjecture would require modifying quantum mechanics
such that it works differently when time travel is involved, similar
to Deutsch's modification of quantum mechanics in the case of multiple
timelines. Let us now see one example of such a modification.

\subsubsection{The Postselected Quantum Time Travel Model\protect\footnote{The author would like to thank Daniel Gottesman and Aephraim Steinberg
for discussions which proved helpful in writing this section.}}

Deutsch's model represents one possible way of making quantum mechanics
consistent in the presence of CTCs. In addition to his model, which
is referred to as \emph{D-CTCs}, there is another model called ``postselected''
CTCs or \emph{P-CTCs}, described by Svetlichny \cite{svetlichny2011time}
and by Lloyd et al. \cite{lloyd2011closed,lloyd2011quantum}. In this
model, one \textbf{simulates }time travel using a \emph{quantum teleportation
}protocol. Let us recall how that works \cite{nielsen2002quantum}.

At time $t=0$, a \emph{Bell state} (or \emph{EPR pair}) of entangled
qubits is created:
\[
\left|\beta_{00}\right\rangle \equiv\frac{1}{\sqrt{2}}\left(\left|0\right\rangle \otimes\left|0\right\rangle +\left|1\right\rangle \otimes\left|1\right\rangle \right)\equiv\frac{1}{\sqrt{2}}\left(\left|00\right\rangle +\left|11\right\rangle \right),
\]
where we used the shorthand notation $\left|xy\right\rangle \equiv\left|x\right\rangle \otimes\left|y\right\rangle $.
Alice takes the first qubit, Bob takes the second, and they go their
separate ways. Later, at time $t=1$, Alice receives some qubit $\left|\psi\right\rangle $
which she knows nothing about, and she needs to transfer it to Bob
using only classical channels.

Since Alice only has one copy of this qubit, and cloning a quantum
state is forbidden by the no-cloning theorem, she can't learn anything
about the qubit without destroying it. Furthermore, even if she knew
the precise state of the qubit, she wouldn't be able to relay that
information classically to Bob, since a qubit is described (up to
overall phase) by a complex number, and a general complex number is
described by an infinite number of classical bits.

However, Alice can use the fact that she has a qubit which is entangled
with Bob's. If
\[
\left|\psi\right\rangle =\alpha\left|0\right\rangle +\beta\left|1\right\rangle ,
\]
then all three qubits are represented together by the state
\[
\left|\Psi\right\rangle \equiv\left|\psi\right\rangle \otimes\left|\beta_{00}\right\rangle =\frac{1}{\sqrt{2}}\left(\alpha\left|0\right\rangle +\beta\left|1\right\rangle \right)\otimes\left(\left|00\right\rangle +\left|11\right\rangle \right),
\]
where the first qubit is $\left|\psi\right\rangle $, the second is
Alice's, and the third is Bob's. Alice now sends both qubits through
a CNOT gate and the first qubit through a Hadamard gate (for our purposes
it doesn't matter what these gates do exactly). After passing through
these gates, the state of the three qubits becomes
\begin{equation}
\left|\Psi\right\rangle \mt\hf\left(\left|00\right\rangle \left(\alpha\left|0\right\rangle +\beta\left|1\right\rangle \right)+\left|01\right\rangle \left(\alpha\left|1\right\rangle +\beta\left|0\right\rangle \right)+\left|10\right\rangle \left(\alpha\left|0\right\rangle -\beta\left|1\right\rangle \right)+\left|11\right\rangle \left(\alpha\left|1\right\rangle -\beta\left|0\right\rangle \right)\right).\label{eq:teleport}
\end{equation}
Now Alice performs a measurement on her two qubits, and she will obtain
one of four results: 00, 01, 10, or 11. These are two classical bits,
which she can send to Bob. With this information, Bob can read from
(\ref{eq:teleport}) exactly which operations he has to perform in
order to obtain $\left|\psi\right\rangle $ from the qubit he has.

In particular, if Bob receives the bits 00, then we can see from (\ref{eq:teleport})
that he \textbf{already had }the qubit $\left|\psi\right\rangle =\alpha\left|0\right\rangle +\beta\left|1\right\rangle $
at time $t=0$, even \textbf{before }Alice obtained it at time $t=1$!
The idea of P-CTCs is that we \emph{postselect }the 25\% of the cases
where the classical bits happened to be 00, ignoring the other cases.
While this is not actually time travel, it can be used to \textbf{simulate
}time travel; in the case where the bits are 00, the state Bob has
in the past is the same state Alice will get in the future, so it's
as if Alice's state ``went back in time'' through a time machine.
Furthermore, one may now conjecture that any time machine behaves,
in a sense, like a quantum teleportation circuit which automatically
projects the result of Alice's measurement on 00.

Both D-CTCs and P-CTCs suggest different ways in which quantum mechanics
may be modified\textbf{ }in the presence of CTCs, and they are both
incompatible with the standard formulation of quantum mechanics, for
example because the evolution is non-unitary -- in D-CTCs due to
the partial trace and in P-CTCs due to the projection, both of which
happen at the CTC itself. One of them could be true, or neither of
them could be true; probably the only way to know is to actually build
a time machine.

While D-CTCs deal with paradoxes using multiple timelines, where the
different timelines correspond to different ``worlds'' in the ``many-worlds''
interpretation, P-CTCs deal with paradoxes by assuming Novikov's self-consistency
conjecture. This can be described in a simplistic way as follows.
The grandfather paradox can be modeled by having the state $\left|\psi\right\rangle $
enter the time machine as $\left|1\right\rangle $, which represents
the time traveler being alive, and, at some point between $t=0$ and
$t=1$, turning it into $\left|0\right\rangle $, which represents
the time traveler being dead. In this case, there is no measurement
where Bob's state $\left|0\right\rangle $ is the same as Alice's
state $\left|1\right\rangle $; the probability for the measurement
to yield 00 is exactly zero.

Since the probability is zero, postselection doesn't work -- you
can't postselect an event with zero probability\footnote{Recall that the \emph{conditional probability }of an event $A$ given
an event $B$ is defined as
\[
P\left(A|B\right)\equiv\frac{P\left(A\cap B\right)}{P\left(B\right)}.
\]
If $P\left(B\right)=0$, this expression is undefined, and therefore
we are not allowed to postselect for $B$; this will make all other
probabilities in our system undefined.}. However, one can then add small perturbations (or quantum fluctuations)
to the system. This will make the probability non-zero, even if still
extremely small. Since the CTC projects the state on 00, even an extremely
small probability for the state we are projecting on suddenly becomes
probability 1! This is exactly the idea behind Novikov's conjecture
-- events with extremely small probabilities, such as my gun getting
jammed every single time I try to kill my past self and create a paradox,
somehow get selected over events with much higher probability. Of
course, all this relies on time machines actually working in the same
way as P-CTCs do, which is something that at the moment we have no
reason to believe. Still, this provides a possible mechanism for Novikov's
conjecture to be realized in nature.

\subsection{Conclusions}

As we have seen, the possibility of time travel has, thus far, not
been ruled out by known physics. If Hawking's conjecture is wrong,
and time travel is indeed possible, then it seems that paradoxes inevitably
occur. Given that we want reality to make sense, there must be a way
in which the universe avoids these paradoxes.

If our universe can have multiple timelines, then we must radically
modify our laws of physics to accommodate this. Perhaps this modification
can be made at the classical level, by replacing the spacetime manifolds
we know and love with something more complicated, such as branching
manifolds. If the ``many-worlds'' interpretation is true, it could
provide a natural way for additional timelines to be created, using
a modification of quantum mechanical such as the D-CTC model.

If there is only one timeline in our universe, then it seems that
the Novikov conjecture must be true (or at least, we have not come
up with any alternatives so far). Perhaps self-consistent solutions
can always be found for any conceivable physical system at the classical
level, although given Krasnikov's unsolvable paradox, this would probably
require some modification of our laws of physics as well. A modification
of quantum mechanics, as suggested in the P-CTC model, could provide
a mechanism for events with infinitesimal probability to nonetheless
always occur if that is required for the system's consistency.

Hawking famously held a party for time-travelers, to which no one
showed up, and sent out invitations only after the party was over.
He jokingly referred to this as ``experimental evidence'' for his
conjecture. However, it is interesting to note that the alternatives
we discussed above may also accommodate this experimental result.
Novikov's conjecture would simply prevent the time travelers from
attending in order to preserve consistency (maybe they all inevitably
get stuck in traffic?), while multiple timelines will result in the
party being full of time travelers, but only in a timeline different
from the one we currently live in.

\section{Further Reading}

For more information about faster-than-light travel and time travel,
the reader is referred to the books by Visser \cite{Visser}, Lobo
\cite{Lobo} and Krasnikov \cite{Krasnikov}. For a discussion of
time travel paradoxes from the philosophical point of view, see the
book by Wasserman \cite{wasserman2018paradoxes}. For qualitative
discussions at the popular-science level, see the book by Everett
and Roman \cite{EverettRoman} and the three books by Nahin \cite{nahin2001time,nahin2011time,nahin2016time}.

\section{Acknowledgments}

The author would like to thank the students who attended this course
for asking interesting questions and providing insightful comments
which helped improve and perfect these notes. The author would also
like to thank the referee Aron Wall for helpful comments and suggestions.

Research at Perimeter Institute is supported in part by the Government
of Canada through the Department of Innovation, Science and Economic
Development Canada and by the Province of Ontario through the Ministry
of Economic Development, Job Creation and Trade.

\bibliographystyle{Utphys}
\phantomsection\addcontentsline{toc}{section}{\refname}\bibliography{FTL_TT}

\end{document}